\documentclass[journal, two column]{IEEEtran}

\UseRawInputEncoding

\usepackage{amsmath,amssymb}
\usepackage[dvips]{graphicx}
\usepackage{amsfonts}
\usepackage[mathscr]{eucal}
\usepackage{latexsym}
\usepackage{amsthm}
\usepackage{exscale}
\usepackage[mathscr]{eucal}
\usepackage{bm}
\usepackage[dvipsnames]{color}
\usepackage{cases}
\usepackage{color}
\usepackage{epsfig}
\usepackage{algorithm}
\usepackage{algorithmic}
\usepackage[verbose,nospace,sort]{cite}
\usepackage{tabularx}
\usepackage{multirow}
\usepackage{multicol}
\usepackage{balance}
\usepackage{caption}
\usepackage{subcaption}
\usepackage{setspace}
\usepackage{url}
\usepackage{enumitem}

\graphicspath{{./figs/}}

\usepackage[acronym]{glossaries}
\newcommand{\newac}{\newacronym}
\newcommand{\ac}{\gls}
\newcommand{\Ac}{\Gls}
\newcommand{\acpl}{\glspl}

\makeglossaries

\newac{uav}{UAVs}{unmanned aerial vehicles}
\newac{mimo}{MIMO}{multiple-input multiple-output}
\newac{xlmimo}{XL-MIMO}{extremely large-scale MIMO}
\newac{irs}{IRS}{intelligent reflecting surface}
\newac{ris}{RIS}{reconfigurable intelligent surface}
\newac{ofdm}{OFDM}{orthogonal frequency-division multiplexing}
\newac{otfs}{OTFS}{orthogonal time frequency space}
\newac{ici}{ICI}{inter-carrier interference}
\newac{ddam}{DDAM}{delay-Doppler alignment modulation}
\newac{ilsac}{ILSAC}{integrated localization, sensing, and communication}
\newac{dfrc}{DFRC}{dual functional radar and communication}
\newac{iot}{IoT}{internet-of-things}
\newac{bs}{BS}{base station}
\newac{ue}{UE}{user equipment}
\newac{adc}{ADC}{analog-to-digital converter}
\newac{music}{MUSIC}{multiple signal classification}
\newac{ml}{ML}{machine learning}
\newac{tdd}{TDD}{time division duplexing}
\newac{fdd}{FDD}{frequency division duplexing}
\newac{dft}{DFT}{discrete Fourier transform}
\newac{fmcw}{FMCW}{frequency-modulated continuous wave}
\newac{rsrq}{RSRQ}{reference signal received quality}
\newac{cqi}{CQI}{channel quality indicator}
\newac{sinr}{SINR}{signal to interference and noise ratio}
\newac{cam}{CAM}{channel angle map}
\newac{esprit}{ESPRIT}{estimation of signal parameters via rotational invariance techniques}
\newac{gnss}{GNSS}{global navigation satellite system}
\newac{toa}{ToA}{time of arrival}
\newac{tdoa}{TDoA}{time difference of arrival}
\newac{rss}{RSS}{received signal strength}
\newac{pu}{PU}{primary user}
\newac{su}{SU}{secondary user}
\newac{lsa}{LSA}{licensed shared access}
\newac{rkhs}{RKHS}{reproducing kernel Hilbert space}
\newac{lmmse}{LMMSE}{linear minimum mean-square error}
\newac{psd}{PSD}{power spectrum density}
\newac{wlog}{W.l.o.g.}{Without loss of generality}
\newac{dnn}{DNN}{deep neural network}
\newac{cnn}{CNN}{convolutional neural network}
\newac{los}{LoS}{line-of-sight}
\newac{nlos}{NLoS}{none-line-of-sight}
\newac{mse}{MSE}{mean squared error}
\newac{mae}{MAE}{mean absolute error}
\newac{aoa}{AoA}{angle of arrival}
\newac{aod}{AoD}{angle of departure}
\newac{lte}{LTE}{long term evolution}

\begin{document}
\title{A Tutorial on Environment-Aware Communications via Channel Knowledge Map for 6G}
\author{Yong Zeng,~\IEEEmembership{Senior Member, IEEE}, Junting Chen,~\IEEEmembership{Member, IEEE}, Jie Xu,~\IEEEmembership{Senior Member, IEEE}, \\Di Wu,~\IEEEmembership{Student Member, IEEE}, Xiaoli Xu,~\IEEEmembership{Member, IEEE}, Shi Jin,~\IEEEmembership{Fellow, IEEE}, \\Xiqi Gao,~\IEEEmembership{Fellow, IEEE},  David Gesbert,~\IEEEmembership{Fellow, IEEE}, Shuguang Cui,~\IEEEmembership{Fellow, IEEE}, and Rui Zhang,~\IEEEmembership{Fellow, IEEE} \\
\thanks{Y. Zeng, D. Wu, X. Xu, S. Jin, and X. Gao are with the National Mobile Communications Research Laboratory, Southeast University, Nanjing 210096, China. Y. Zeng and S. Jin are also with Frontiers Science Center for Mobile Information Communication and Security, Southeast University. Y. Zeng and X. Gao are also with Purple Mountain Laboratories, Nanjing 211111, China (e-mail: \{yong\_zeng, xiaolixu, studywudi, shijin, xqgao\}@seu.edu.cn). (\emph{Corresponding author: Yong Zeng.})

J. Chen, J. Xu, and S. Cui are with the School of Science and Engineering and Future Network of Intelligence Institute (FNii), The Chinese University of Hong Kong, Shenzhen, Shenzhen 518172, China. S. Cui is also with Peng Cheng Laboratory, Shenzhen 518066, China. (email: \{juntingc, xujie, shuguangcui\}@cuhk.edu.cn).

D. Gesbert is with the Communication Systems Department, EURECOM, France (email: david.gesbert@eurecom.fr).

R. Zhang is with School of Science and Engineering, Shenzhen Research Institute of Big Data, The Chinese University of Hong Kong, Shenzhen, Shenzhen 518172, China (e-mail: rzhang@cuhk.edu.cn). He is also with the Department of Electrical and Computer Engineering, National University of Singapore, Singapore 117583.

This work was supported  in part by the National Key R\&D Program of China with Grant number 2019YFB1803400; in part by the National Natural Science Foundation of China under Grant 62071114 and 6504009712; in part by the Fundamental Research Funds for the Central Universities 2242022k60004; 
in part by the National Natural Science Foundation of China under grants No. U2001208 and No. 62171398, by Guangdong Research Project No. 2019QN01X895; 
in part by the Jiangsu Province Basic Research Project under Grant BK20192002, and the Fundamental Research Funds for the Central Universities under Grant 2242022k60007;
in part by NSFC with Grant No. 62293482, the National Key R\&D Program of China with grant No. 2018YFB1800800, the Shenzhen Outstanding Talents Training Fund 202002, the Guangdong Research Projects No. 2017ZT07X152 and No. 2019CX01X104, the Guangdong Provincial Key Laboratory of Future Networks of Intelligence (Grant No. 2022B1212010001), the Shenzhen Key Laboratory of Big Data and Artificial Intelligence (Grant No. ZDSYS201707251409055), and the Key Area R\&D Program of Guangdong Province with grant No. 2018B030338001.

}
}
\maketitle

\begin{abstract}
Sixth-generation (6G) mobile communication networks are expected to have dense infrastructures, {large antenna size, wide bandwidth}, cost-effective hardware, diversified positioning methods, and enhanced intelligence. Such trends bring both new challenges and opportunities for the practical design of 6G. On one hand, acquiring channel state information (CSI) in real time for all wireless links becomes quite challenging in 6G. On the other hand, there would be numerous data sources in 6G containing high-quality location-tagged channel data, {e.g., the estimated channels or beams between base station (BS) and user equipment (UE),} making it possible to better learn the local wireless environment. {By exploiting this new opportunity} and for tackling the CSI acquisition challenge, there is a promising paradigm shift from the conventional environment-unaware communications to the new environment-aware communications based on the novel approach of channel knowledge map (CKM). This article aims to provide a comprehensive overview on environment-aware communications enabled by CKM to fully harness its benefits for 6G. {First, the basic concept of CKM is presented, followed by the comparison of CKM with various existing channel inference techniques.} Next, the main techniques for CKM construction are discussed, including both {environment model-free and environment model-assisted} approaches. Furthermore, a general framework is presented for the utilization of CKM to achieve environment-aware communications, followed by some typical CKM-aided communication scenarios. Finally, important open problems in CKM research are highlighted and potential solutions are discussed to inspire future work.
\end{abstract}


\section{Introduction}
\subsection{6G KPIs and Potential Technologies}
Sixth-generation (6G) mobile communication networks are expected to be ready for commercial deployment around 2030. Driven by various emerging applications such as fully-autonomous vehicles, network-connected robots, and mixed reality, which are difficult to be fully supported by the current fifth-generation (5G) wireless networks, wireless researchers have envisioned various 6G blueprints, such as ``ubiquitous wireless intelligence''\cite{1205}, ``global coverage, all spectra, full applications, strong security''\cite{3021}, and ``from connected people and things to connected intelligence'' \cite{3022}. Recently,  International Telecommunication Union (ITU)  completed a draft new recommendation on framework and overall objectives of the future development of international mobile telecommunications (IMT) for 2030 and beyond, which specifies six  usage scenarios, including immersive communication, massive communication, hyper reliable \& low-latency communication (HRLLC), ubiquitous connectivity, integrated artificial intelligence (AI) and communication, and integrated sensing and communication (ISAC) \cite{4044}. Though the standard for 6G is yet to be developed, there is more or less a consensus on its key performance indicators (KPIs), such as a connection density of at least $10$ million devices per km$^2$ \cite{3021,3022,3025,4044}, a peak data rate of Tera bits per second (Tbps) \cite{1205,3021,3023}, latency of less than 1 ms \cite{1205,3021,3023,3025}, energy efficiency of at least 10 times that of 5G networks \cite{1205,3021,3023,3025}, and positioning accuracy of 1-10 centimeter (cm) \cite{1205,3025,4044}.

To fulfill the above ambitious KPIs, various key technologies have been identified and studied from different aspects. At the network level,  cell-free massive \ac{mimo} has been proposed as a promising architecture towards user-centric networks \cite{3027,4020,4021}, which blurs the traditional cell boundaries and may provide uniformly good services everywhere. Besides, 6G is also expected to seamlessly integrate with non-terrestrial networks (NTNs), which may not only extend the network coverage to the three-dimensional (3D)  space for serving aerial users such as \ac{uav}, but also provide wireless connectivity from the sky via \ac{uav} or satellite mounted \acpl{bs}/relays \cite{1095}. This may greatly reduce the cost to achieve truly global coverage, and also enhance the network resilience to {natural calamities} such as earthquakes and floods.

 At the radio access level,  {the research has advanced in the areas of new spectrum, new MIMO, as well as new modulations and waveforms over the past few years.} For example, {to use the wide} bandwidth and realize super temporal resolution that are necessary to achieve ultra-high data rate and positioning/sensing accuracy, above-6 GHz centimeter wave (cmWave), millimeter wave (mmWave) \cite{569}, and Terahertz \cite{3028} technologies are being extensively studied. In the meantime, with massive MIMO successfully implemented in 5G \cite{3029,3030}, MIMO technologies have been further developed following different pathways. For example, by dramatically increasing the array size beyond massive MIMO, \ac{xlmimo} \cite{3032,4022,4023,4024} or extremely large-scale aperture array (ELAA) \cite{3029} is expected to achieve super spatial resolution, beamforming gain, and ultra-high spectral efficiency.  On the other hand, instead of using the conventional discrete antenna arrays, the alternative continuous surfaces such as large intelligent surfaces (LIS) \cite{3033} or holographic MIMO \cite{3034} are being studied for 6G. Moreover, unlike conventional MIMO technologies that mainly optimize  the transmitters or receivers, \ac{ris} \cite{3037} or \ac{irs} \cite{3035,3036} are regarded as the cost-effective and energy-efficient technologies to dynamically reconfigure the wireless propagation environment.

 Besides new spectrum and MIMO evolution, novel modulations and waveforms beyond the dominating \ac{ofdm} technologies are also studied. For example, to address the \ac{ici} issue suffered by \ac{ofdm} in high mobility scenarios, \ac{otfs} modulation has been recently proposed to modulate information in the delay-Doppler domain \cite{4019}. Besides, by exploiting the super high spatial resolution of large antenna arrays and multi-path sparsity of mmWave and Terahertz channels, a novel \ac{ddam} technique was recently proposed \cite{3039,4016}, which is able to resolve the inter-symbol interference (ISI) issue or manipulate the channel delay spread using path-based beamforming and delay/Doppler alignment, without relying on conventional channel equalization or multi-carrier transmission.

 Besides, 6G is also expected to achieve the harmonic integration of the conventional active communication with the new passive communication. For example, backscattering and symbiotic radio \cite{3040} technologies enable the passive secondary devices to reuse not only the spectrum, but also the energy of the primary active communications. In return, the scattered signals by the passive devices may create the additional multi-path signal components to enhance the primary communication, which can be exploited  to achieve the mutualism in symbiotic radios. Another aspect of integration towards 6G is the integrated localization, sensing, and communication (ILSAC) \cite{3041,4017,4018}, which integrates the three functionalities to best utilize the network infrastructure, hardware components, radio resources, and signal processing modules, etc. In particular, with ILSAC, the localization and sensing outputs may be exploited to  enhance the communication performance, leading to the paradigms of location-aware communications \cite{3042} or sensing-assisted communications \cite{3043,3045}.

\begin{figure*}
\centering
\includegraphics[scale=0.35]{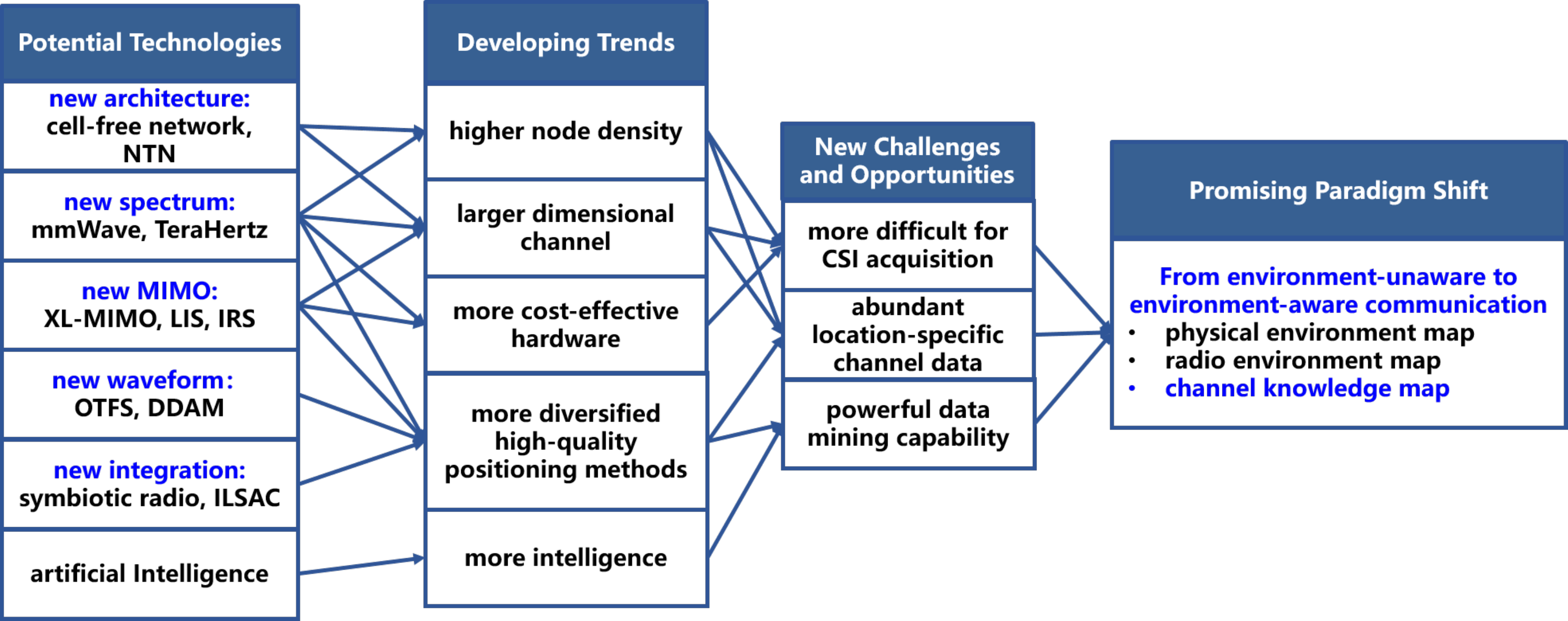}
\caption{6G potential technologies, important trends, and a promising paradigm shift from environment-unaware to  environment-aware communication.}\label{F:6GTechTrendOpp}
\end{figure*}

In addition, 6G is envisioned to be integrated with edge computing \cite{4028} and AI \cite{4029} technologies for achieving distributed computing and network AI. In particular, edge devices such as cellular BSs and \acpl{ue} can be incorporated with cloud-like computing capabilities, such that the rich sensing data generated by these edge devices can be processed swiftly in a distributed manner for acquiring intelligence from them. For instance, UEs and BSs can be coordinated in both edge learning and edge inference, in which these network nodes can use their distributed local data for cooperatively training shared AI models and cooperatively accomplishing AI tasks, respectively \cite{4029}. With such integrated communication and computation paradigm, future 6G will evolve from the conventional data-delivery communication networks to the new task-oriented and semantics-aware communication networks for supporting new AI applications at network edge.

The aforementioned technologies render 6G networks exhibiting several important developing trends, such as denser wireless nodes, {larger antenna size, wider spectrum}, lower-cost hardware,  more diversified  high-quality positioning techniques, and more intelligence,  which are illustrated in Fig.~\ref{F:6GTechTrendOpp} and further elaborated in the following.

\subsection{Important Trends Towards 6G}
\subsubsection{Higher node density}
Compared with 5G, 6G networks are expected to have much higher node density, {in terms of both user and network infrastructure densities.}

From the user side perspective, it is generally agreed that a connection density of at least 10 million devices per km$^2$ needs to be supported by 6G \cite{1205}. Besides, supporting low-altitude aerial users such as flying robots, UAVs, or UAV swarms are likely to become routine operations in 6G, which expands 6G users from the conventional 2D ground surface to 3D space.

From the infrastructure side perspective, as higher frequencies such as above-6GHz cmWave, mmWave and Terahertz bands are used more extensively, cell size is likely to continuously shrink for enabling denser \ac{bs} deployment in 6G.  In fact,  denser network infrastructure is also a key feature of cell-free networking architectures, where massive access points (APs) with density even comparable to user density could be deployed to enable user-centric networks. Besides,  since the emerging passive and semi-passive infrastructures such as  \ac{irs} \cite{3037,3035,3036} or dedicated metal reflectors \cite{3044} involve at least double-hop signal attenuation, {i.e., the signal needs to travel two hops to reach the destination,} they typically need to be densely deployed in order to achieve significant performance enhancement. Similarly, double-hop signal attenuation is also intrinsic for radar sensing, which magnifies the trend of network densification. For example, besides co-sited \ac{dfrc} BSs, additional dedicated radar transmitter and/or receiver may need to be deployed in order to achieve ubiquitous sensing service coverage in 6G \cite{3046}.

\subsubsection{Larger dimensional channel}\label{sec:largerDim}
 The channel dimension of a wireless communication link will dramatically increase with the use of larger antenna arrays and wider bandwidth, as well as joint processing between dense nodes together with higher user mobility in more complex environment. Within each channel coherence block, the wireless channel for a MIMO system with $M_{\mathrm{UE}}$ antennas at the \ac{ue} and  $M_{\mathrm{BS}}$ antennas at the  \ac{bs} can be described by $N_{\mathrm{SC}}$ matrices, each of dimension $M_{\mathrm{UE}}\times M_{\mathrm{BS}}$, where $N_{\mathrm{SC}}$ denotes the number of subcarriers that typically increases with the available bandwidth.

 In the spatial domain, with the evolution from massive MIMO in 5G to XL-MIMO in 6G, $M_{\mathrm{BS}}$ at the BS is expected to increase from today's typical value of $64$ to hundreds or even thousands. A similar increase for $M_{\mathrm{UE}}$ at the UE side is also possible with the use of higher frequency and the advancement of antenna manufacturing technology. In the {frequency domain}, the system bandwidth is expected to increase from hundreds of MHz in 5G to GHz in 6G, with aggregated frequency bands from the conventional sub-6 GHz to mmWave and Terahertz. This may lead to an order-of-magnitude increase in $N_{\mathrm{SC}}$. By combining the joint effects of spatial and bandwidth expansion in 6G, the channel dimension may increase by two or even three orders of magnitude. The situation is exacerbated when higher user mobility and more complex environment are considered, for which channels fluctuate more frequently. This renders it very costly, if not impossible, to estimate the complete channel matrices. %One effective way to mitigate such issues is to exploit the array structure and multi-path sparsity, by estimating a limited number of channel paths for channel reconstruction. However, as the spatial and temporal resolution increases,  the number of resolvable channel paths to be estimated also increases.

\subsubsection{More cost-effective hardware}\label{sec:simplerHardware}
The seemingly conflicting 6G objectives of increasing energy/cost efficiency while using higher frequencies, larger antenna arrays and wider bandwidth make it especially important to develop more cost-effective hardware. As radio frequency (RF) chains are believed to be the most costly and power-hungry modules, two effective ways to reduce hardware and energy cost are using fewer RF chains and using more cost-effective RF components \cite{851}, such as low-resolution \ac{adc}.

Analog beamforming is an effective technique to reduce RF chain cost \cite{574,575,573}, where one RF chain is shared by all array elements and dynamic beamforming is achieved in the analog domain by phase shifters. To achieve spatial multiplexing, the more generic hybrid analog/digital beamforming has been extensively studied \cite{576,578}, where RF chains lesser than the number of antenna elements are used so as to achieve a trade-off between cost and performance. One effective implementation of hybrid beamforming is lens antenna array \cite{823,1026}, by exploiting the direction-dependent energy focusing capability of electromagnetic (EM) lenses. With the similar objective to reduce RF chains, the semi-passive IRS \cite{3037,3035,3036} or the fully passive metal reflector \cite{3044} can be used as an alternative to the conventional active relays by creating virtual \ac{los} links. Similarly, with the integration of passive communication technologies like backscattering or symbiotic radio, RF chain-free communication can be achieved for Internet-of-Things (IoT) devices with low rate requirement. On the other hand, using low-resolution ADC is an alternative approach to reduce the RF chain cost while still maintaining one RF chain for each antenna element \cite{828,824}. One extreme case is to use only one-bit ADC \cite{3047}, i.e., by only keeping the signs of the signals for each I/Q component. As more severe quantization errors are resulted, low-resolution ADCs  would require more advanced signal processing techniques to recover the performance loss.

\subsubsection{More diversified high-quality positioning methods}\label{sec:more_accessible_loc}
Compared with today's cellular networks, the location information of UEs, reflectors, and obstacles will  not only be more readily obtainable in 6G, but also with much higher quality, e.g., in terms of accuracy and {location updating rate}. This will be driven by the advancement of both cellular and non-cellular based localization/sensing technologies.

Cellular-based localization has been an integral service of mobile communication networks since 2G \cite{3041}. Over the past few decades, cellular localization technologies have been tremendously advanced, from the coarse techniques like proximity or \ac{rss}-based positioning to advanced techniques like \ac{toa}/\ac{tdoa}/\ac{aoa} based positioning. It is generally agreed that 6G networks should provide submeter level or even centimeter level localization accuracy. Fortunately, such ambitious goals are possible with the use of XL-MIMO and mmWave/Tertahertz technologies, which provide super resolution in the spatial and temporal domains. Besides, the localization accuracy can be further improved by utilizing advanced signal processing techniques such as super-resolution algorithms, including \ac{music}\cite{4025}, \ac{esprit}. Localization in the challenging \ac{nlos} environment or even 6 dimensional localization (both 3D position and orientation) have also been extensively studied \cite{3049,3048}. In terms of location update rate, millisecond update rate has been envisioned in \cite{3021}.  Besides conventional localization for which the targets to be localized are fully cooperative, localization based on radar sensing has received extensive research attention recently, mostly in the framework of ISAC \cite{4017}.

Besides cellular-based localization and sensing, the multitude of sensory devices equipped by future UEs also facilitate the acquisition of location information. For example, satellite-based localization such as global positioning system (GPS) or Beidou are readily available nowadays. Laser or camera based sensing has been tremendously advanced over the past years and now become essential components of certain UEs such as drones and vehicles. Besides, inertial measurement unit (IMU) such as accelerometer and gyroscope provide alternative ways for location estimation without relying on external signals. All such non-cellular based localization technologies provide effective complementary means to improve the accessibility and quality of location information in future 6G networks.

\subsubsection{More intelligence}
6G is expected to be more intelligent and autonomous than 5G, enabled by AI and other advanced computation and data mining technologies \cite{1205}. In particular, 6G  will offer super capacity with ultra dense device connectivity, rendering it possible to fuse massive amount of data from various sources, including sensors, UEs, and IoT devices. The advancement of AI and machine learning may enable the network to efficiently analyze, interpret, and make sense of such huge amount of data in real-time. Besides,  with the advancement of edge computing\cite{4003}, data can be processed closer to the source, which can significantly reduce the data processing latency. Furthermore, the intelligence of 6G is  expected to be further improved by leveraging  techniques like blockchain \cite{4005} and over-the-air computing \cite{4032}. Thus, with higher data exchange rate, lower data mining latency, and more powerful computation capability, 6G is expected to enable a new era of intrinsic intelligence.

\subsection{New Challenges and Opportunities}
The developing trends discussed above bring both new challenges and opportunities for the practical design  of 6G networks. On one hand, the trend of network densification and larger dimensional channel make it more difficult to acquire real-time CSI for all wireless links, and this issue is even exacerbated with the desire to use simpler hardware for energy/cost-effective implementation. On the other hand, higher node density and larger dimensional channel also imply that there would be more sources of data containing high-quality location-specific channel knowledge that reflects the actual wireless propagation environment. With more diversified localization methods and the powerful data mining capability enabled by machine learning and advanced AI techniques, such location-specific data, with the location tagged properly, could be exploited to enable a paradigm shift from the conventional environment-unaware communications to environment-aware communications. An illustrative diagram for such new challenges and opportunities, together with the promising paradigm shift towards environment-aware communications, is also shown in Fig.~\ref{F:6GTechTrendOpp} and further elaborated as follows.
\subsubsection{More difficult for CSI acquisition} In contemporary wireless networks, acquiring transmitter and receiver-side CSI is essential for achieving high  rate communications reliably by adapting to the time-varying channel conditions. At the transmitter side, CSI is usually needed for beamforming/precoding design, interference avoidance, adaptive modulation and coding, power control, channel assignment, etc. At the receiver side, CSI is typically needed for tasks such as receive beamforming, interference cancellation, demodulation, etc. 

One major approach for CSI acquisition is pilot-based channel training, which estimates the channels in the spatial, frequency, and time domain via sending pilot sequences {\it a priori} known to both transmitter and receiver. {For example, for a frequency-flat MIMO channel with $M_r$ receive antennas and $M_t$ transmit antennas, to avoid solving an under-determined linear system for channel estimation, the number of signal measurements at the receiver should be no smaller than the number of channel coefficients to be estimated, which is $M_rM_t$. Since the receiver has only $M_r$ measurements during each training interval, at least $M_t$  training intervals are needed.} As the minimum required number of training intervals is independent of $M_r$, massive MIMO systems typically use uplink training for downlink channel estimation \cite{373}, when uplink-downlink reciprocity holds in \ac{tdd}. However, for \ac{fdd}, the required downlink channel training overhead is prohibitive when BS has large antennas. For mmWave massive MIMO systems when there is spatial sparsity, extensive research efforts have been devoted to reducing training overhead by techniques like compressive sensing-based channel estimation or channel extrapolation. More discussions on such techniques are given in Section~\ref{sec:relevantTechniques}.

For the particular task of MIMO beamforming or beam alignment, estimating the complete channel matrices is not a must. Instead, beam sweeping method is usually applied, by using codebook-based beams \cite{3050}. To achieve the desired beam resolution, the number of beams  in the codebook usually increases with the number of antennas. Therefore,  for an $M_r\times M_t$ MIMO system, an exhaustive search of beam sweeping would require an overhead on the order of $M_rM_t$. To reduce such overhead, various beam codebook designs such as hierarchical codebooks have been  studied \cite{579,3051}.

The above discussions reveal that for both pilot-based channel training and beam sweeping methods, from the link perspective, the required training overhead increases with the channel dimensions. From the network perspective, as both users and network infrastructure are more densely located, CSI acquisition is further complicated by practical issues like co-channel interference (CCI) or pilot contamination \cite{373,470}. The situation is worsened when traditional fully digital MIMO communications are replaced by systems with simpler hardware as discussed in Section~\ref{sec:simplerHardware}. For example, for systems with analog or hybrid beamforming, the baseband signal that is directly accessible to the transmitter/receiver has much lower dimension than the channel matrix to be estimated. This renders it necessary to use longer training durations. For example,  for hybrid beamforming  with $M_{\mathrm{RF}}$ chains and $M_{\mathrm{BS}}$ antennas, where $M_{\mathrm{RF}}<M_{\mathrm{BS}}$, in order to get as many channel measurements as channel coefficients, the required number of training durations would be increased by $\frac{M_{\mathrm{BS}}}{M_{\mathrm{RF}}}$ times as compared with the fully digital systems. As a concrete example, for analog beamforming with $M_{\mathrm{BS}}=64$ antennas at the BS and $M_{\mathrm{UE}}=16$  antennas at the UE, it may take seconds to complete the sweeping of all the transmit and receive beams \cite{3050}, which makes it infeasible for practical use. The training overhead is even more significant for passive communication systems like IRS-assisted communications. Though various channel estimation techniques have been proposed for such systems \cite{3052},  significant overhead and signal processing complexity are still needed to unlock the full potential of future 6G systems with much larger dimensional channel and denser wireless  nodes.

\subsubsection{Abundant location-specific channel data}
The major developing trends of denser wireless nodes and larger dimensional channel, together with the more diversified  positioning methods, imply that there would be abundant sources of location-specific channel data to be exploited in 6G. Such data includes received signal strength, channel gains, \ac{aoa}, \ac{aod}, etc. In fact, from the environment reconstruction perspective, the BSs and UEs can be regarded as fixed and mobile sensors, respectively. Each time wireless signal propagating between a pair of BS and mobile device, or between different devices for device-to-device (D2D) communications, we would be able to obtain one realization of location-specific data sample that reflects the actual radio propagation environment. The denser the network nodes are, the finer granularity of the spatial samples generating such location-specific data. {For example, with a density of 10 devices per m$^2$, each device would span an area of 0.1m$^2$ if the devices are uniformly scattered, and the average distance between adjacent devices (or the spatial samples)  would be roughly $30$ cm, which could be smaller than the coherence distance of channel shadowing \cite{ShadowingModel}.}  The distance for spatial samples would be even smaller in practice, since multitude of UEs would repeatedly appear at locations such as stadiums, roads, and shopping malls. This implies that if relevant channel data is properly measured and stored each time when wireless signal propagates between transmitter-receiver pairs, we would be able to obtain a data pool of sufficient fine granularity to accurately learn the local wireless environment.

\subsubsection{Powerful data mining capability}
Data mining refers to the process of extracting useful insights and knowledge from large datasets. It involves the use of various statistical, mathematical, and machine learning techniques to identify patterns, correlations, and trends in the data. The goal of data mining is to uncover hidden information that can be used to make better decisions or predictions \cite{4006}. In 6G, data mining is expected to become extremely powerful, thanks to the advancements of AI and machine learning techniques. With the massive amount of data collected and exchanged by 6G networks, data mining will play a critical role in extracting valuable insights and driving innovations in a wide range of applications. In particular, by mining the location-specific channel data, 6G networks may efficiently adapt communication strategies to the local wireless propagation environment, leading to improved spectral efficiency, energy efficiency, and reliability. Besides, data mining capabilities can facilitate automated feature selection and dimensionality reduction, which can improve the efficiency of data storage and processing.

%There are several approaches to data mining, including model-based and model-free approaches \cite{4007,4008}. In model-based approach, statistical models are created based on the data to explain the relationships and patterns observed in the dataset. The models are then used to make predictions or to identify patterns in new data.
%Conversely, model-free approach does not rely on a pre-existing model to identify patterns and relationships in the data. Instead, it uses algorithms such as clustering and association rules to discover patterns and relationships in the data. Model-free approaches are often used in cases where the underlying relationship between the variables is not well understood.
%Additional approaches to data mining include rule-based approaches, decision trees, neural networks, and genetic algorithms. Rule-based approaches employ if-then rules to extract patterns from the data, whereas decision trees use a tree-like structure to represent a set of decisions and their possible consequences. Neural networks, on the other hand, draw inspiration from the structure and function of the human brain and are useful in learning complex patterns in the data. Finally, genetic algorithms are based on the principles of natural selection and are employed to find optimal solutions to complex problems.

\subsection{Toward Environment-Aware Communications via CKM}\label{sec:towardEnvironmentAware}
The aforementioned new challenges and opportunities bring a promising paradigm shift from the conventional environment-unaware communications to the new environment-aware communications, which fully exploits {\it a priori} knowledge of the actual local environment for the design and optimization of communication networks. Environment-aware communication can be regarded as a further advancement of the conventional location-aware communication \cite{3042}, since it not only utilizes the location information of the mobile devices, but also the actual local environment information where the communication takes place. On one hand, environment-aware communication is a promising technique to resolve new challenges faced by 6G, such as the difficulty to acquire real-time CSI with the increasing of network density and channel dimensions. On the other hand, environment-aware communication is becoming more feasible than ever before, thanks to the more accessibility of high-quality location information, abundant sources of location-specific data, and more powerful data mining capabilities to learn the environment.

\begin{figure*}
\centering
\includegraphics[scale=0.85]{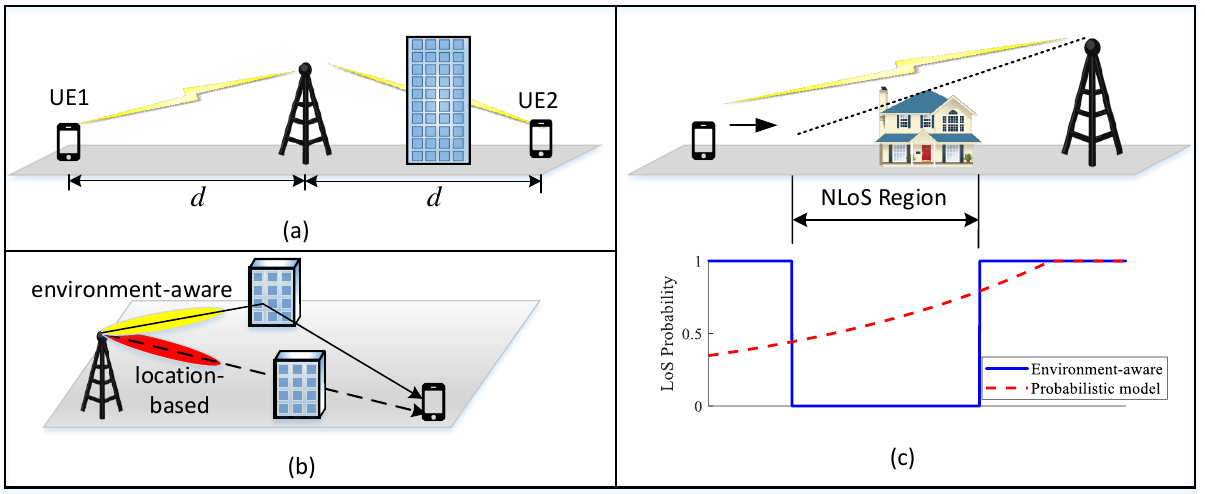}
\caption{Examples illustrating the significance of environment-aware communications \cite{2101}. (a) channel gain prediction; (b)	training-free beamforming; (c) LoS prediction.}\label{F:ToyExamples}
\end{figure*}

To illustrate the significance of environment-aware communication, Fig.~\ref{F:ToyExamples} shows three toy examples excerpted from \cite{2101} for channel gain prediction, beamforming, and LoS prediction, respectively. Consider the scenario shown in Fig.~\ref{F:ToyExamples}(a), where two UEs have equal distance from the BS. With the conventional environment-unaware communication, it tends to make the prediction that the two UEs have similar {channel path loss}. By contrast, with the new paradigm of environment-aware communications, the more accurate prediction that UE1 has better channel than UE2 could be made. {Similar observations can be obtained for training-free beamforming and  LoS prediction shown in Figs.~\ref{F:ToyExamples}(b) and (c), respectively. Specifically, in Fig.~\ref{F:ToyExamples}(b), the location-based beamforming simply directs the signals toward the UE location, while the environment-aware beamforming can beam toward the potential reflectors/scatterers that would eventually direct the signal to the UE \cite{2101}. In Fig.~\ref{F:ToyExamples}(c), the existence/absence of an LoS link determined from the propagation environment is more accurate than the probabilistic LoS model, which is meaningful only by considering a large number of realizations of similar communication environments \cite{2101}.} In other words, compared with the conventional environment-unaware communication, the awareness of the actual wireless environment is able to significantly improve the accuracy of channel knowledge prediction before any sophisticated real-time channel training is applied.

Perhaps the most straightforward approach to realize  environment-aware communication is to use the physical environment maps \cite{1203,1063,2132}, such as the 3D city/terrain map or scatterer map. {Though readily available in certain areas such as city downtown, such physical environment maps fail to directly reflect the intrinsic channel characteristics, and additional parameters such as the dielectric properties of the environment need to be specified. Besides, even with such data, computation-intensive processing like ray tracing is needed to convert the physical environment map into desired site-specific channel environment map \cite{1204}. Therefore, using physical environment maps for environment-aware communication is more suitable for offline simulation, rather than real-time environment-aware channel inference. }On the other hand, if sufficient labelled data is available, deep neural networks such as RadioUNet  can be trained to infer the channel path loss by using the transmitter location and morphological images of the urban geometry as the input of the neural network \cite{2008}. However, high-quality images may not always be available  and the performance critically depends on the quality of data and neural network configurations.

An alternative technique towards environment-aware communication is radio environment map (REM) \cite{2100,2096,2097,1067}. According to \cite{2100}, REM is a spatial-temporal database of real-world radio scenarios that provides multi-domain environment information, such as geographical features, spectral regulations, and RF emissions. One of the main focuses of REM is the spectrum usage information, which is useful for interference prediction and avoidance. Therefore, REM was mostly studied for interference management or resource allocation in cognitive radio systems \cite{2100,2096}. Similar to physical environment map, REM does not directly reflect the intrinsic channel characteristics, since it critically depends on the transmitter setup and activities, such as its occupied spectrum and transmitting power.

\begin{table*} \footnotesize
\caption{Comparison of different {techniques} for realizing environment-aware communications.}\label{table:maps}
\centering
\begin{tabular}{|p{2cm}|p{3.5cm}|p{5cm}|p{5cm}|}
\hline
{\bf     ~~~~~Map} & {\bf     ~~~~~Main characteristics} & {\bf     ~~~~~Pros} & {\bf     ~~~~~Cons}\\
\hline\vspace{0.1in}
Physical environment map\cite{1203,1063,2132}
& \begin{itemize}[leftmargin=1em]
\item 3D city/terrain map \item Scatterer map
\end{itemize}
&\begin{itemize}[leftmargin=1em]
\item  	Map is available in many areas
\end{itemize} &
\begin{itemize}[leftmargin=1em]
\item Cannot directly reflect intrinsic channel characteristics
\item Requires additional parameter specification (such as dielectric properties of environment) and complicated processing (like ray tracing)
\item Unsuitable for real-time channel inference
\end{itemize}\\
\hline\vspace{0.1in}
Radio environment map (REM)\cite{2100,2096,2097,1067} &
\begin{itemize}[leftmargin=1em]
\item Focuses on spectrum usage information with coarse granularity
\end{itemize} &
\begin{itemize}[leftmargin=1em]
\item Effective for interference prediction
\item Extensively studied for cognitive radio
\end{itemize} &
\begin{itemize}[leftmargin=1em]
\item Does not directly reflect intrinsic channel characteristics
\item Depends on the transmitter setup and activities
\item Requires offline REM construction
\end{itemize}\\
\hline\vspace{0.1in}
Channel knowledge map (CKM)\cite{2101} &
\begin{itemize}[leftmargin=1em]
\item Site-specific database to provide location-specific channel knowledge
\item Facilitates or even obviates real-time CSI acquisition
 \end{itemize} &
 \begin{itemize}[leftmargin=1em]
 \item Directly offers location-specific channel knowledge
\item Without requiring additional parameter specification (such as dielectric properties of environment) or complicated processing (like ray tracing)
\item Independent of transmitter/receiver setup or activity
\item Suitable for real-time channel knowledge inference
\end{itemize}&
\begin{itemize}[leftmargin=1em]
\item Requires offline CKM construction
\item CKM in dynamic environment (difficulty of maintaining CKM)
\end{itemize}\\
\hline
\end{tabular}
\end{table*}

To address the above issues, a novel concept termed {\it channel knowledge map} (CKM) was recently proposed \cite{2101}. Different from physical environment map or REM, CKM aims to directly reflect the intrinsic  wireless channel  properties of the local environment. Specifically, according to \cite{2101}, CKM is a site-specific database, tagged with the locations of the transmitters and/or receivers, which provides location-specific channel knowledge useful to enhance environment-awareness and facilitate or even obviate sophisticated real-time CSI acquisition. The abstractive word ``knowledge'' is deliberately chosen so as to emphasize that any information that is relevant to the local wireless channel could be the desired query output of CKM, including but not limited to the region-specific channel modelling parameters such as path loss exponent, Rician factor, and delay/Doppler/angular spread, as well as the location-specific channel knowledge such as channel gains, delays/Dopplers/AoAs/AoDs of multi-paths, or even the complete channel impulse response.  More detailed discussions on the CKM  categories are given in Section~\ref{sec:classification}.

As CKM is mainly maintained offline and UEs can be localized by systems like GPS, laser, IMU or ISAC that do not consume the precious communication resources, the {\it a priori} location-specific channel knowledge enabled by CKM {does not incur additional cost} to wireless communication networks. This thus opens many new promising opportunities and may bring paradigm shifts for the design and optimization of wireless communication networks, e.g., from site-specific to location-specific channel knowledge inference, from environment-unaware to environment-aware communication and sensing, from discarding and under-utilizing valuable location-specific channel data to exploiting its full benefit, from heavy real-time channel training to light training or even training-free communications. Table~\ref{table:maps} compares the three different types of maps for realizing environment-aware communications. {It is worth mentioning that techniques like ray tracing are unnecessary for CKM during the online channel knowledge inference phase, but they can be exploited during the CKM construction phase. For example, if the local wireless environment information (such as buildings and their dielectric properties) is available, ray tracing could be applied offline to generate wireless channel knowledge that is highly related to the static wireless environment, and such knowledge could provide an initial data source for CKM construction. Besides, when additional online measurements are available, the obtained data can be fused with the ray tracing data to update the CKM.}

In the literature, various terminologies are used to emphasize different aspects related to CKM. For example, as discussed above, the term REM has been widely used in cognitive radio systems for database covering multi-domain information, such as the geographic information and activities of the available radios \cite{2096,2100,1210,2097}. Similarly, the term radio map has been mainly used to refer to databases of \ac{rss} or large-scale channel gains (path loss and shadowing), which are essential for RSS fingerprinting-based localization \cite{2067,2069,2071,2072} or environment-aware wireless communications with path loss and shadowing prediction \cite{1062,1067,2059,2039,2005,2074,2008,2105,3001,2107}. Alternative terminologies for databases of large-scale channel gains include {channel gain map} \cite{1206,2013}, {channel gain cartography} \cite{2038,2042,2043}, or location-based channel database \cite{2052}.  {Rate map} \cite{2009} and {coverage map} \cite{2026,2046} have been used for average communication rate and coverage prediction, respectively. To enable the location-specific interference level prediction for applications like cognitive radio, concepts like  interference cartography \cite{2014}, spectrum cartography \cite{2015}, and power spectral density (PSD) map \cite{2019,2107} have been developed. For environment-aware mmWave beam alignment, databases like beam index map (BIM) \cite{2135,CKM_Utilization_Yong_Hybrid_BF,CKM_Utilization_Yong_IRS_BF}, multipath fingerprint\cite{2078} or fingerprint-based database \cite{2076} are studied, where the transmit and receive beam pairs are learned for each location. 
In this article, unless otherwise stated, we will use the terminology CKM to emphasize the environment-aware database that offers location-specific channel knowledge whatever related to the intrinsic property of the wireless channel, while irrespective of the transmitter or receiver activities. 

\begin{figure*}[htb]
\centering
\includegraphics[scale=0.4]{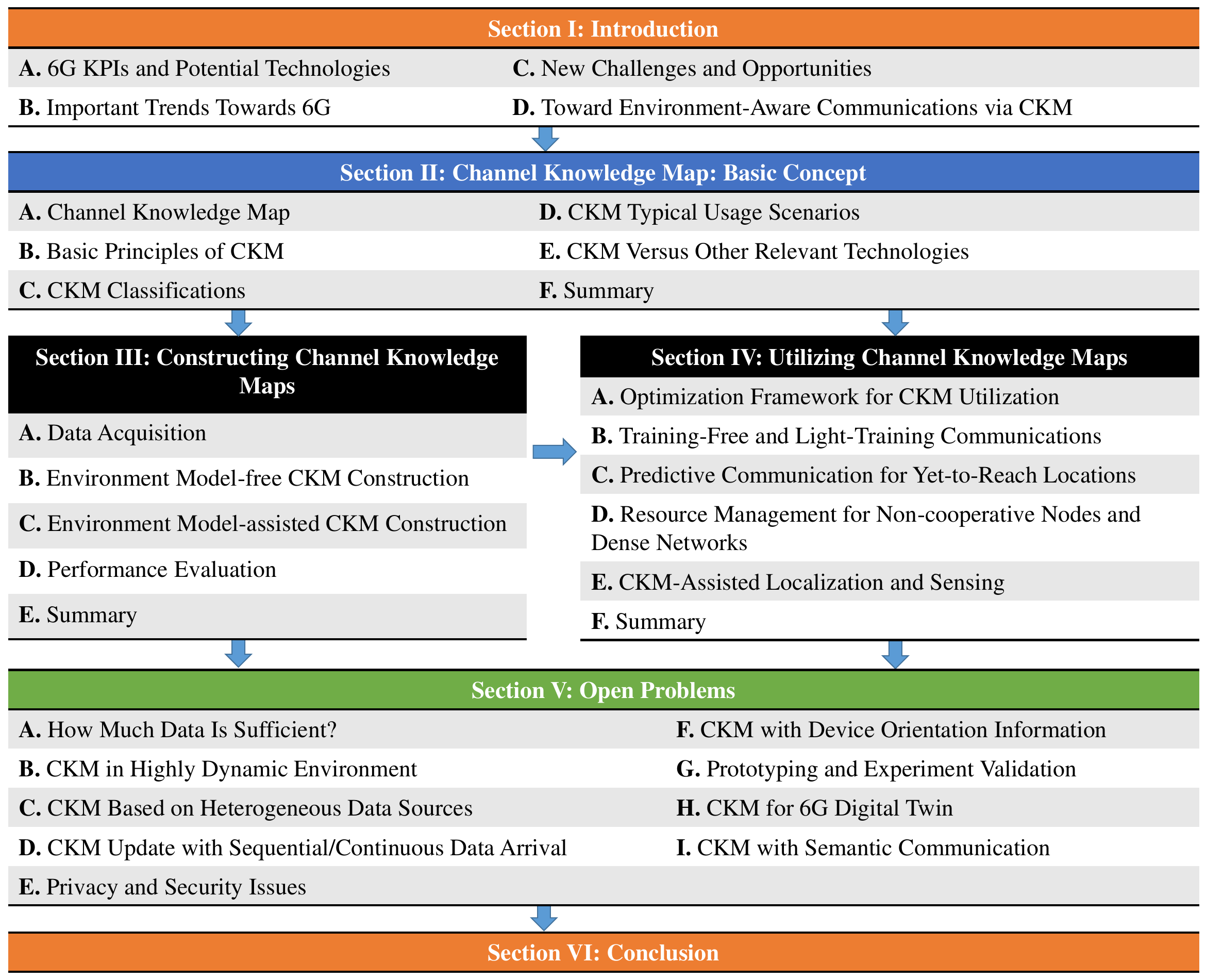}
\caption{Organization of this paper.}\label{F:articleOrganization}
\end{figure*}

In this paper, we aim to present a comprehensive tutorial
overview on the emerging CKM-enabled environment-aware
communications. While there are a few relevant overview papers in the literature, most of them are magazine papers \cite{2097,2096,1067} discussing the particular REM or radio map, rather than the novel CKM that reflects the intrinsic wireless channel  features. In \cite{2046}, the authors provide a survey focusing on the particular coverage map that mainly depends on the large-scale channel path loss. In the magazine paper \cite{2101}, the concept of CKM is proposed, but only a very brief introduction is given there. To our best knowledge, a comprehensive and in-depth tutorial overview on CKM that clearly explains its major concept as well as efficient map construction and utilization methods, is still lacking. This thus motivates our current article. The specific contributions of this article are summarized as follows:       
\begin{itemize}
    \item 
First,  we explain in details the basic concept of CKM, discuss the rationale why CKM works and provide its general classifications. Furthermore, several typical usage scenarios of CKM are presented and a comprehensive comparison of CKM with relevant channel inference techniques is provided. 
\item Next, we present the key techniques for constructing CKM. To this end, various data acquisition methods are discussed, followed by the presentation of both {environment} model-free and {environment} model-assisted CKM construction methods. 
\item Furthermore, we discuss how CKM can be utilized to enable environment-aware communications efficiently. A generic framework of CKM-enabled wireless networks is firstly discussed, followed by some use case studies, including CKM-enabled training-free and light-training communications, predictive communications, and environment-aware localization and sensing.
\end{itemize}

{As shown in Fig. \ref{F:articleOrganization}, the rest of this paper is organized as follows.} Section II introduces the concept of CKM and its classifications. Section III presents the CKM construction techniques, and the utilization of CKM is discussed in Section IV.  Section~\ref{sec:openProblems} points out some future directions worthy for further investigation and Section~\ref{sec:conclusion} concludes the article. Some of the acronyms used in this paper are summarized in Table~\ref{tb:acronyms}. 
\begin{table}
\centering
\caption{List of Acronyms}
\label{tb:acronyms}
\begin{tabular}{|l|l|}
     \hline
     A2G & Air-to-ground\\
     \hline
     ADC & Analog-to-digital converter\\
     \hline
     AI & Artificial intelligence\\
     \hline
     ALS & Alternating least square \\
     \hline
     AoA & Angle of arrival \\
     \hline
     AoD & Angle of departure\\
     \hline
     B2X & BS-to-any\\
     \hline
     BIM & Beam index map\\
     \hline
     BS & Base station \\
     \hline
     CAM & Channel angle map\\
     \hline
     CKM & Channel knowledge map \\
     \hline
     CMM & Channel matrix map\\
     \hline
     CNN & Convolutional neural network\\
     \hline
     CPM & Channel path map\\
     \hline
     CRLB & Cram\'{e}r-Rao lower bound\\
     \hline
     CSI & Channel state information\\
     \hline
     D2D & Device-to-device\\
     \hline
     DDAM & Delay-Doppler alignment modulation\\
     \hline
     DFRC & Dual functional radar and communication\\
     \hline
     GBS & Ground base station\\
     \hline
     GPR & Gaussian process regression\\
     \hline
     GPS & Global positioning system\\
     \hline
     IDW & Inverse-distance-weighted\\
     \hline
     ILSAC & Integrated localization, sensing and commun.\\
     \hline
     IMU & Inertial measurement unit \\
     \hline
     IoT & Internet-of-things\\
     \hline
     IRS & Intelligent reflecting surface\\
     \hline
     ISAC & Integrated sensing and communication\\
     \hline
     KNN & $K$-nearest-neighbor\\
     \hline
     LoS & Line-of-sight\\
     \hline
     MDT & Minimization of drive tests\\
     \hline
     MIMO & Multiple-input multiple-output\\
     \hline
     MSE & Mean square error\\
     \hline
     MUSIC & Multiple signal classification \\
     \hline
     NLoS & Non-line-of-sight\\
     \hline
     NTN & Non-terrestrial network\\
     \hline
     OFDM & Orthogonal frequency-division multiplexing\\
     \hline
     OTFS & Orthogonal time frequency space\\
     \hline
     PAS & Power angular spectrum\\
     \hline
     PSD & Power spectrum density\\
     \hline
     REM & Radio environment map \\
     \hline
     RKHS & Reproducing kernel Hilbert space\\
     \hline
     RNN & Recurrent neural network\\
     \hline
     RIS & Reconfigurable intelligent surface\\
     \hline
     RSS & Received signal strength\\
     \hline
     SINR & Signal-to-interference-plus-noise ratio\\
     \hline
     ToA & Time of arrival \\
     \hline
     TDoA & Time difference of arrival\\
     \hline
     TPS & Thin plate spline\\
     \hline
     UAV & Unmanned aerial vehicle\\
     \hline
     UE & User equipment \\
     \hline
     WPT & Wireless power transfer\\
     \hline
     X2X & Any-to-any\\
     \hline
     XL-MIMO & Extremely large-scale MIMO\\
     \hline
\end{tabular}
\end{table}

{{\it Notations:} Boldface lower- and upper-case letters denote vectors and matrices, respectively. 
$ \mathbb{C}^{M\times N} $ and $ \mathbb{R}^{M\times N} $ denote the spaces of $ M \times N $ complex and real matrices, respectively.
$ \mathbf{A}^{T},\mathbf{A}^{*},\mathbf{A}^H $ denote the transpose, conjugate, and conjugate transpose of the matrix $ \mathbf{A} $, respectively.
$ ||\mathbf{A}||_F $, $ \mathrm{tr}(\mathbf{A}) $, and $ |\mathbf{A}| $ denote the Frobenius norm, trace, and determinant of $\mathbf{A}$, respectively.
$\mathbf{I}_N$ is the $ N\times N $ identity matrix.
For a set $\mathcal{A} $, $ |\mathcal{A}| $ denotes its cardinality. 
Expectation is denoted by $ \mathbb{E}[\cdot] $.}

\section{Channel Knowledge Map: Basic Concept}\label{sec:concept}

\subsection{Channel Knowledge Map}
Let $\mathbf q \in \mathbb{R}^{D}$ be a vector containing the locations of the transmitter and/or receiver of a wireless link, whose dimension $D$ varies for different application scenarios. For example, for BS-centric CKM as discussed in Section~\ref{sec:classification} where the BS location is fixed, $\mathbf q$ only needs to include the UE location, for which $D=2$ for typical ground UEs and $D=3$ for UEs in high-rise buildings or cellular-connected aerial UEs \cite{1095}. Further denote by $\mathbf z\in \mathbb{C}^{J}$ the channel knowledge that is of interest, whose dimension $J$ depends on the channel knowledge type and system configurations. For example, consider a BS-centric CKM where the BS and UE have $M_{\mathrm{BS}}$ and $M_{\mathrm{UE}}$ antennas, respectively. If the complex-valued MIMO channel matrices of all the $N_{\mathrm{SC}}$ subcarriers are the desired map output, which correspond to {\it channel matrix map} (CMM), we would require $J_{\mathrm{CMM}}=M_{\mathrm{BS}}M_{\mathrm{UE}}N_{\mathrm{SC}}$, which could be a huge number for wideband massive MIMO or  XL-MIMO communication systems. On the other hand, if we are interested in the key parameters (i.e., path gain, delay, azimuth/elevation AoA/AoD, and Doppler) of the $L$ dominating multi-path components, which correspond to {\it channel path map} (CPM) \cite{2135,CKM_Utilization_Yong_Hybrid_BF}, we have $J_{\mathrm{CPM}}=7L$. Obviously, $J_{\mathrm{CPM}}\ll J_{\mathrm{CMM}}$ for massive MIMO systems with multipath sparsity and large antenna arrays since $L\ll M_{\mathrm{BS}}$.

In a broad sense, CKM can be regarded as a mapping $\mathcal{M}$ from the location vector $\mathbf q\in \mathbb{R}^{D}$ to the channel knowledge vector $\mathbf z\in \mathbb{C}^{J}$, i.e.,
\begin{align}\label{eq:mapping}
\mathcal M: \mathbb{R}^{D}\rightarrow \mathbb{C}^{J}.
\end{align}
Take BS-centric communication as an example. Let $K$ denote the number of UEs under consideration, and $\mathbf q_k[n]\in \mathbb{R}^{D}$ with $1\leq n\leq N$ denote the trajectory of UE $k$ over a period of $N$ time epochs, $1\leq k\leq K$. If the trajectories $\mathbf q_k[n]$'s are known, which can be achieved by a multitude of localization and sensing technologies as  discussed in Section~\ref{sec:more_accessible_loc}, then with the CKM $\mathcal M$ maintained by each BS, the corresponding location-specific channel knowledge $\mathbf z_k[n]\in \mathbb{C}^{J}$ of all the $K$  UEs across all $N$ time epochs can be directly inferred, before sophisticated real-time CSI acquisition methods are applied.

\subsection{Basic Principles of CKM}
Intuitively, CKM is able to achieve environment-awareness and offer location-specific  channel knowledge because a device arriving at the same location that was previously visited, either by itself or other devices, is likely to experience a quite similar wireless environment. Note that although wireless channel is time-varying in nature, one major cause of channel variations is due to UE mobility itself, which can be fully characterized by its moving trajectory. As such, by fully utilizing the UE trajectory and the surrounding environment information, CKM may significantly reduce the channel uncertainty and achieve much more accurate channel inference than the conventional environment-unaware counterparts.

 For the purpose of further illustration, let $E(t)$ denote the abstractive wireless environment for a particular local area, which includes both the static environment $E_s$ such as terrains and surrounding buildings, as well as the dynamic environment $E_d(t)$ such as those passing-by vehicles and pedestrians. Let $\mathbf q(t)\in \mathbb{R}^{D}$ denote the trajectory of a UE that wishes to communicate with the BS, and $\mathbf z(t)\in \mathbb{C}^{J}$ be the channel knowledge that we are interested in. At an abstract level,  the wireless channel and all its related channel knowledge $\mathbf z(t)$ is fully determined by the radio wave property (such as carrier frequency), the locations of the UE $\mathbf q(t)$, and the radio propagation environment $E(t)$. Thus, for each specified radio wave, we have
\begin{align}\label{eq:abstract_function}
\mathbf z(t)=f(\mathbf q(t), E(t)),
\end{align}
where $f(\cdot,\cdot)$ is some function that maps the UE location and the radio environment to the channel knowledge. Therefore, if both $\mathbf q(t)$ and $E(t)$  are known, the channel knowledge $\mathbf z(t)$ should be in principle obtained based on \eqref{eq:abstract_function}. However, in practice, directly applying \eqref{eq:abstract_function} for channel knowledge inference is hindered by two major challenges. Firstly, due to the complicated interaction between radio waves and the radio environment, it is very difficult, if not impossible, to find the mathematically tractable function $f(\cdot,\cdot)$ that provides universally good approximation to all environments. Therefore, in practice, numerical methods such as ray tracing are usually used to obtain location-specific and environment-aware channels. However, due to the extensive computation complexity, ray tracing is only suitable for offline simulations, rather than real-time online channel knowledge inference. Another challenge for directly applying \eqref{eq:abstract_function} is that the radio environment $E(t)$ is very difficult to be mathematically represented. Note that $E(t)$ should not only include the 3D physical environment such as the terrains, locations and heights of surrounding buildings, but also their dielectric properties that would have crucial impact on radio wave propagation.

The two aforementioned  challenges can be effectively resolved by the concept of CKM, which essentially exploits the historical data of the location-channel knowledge pair $(\mathbf q, \mathbf z)$ to make new channel knowledge inference, without attempting to explicitly characterize the function $f(\cdot,\cdot)$ nor the environment $E(t)$. For ease of illustration, consider a quasi-static environment where $E(t)$ is approximately unchanged over the time duration of interest, i.e., $E(t)\approx E$, $\forall 0\leq t \leq T$. Thus, \eqref{eq:abstract_function} reduces to
\begin{align}\label{eq:abstract_function_static}
\mathbf z(t)=f(\mathbf q(t), E),  0 \leq t \leq T.
\end{align}
Let $\bar {\mathbf Q}=[\mathbf q_1,...,\mathbf q_Q]\in \mathbb{R}^{D\times Q}$ and $\bar {\mathbf Z}=[\mathbf z_1,...,\mathbf z_Q]\in \mathbb{C}^{J\times Q}$ be the $Q$ historically visited locations and their corresponding measured channel knowledge data, respectively. Note that the data $(\bar {\mathbf Q}, \bar{\mathbf Z})$ essentially corresponds to $Q$ sample realizations of the mapping \eqref{eq:abstract_function_static}, which can be obtained by different devices. Therefore, if the channel knowledge type is properly designed and sufficient locations have been visited, it follows from \eqref{eq:abstract_function_static} that the environment $E$ can be in principle reconstructed based on the data, represented as
\begin{align}\label{eq:E}
E=g(\bar{\mathbf  Q}, \bar {\mathbf Z}),
\end{align}
for some function $g(\cdot,\cdot)$. Therefore, during the online communication process, if the channel knowledge $\mathbf z$ of a new location $\mathbf q$ needs to be inferred, by applying \eqref{eq:abstract_function_static} and substituting $E$ with \eqref{eq:E}, we have
\begin{align}
\mathbf z& =f(\mathbf q, g(\bar{\mathbf  Q}, \bar {\mathbf Z}))\\
& =h(\mathbf q;\bar{\mathbf  Q}, \bar {\mathbf Z}), \label{eq:z}
\end{align}
where $h(\cdot;\cdot)$ is the function taking the composition of functions $f(\cdot,\cdot)$ and $g(\cdot,\cdot)$.
The relationship \eqref{eq:z} indicates that based on the historical data $(\bar {\mathbf Q}, \bar{\mathbf Z})$, i.e., channel knowledge tagged with transmitter/receiver locations, the CKM mapping $\mathcal M: \mathbb{R}^{D}\rightarrow \mathbb{C}^{J}$ as defined in \eqref{eq:mapping} can be in principle constructed to fully reflect the actual local wireless environment. Note that as will be further discussed in Section~\ref{sec:construc}, the historical data $(\bar {\mathbf Q}, \bar{\mathbf Z})$ does not necessarily come from dedicated site survey and measurements. It may also be obtained during the actual online  communication process or offline numerical computations. Besides, various techniques for CKM construction based on \eqref{eq:z} are discussed later in Section~\ref{sec:construc}.

\begin{figure*}
\centering
\includegraphics[scale=0.50]{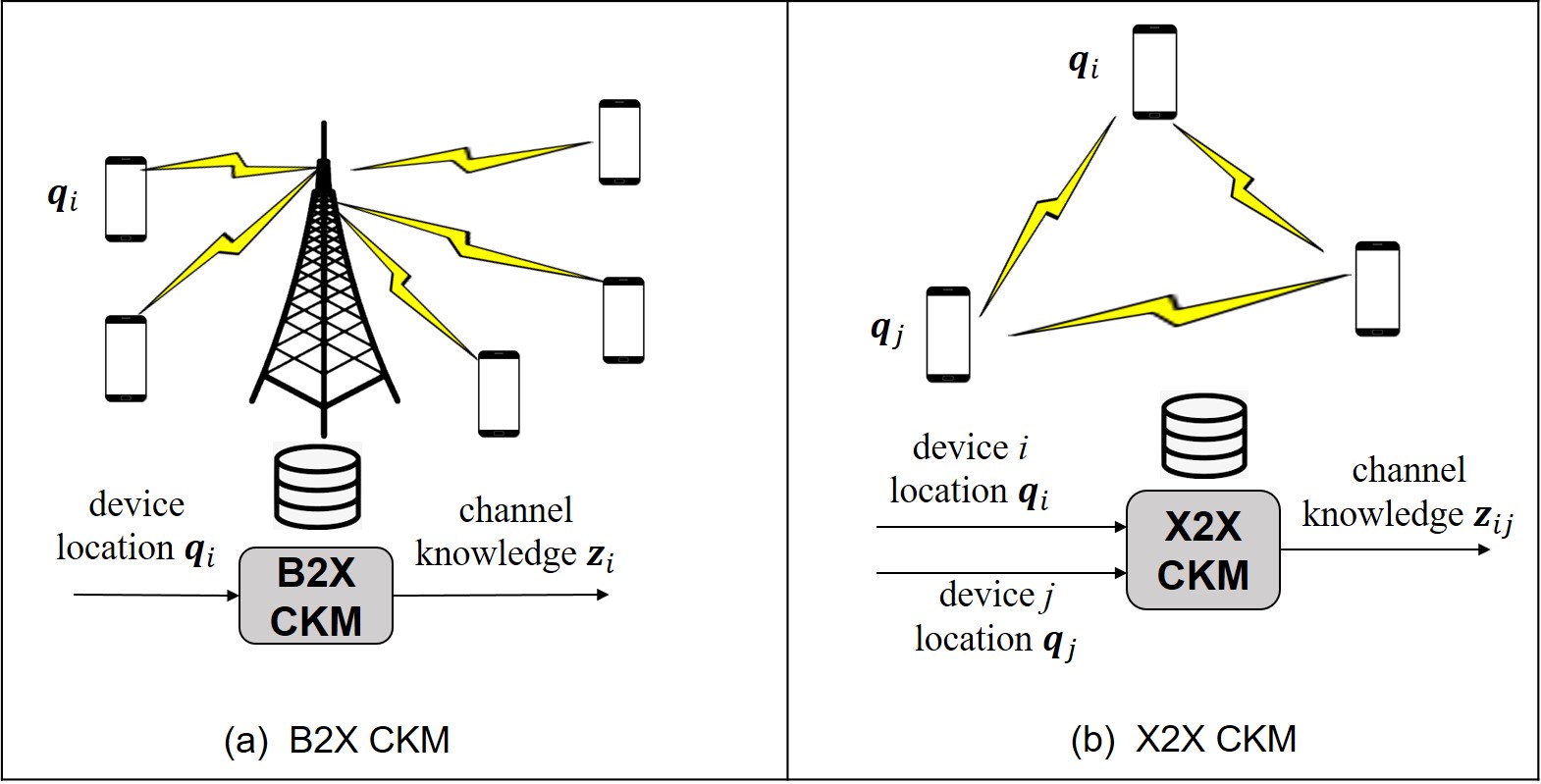}
\caption{An illustration of B2X CKM  and X2X CKM.}\label{F:CKMTypes}
\end{figure*}

\subsection{CKM Classifications}\label{sec:classification}
CKM can be classified based on different criteria, such as communication modes and channel knowledge types.
\subsubsection{Classification based on communication modes}
Two most common communication modes for cellular networks are BS-centric communication and D2D communication, which correspond to  {\it BS-to-any (B2X)} CKM and {\it any-to-any (X2X)} CKM, respectively. 

{\bf B2X CKM:}  As illustrated in Fig.\ref{F:CKMTypes}(a), B2X CKM aims to enable environment-aware communications between each BS and UEs at all locations within its coverage. Thus, one natural way is for each BS to maintain its own CKM, and if necessary, CKMs of different BSs are shared via backhaul links (say during off-peak hours) to enable cooperative channel knowledge inference. As the locations of BSs are fixed, for each BS, B2X CKM only requires the UE locations for map construction and data query. Thus, the input dimension $D$ of the mapping in \eqref{eq:mapping} is $D=2$ for typical ground UEs and $D=3$ for UEs in high-rise buildings or aerial UEs. On the other hand, with the increasing densification of cellular BSs, it is likely that CKMs associated with different BSs involve overlapping coverage areas. In this case, the BS ID may also be  needed for CKM construction and data query.

{\bf X2X CKM:} As illustrated in Fig.\ref{F:CKMTypes}(b), X2X CKM is useful for D2D communication, which is an effective technology for BS offloading, by transmitting signals directly between those nearby communication devices without having to be forwarded by the BS. In this case, both the transmitter and receiver can be mobile. As a result, the location vector $\mathbf q\in \mathbb{R}^{D}$ for X2X CKM needs to include locations of both transmitter and receiver, and the dimension $D$  varies from $D=4$ for  ground-ground communication,  $D=5$ for air-ground communication, to $D=6$ for air-air communication. Therefore, X2X CKM has larger dimension than B2X CKM, which implies that more efforts on data acquisition, storage, and mining are needed for X2X CKM. Similar to B2X CKM,  X2X CKM may be maintained by BSs, since D2D links are usually established with the assistance of BSs.

\subsubsection{Classification based on channel knowledge types}
Depending on the practical application scenarios, various CKMs can be constructed to enable environment-aware inference for different channel information, including the region-specific channel modelling parameters, as well as location-specific large-scale and small-scale channel knowledge.

{\bf CKM for region-specific modelling parameters:} Stochastic-based channel modelling is an efficient way to provide mathematically tractable and generalizable characterization for wireless channels \cite{1203}. Built on expert knowledge gained from the extensive research on wireless channels over the past few decades, stochastic channel models try to characterize the wireless channels such as their path loss, multi-path AoA/AoD, and delays, in a stochastic manner with certain modelling parameters. For example, distance-dependent path loss with log-normal shadowing has been widely used to predict the large-scale channel gain, where the path loss exponent, intercept and shadowing variance are the most important modelling parameters. In multi-antenna systems, the Gaussian or Laplacian distribution based power angular spectrum (PAS) is usually used to model the AoA/AoD of the multi-path signal components, where the modelling parameters are the mean AoA/AoD and their angular spread. However, the modelling parameters in conventional stochastic-based channel modelling methods only depend on the very coarse information of the environment, such as urban,  rural, or sub-rural, while ignoring the specific radio propagation environment where the communication actually takes place. This may lead to inaccurate or spatially inconsistent modelling outputs. For example, measurement results have shown that even in the same area, different streets should have different modelling parameters for more accurate path loss modelling \cite{3053}. Similar issues have also been studied by quasi-deterministic radio channel generator (QuaDRiGa) \cite{3054}. As a result, CKM can be constructed to provide the local region-specific channel modelling parameters to enable environment-aware channel inference based on the well-established expert knowledge \cite{2093}. Some typical examples include CKMs for path loss exponent, path loss intercept, shadowing variance, correlation distance, Rician factor, spatial correlation matrix, mean AoA/AoD,  and delay/Doppler/angular spreads. Note that compared to the two alternative CKM types discussed below, CKMs for region-specific channel modelling parameters are expected to be more stable, and less demanding for localization accuracy and map storage capacity. Besides, both expert knowledge and historical data can be exploited for environment-aware channel knowledge inference. However, the quality of inference may be biased by the pre-assumed mathematical models.

{\bf CKM for location-specific large-scale channel knowledge:} Compared to region-specific modelling parameters, a more aggressive type of CKM is that for location-specific large-scale channel knowledge, such as the presence or absence of LoS link, the large-scale channel gain including path loss and shadowing, the AoAs/AoDs of the deterministic channel multi-paths, and the transmitter/receiver beam indices leading to the highest power, etc. Unlike CKMs for region-specific modelling parameters, those for large-scale channel knowledge do not have to rely on expert knowledge or pre-assumed mathematical models, so that the potential error caused by model bias is avoided. Besides, the non-trivial problem of partitioning the area into several suitable regions is not needed. Note that since large-scale channel knowledge usually varies across correlation distances that are much larger than signal wavelength (say hundreds of wavelength for shadowing), the corresponding CKM can be constructed by discretizing the area/space with  grid size on the order of the correlation distances.

{\bf CKM for location-specific small-scale channel knowledge:} The most ambitious channel knowledge type aims to provide the fine location-specific small-scale channel knowledge, such as the instantaneous channel gains in time-frequency-space domain, the AoAs/AoDs, delays, and gains of those significant channel paths or even the complete channel impulse response. Compared to the two alternative channel knowledge types discussed above, CKM for small-scale channel knowledge offers much richer information. However, it is also the most demanding in terms of the localization accuracy, data acquisition, storage, and mining. For example, since the small-scale channel gain varies over a displacement on the order of signal wavelength, a localization accuracy and site survey on the similar scale are needed. Though 6G is envisioned to be able to achieve centimeter (cm) or even millimeter(mm)-level localization accuracy, its practical usage for highly accurate CKM construction is yet to be studied. Besides, the highly dynamic nature of small-scale channel knowledge also needs to be considered for its map construction, update, and data query.

{Note that sufficient storage is generally needed for CKM construction,  and such an issue could be alleviated by various ways. For example, instead of maintaining a giant CKM for the whole wireless network, it is more feasible to maintain CKM in a hierarchical manner, i.e., different BSs may maintain their respective CKMs and higher-level CKMs with lighter information could be maintained for inter-cell coordination. Another method for alleviating the storage issue is  to use various machine learning or data mining techniques to extract the key features of raw data to generate CKM, instead of directly storing the raw data, as will be discussed in Section~\ref{sec:construc}.}

\begin{figure*}
\centering
\includegraphics[scale=0.55]{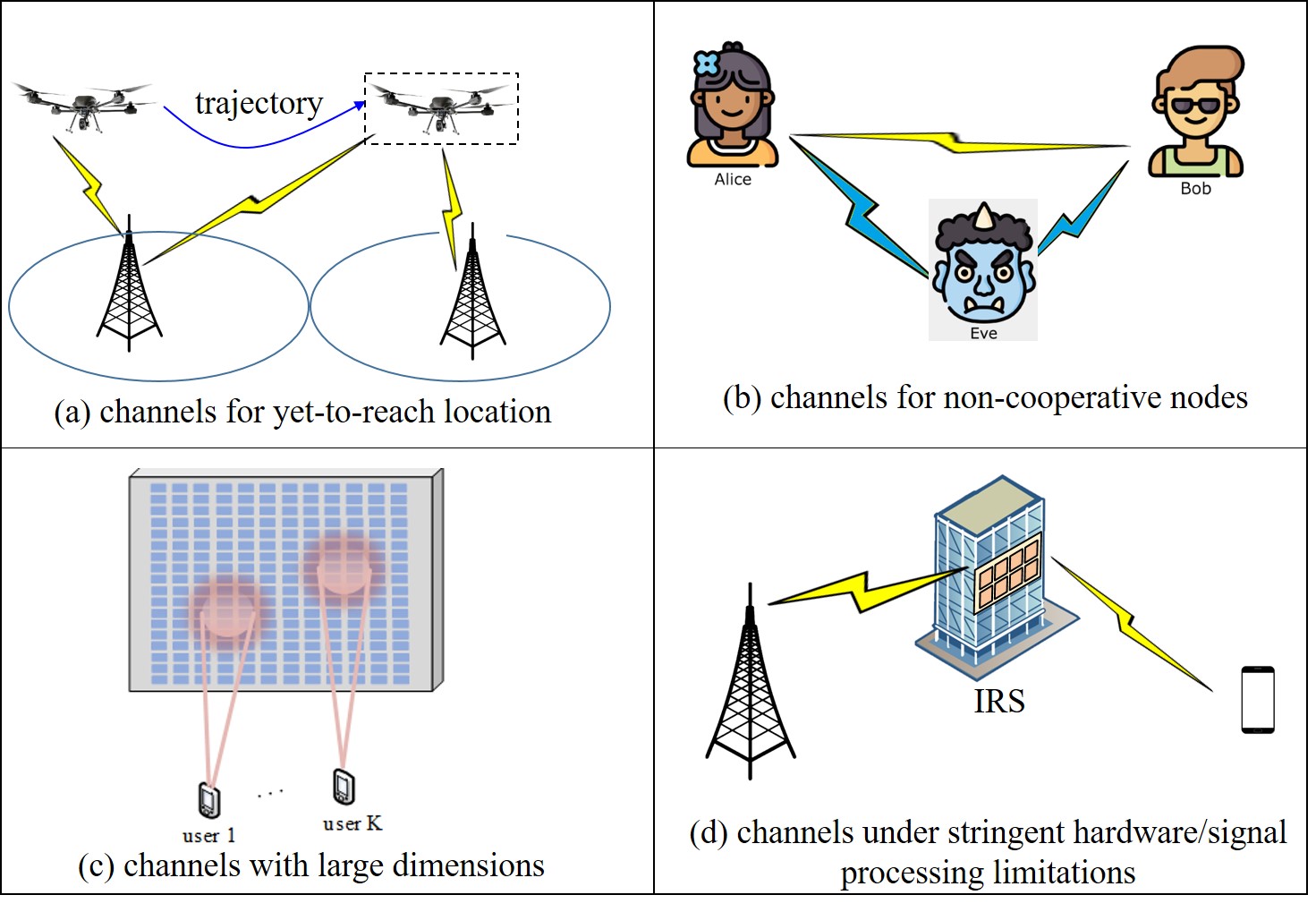}
\caption{An illustration of four typical use  cases for CKM-enabled  environment-aware communications.}\label{F:useCases}
\end{figure*}

\subsection{CKM Typical Usage Scenarios}
CKM is expected to play an important role whenever channel information is needed, but conventional environment-unaware channel acquisition  methods are either infeasible or too costly to implement. In particular, as illustrated in Fig.~\ref{F:useCases},  there are four typical usage scenarios for CKM, namely predicting channels for yet-to-reach or never-to-reach locations, channels for non-cooperative nodes, channels with large dimensions, and channels in scenarios under stringent hardware/signal processing limitations \cite{2101}.

\subsubsection{Channels for yet-to-reach or never-to-reach locations}
Future wireless networks are expected to make more intelligent and foresighted decisions, in a more proactive and responsive manner. This usually requires that the network has a global view of the wireless environment, i.e., knowing the channel knowledge for locations where no terminals currently exist. In this case, conventional CSI acquisition methods that purely rely on real-time channel training are not applicable, since channel training is implementable only if the devices reach the corresponding locations. This issue can be effectively addressed by CKM-enabled environment-aware networks. For example, for network-connected ground or aerial robots, CKM may enable communication-aware motion/trajectory planning to avoid entering the blind coverage zones. Besides, CKM is also useful for predictive resource allocation, foresighted handover to avoid ``ping-pong'' phenomenon, and intelligent node sleeping and wake-up without periodic channel training \cite{2052}.

\subsubsection{Channels for non-cooperative nodes}
Training-based channel acquisition typically requires that both transmitter and receiver are cooperative, which may not be true in many scenarios.  For example, for eavesdropping systems, the malicious eavesdropper is certainly unwilling to assist the legitimate transmitter to estimate its channel. As another example, for cognitive radio systems, there is no much incentive for the primary receiver to cooperate with the secondary transmitter to estimate their channel. However, such channels for non-cooperative nodes are essential to implement techniques like zero-forcing beamforming to enable physical layer security or interference temperature control for cognitive radio. CKM provides an effective technique to resolve such issues, since channels of non-cooperative nodes can be inferred based on their locations. Note that there are many ways to obtain the location information of non-cooperative nodes, say radar/lidar/comera sensing-based localization.

\subsubsection{Channels with large dimensions}
As discussed in Section~\ref{sec:largerDim}, with the use of massive MIMO and mmWave technologies, the dimensions of wireless channels grow dramatically, both in spatial and frequency domains.  Estimating such high-dimensional channels is costly, in terms of training and feedback overhead, as well as the signal processing complexity. The situation deteriorates in high-mobility scenarios where channel coherence time is short, or for low-latency applications. Fortunately, by offering fine location-specific {\it a priori} channel knowledge, CKM is a promising solution to facilitate or even avoid the costly real-time CSI acquisition for channels with large dimensions.

\subsubsection{Channels under stringent hardware/signal processing limitations}
To support diversified applications cost-effectively, future wireless networks are expected to engage massive semi-passive devices like backscatter radios and IRSs. Furthermore, with the continuous increase of antenna size from massive MIMO to XL-MIMO, conventional fully digital array architectures may be replaced by the more energy-efficient analog or hybrid analog/digital beamforming architectures, or low-resolution ADCs, as discussed in Section~\ref{sec:simplerHardware}. Such devices or architectures have very limited hardware or signal processing capabilities, which makes the training-based channel acquisition more difficult and renders CKM-based approach more practically appealing.

%\begin{itemize}
%\item Channels for yet-to-reach or never-to-reach locations: conventional training based CSI acquisition methods are doable only after the communication devices reach the corresponding locations. However, channel knowledge is often useful beforehand, e.g., proactive resource allocation, foresighted handover to avoid ¡°ping-pong¡± phenomenon, communication-aware motion planning for robots [88], aerial node placement and trajectory planning for integrated aerial-ground network [6][7][75].  Similarly, CKM is also useful to predict channels with sleeping BSs [53]or sleeping terminals.
%\item Channels for non-cooperative nodes: When one side of the communication link is non-cooperative, conventional training-based CSI acquisition is difficult, if not impossible. For example, the link from secondary transmitter to primary receiver in cognitive radio systems, the link from legitimate transmitter to eavesdropper for physical-layer security
%\item Channels with large-dimensions: massive MIMO, XL-MIMO, cell-free massive MIMO
%\item Channels with severe hardware/signal processing limitations: analog beamforming, hybrid analog/digital beamforming, low-resolution ADCs, fully passive transmission/reflection with RIS[4], wireless power transfer, backscattering communication¡­
%\end{itemize}
%
%

\subsection{CKM Versus Other Relevant Technologies}\label{sec:relevantTechniques}
In this subsection, we discuss some other relevant technologies for channel inference or extrapolation. The major differences between CKM and such technologies are highlighted. The relation of CKM with fingerprinting-based localization is also discussed.

\subsubsection{Time-domain channel prediction}
{As illustrated in Fig.~\ref{F:Channel_inference}(a)}, the key idea of time-domain channel prediction is to forecast the future CSI based on the past channel observations, so as to avoid the channel aging issue caused by the delays required by channel training and feedback. Channel prediction is feasible since wireless channels are temporally correlated. Some informative references for fading channel prediction can be found in e.g. \cite{3015,3016}. The commonly used channel prediction methods include auto-regressive (AR) model-based prediction, parametric model (PM)-based prediction, and neural networks (NN)-based prediction. In AR model-based prediction, the evolution of wireless channels is treated as a random process, and the future channel sample can be estimated as a linear combination of the $p$ previous channel samples, with $p$ being the AR-model order. The optimal linear coefficients to minimize the \ac{mse} can be obtained based on the auto-correlation function of the random channel process. On the other hand, for PM-based channel prediction, the channel is modelled as a sum of complex exponentials (or sum-of-sinusoids, SOS), each corresponding to one multi-path channel component with parameters including amplitudes, Doppler frequency, and phases. After such parameters are estimated by algorithms like MUSIC or ESPRIT, future channels can be reconstructed. Lastly, NN-based channel prediction is a data-driven approach that avoids the complex algorithms for parameter estimation, and channel prediction is realized by the powerful time-series prediction capability of neural networks, such as recurrent neural networks (RNNs).

While time-domain channel prediction is an effective method to resolve the outdated CSI issue, frequent channel training is still needed, so as to continuously provide accurate past channel observations to make meaningful future predictions. From this perspective, it is unlikely that channel prediction can achieve significant saving of real-time channel training overhead, as achieved by techniques like CKM. Besides, channel prediction only achieves {\it intra-device} channel inference in the time domain, i.e., it only uses the past channel observations of one single device to forecast its own future channel. By contrast, CKM can achieve {\it inter-device} channel inference across time, frequency, and spatial domains, by making use of the historical data of all devices that have visited the same area.

\subsubsection{Channel inference using out-of-band measurements}
As wireless communication systems are expected to concurrently support multi-band operations, inferring channel knowledge using out-of-band measurements has received fast growing attention recently \cite{3005,3006,3008,3007,3009,3010,3013,3014,2124}, {as illustrated in Fig.~\ref{F:Channel_inference}(b).} This is feasible since though channel gains at different bands become uncorrelated if their frequency separation exceeds the channel coherence bandwidth, they may have high spatial correlation since their underlying multi-path signals interact with the same physical objects in the environment. This has been experimentally verified in \cite{3005,3006}. There are two major application scenarios for utilizing out-of-band measurement information. The first one is channel extrapolation for FDD massive MIMO systems \cite{3008,3007,3009,3010,3017,3018}, for which channel reciprocity does not hold and extrapolating the downlink channel from the uplink measurements becomes more challenging than TDD systems. By exploiting the fact that the same underlying physical paths would be traversed by signals of different frequencies, the authors in \cite{3007} proposed a scheme to infer the downlink channel by using uplink measurements, by firstly extracting those frequency-independent parameters of channel paths. The second major use case of out-of-band channel extrapolation is mmWave beam alignment by using sub-6 GHz measurements. While the frequency separation for FDD uplink-downlink channel extrapolation is relatively small, say 30 MHz in \cite{3007}, a more aggressive use of out-of-band measurements is to infer the mmWave beam direction based on sub-6 GHz signal measurement \cite{3005,3006,3013,3014,2124}. For example, \cite{3005} proposed a scheme to infer the 60GHz mmWave beam direction by overhearing the signals from 2.4/5 GHz, based on the estimated angular profile under LoS condition. Similarly, in \cite{3006}, coarse AoA is firstly estimated using sub-6 GHz channel measurements, which is then used to reduce the training overhead for mmWave band. Similar mmWave beam inference or link blockage techniques have been studied for NLoS scenarios \cite{3013,3014,2124}.

Similar to time-domain channel prediction, channel inference using out-of-band measurements only achieves {\it intra-device} channel extrapolation in the frequency domain. Besides, the quality of inference critically depends on the correlation across different bands, which usually becomes weaker as the frequency separation becomes larger. Note that even confronting the same physical objects in the environment, signals differing significantly in frequencies may experience different radio interaction mechanisms. For example, an object that is electrically small for sub-6G signals may appear large for mmWave signals, which may cause diffraction or reflection for sub-6G and mmWave signals, respectively. Besides, the dielectric properties of the same physical material may differ for different frequency bands. Such discrepancies need to be considered if high-quality channel inference using out-of-band measurements are desired. Besides, extensive channel measurements and sophisticated signal processing are still needed for utilizing out-of-band information. By contrast, with CKM, only location information is required, based on which the location-specific channel knowledge can be inferred directly for different frequencies.

\subsubsection{Channel-to-channel mapping} 
\begin{figure*}
\centering
\includegraphics[scale=0.50]{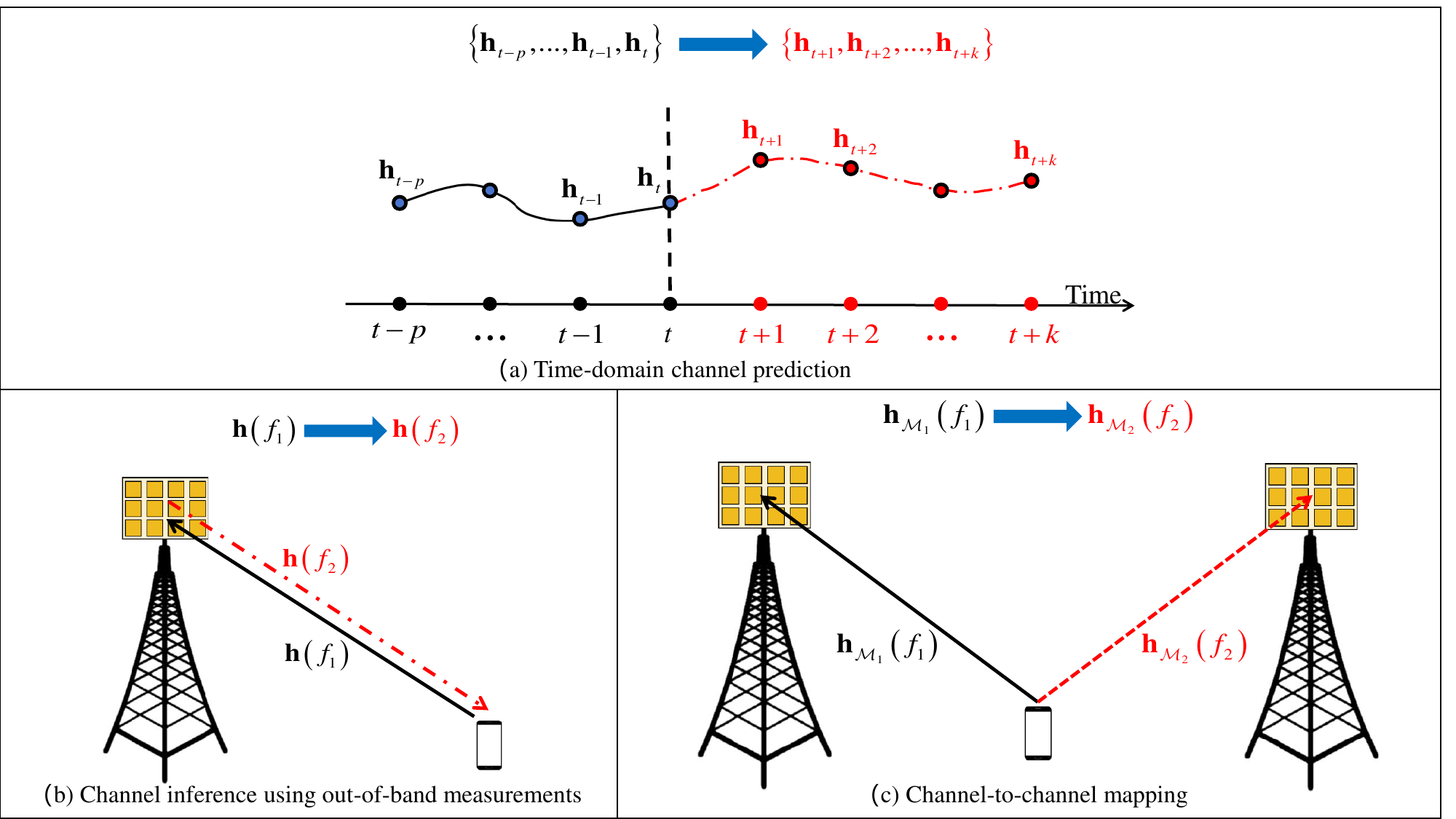}
\caption{Different channel inference technologies.}\label{F:Channel_inference}
\end{figure*}

While the above two techniques try to achieve channel inference in time and frequency domains, respectively, another technique, termed {\it channel-to-channel mapping}, was proposed in \cite{2127} trying to achieve channel inference in both frequency and space domains, {as illustrated in Fig.~\ref{F:Channel_inference}(c)}. Specifically, the authors tried to answer the following question: for a given user, if its CSI $\mathbf h_{\mathcal M_1}(f_1)$ corresponding to one set of BS antennas $\mathcal M_1$ over a frequency $f_1$ is known, is it possible to directly infer its channel $\mathbf h_{\mathcal M_2}(f_2)$ associated  with another set of BS antennas $\mathcal M_2$ over a possibly different frequency $f_2$? The answer is affirmative, if the mapping between user location and CSI is bijective, i.e., every user location in the candidate location set has a unique channel vector $\mathbf h_{\mathcal M_1}(f_1)$. Such a conclusion is intuitively understandable, since under the bijective location-to-channel mapping, it is feasible to actually localize the user based on the CSI $\mathbf h_{\mathcal M_1}(f_1)$. Furthermore, since the mapping from user location to CSI like \eqref{eq:abstract_function_static} exists, a channel-to-channel mapping from $\mathbf h_{\mathcal M_1}(f_1)$ to $\mathbf h_{\mathcal M_2}(f_2)$ is established with the bridging of user location. Since such a channel-to-channel mapping is highly nonlinear and difficult to characterize, the authors in \cite{2127} proposed a deep learning based method to approximate the complex channel mapping functions.

Different from CKM, channel-to-channel mapping does not attempt to explicitly use the user location information, which is appealing from the privacy protection perspective. However, this also limits its saving of the real-time channel training  overhead and restricts its  application scenarios. Specifically, the assumption of bijective location-to-channel mapping implies that the antenna set $\mathcal M_1$ that still requires real-time channel acquisition cannot be too small and they need to satisfy certain geometric relationships. By contrast, CKM does not rely on the assumption of bijective location-to-channel mapping nor real-time CSI for channel knowledge inference, and its potential saving of  channel training overhead is more substantial. Compared to channel-to-channel mapping, CKM essentially bypasses the demanding task to firstly localize the user (implicitly), but instead uses the already available location information for channel prediction, thanks to the abundant advanced localization techniques in contemporary wireless systems, as discussed in Section~\ref{sec:more_accessible_loc}. Of course, if such external localization systems are unavailable, cellular based localization can still be used to provide the location information required by CKM. Note that channel-to-channel mapping relies on deep neural networks to approximate the mapping relationships, which is a black box technique and requires extensive labelled data. By contrast, CKM is more explainable and more flexible, in terms of the channel knowledge to be inferred and the map construction methods to be used. It is also worth remarking that the privacy issue that might be a concern to skeptics of CKM can be effectively addressed by techniques like crowdsourcing and virtual location, as will be elaborated in Section~\ref{sec:conclusion}.

\subsubsection{Radar/Lidar/Vision-aided communication} Wireless communications by using measurement data from various sensors, such as Radar \cite{3003,3002,2112,2119}, Lidar \cite{2116,2118,2129,2131} or camera \cite{2114,2117,2120,2121,126}, have received fast growing attention recently. For example, dedicated radar sensors such as the classical \ac{fmcw} radars can be installed collocated with the communication BS or roadside unit (RSU), and lidar or vision-based sensing have been widely adopted in contemporary vehicles or robots. As such, the sensing signals and communication signals would experience the same wireless environment \cite{3002}, which makes it possible to infer the communication channel knowledge like beam directions based on sensory measurement. There are in general two ways to achieve such goals. The first one is to firstly explicitly estimate the AoA/AoD or location of the communication UE based on the sensory data \cite{3003,3002}, and then infer the beam direction based on the estimated information. The second approach is to directly infer the best beam pair based on the sensory data without trying to estimate the intermediate sensing parameters such as the AoA/AoD or UE location \cite{2112,2119}. %Since such an inference is highly nonlinear in general, deep learning methods are usually used to train artificial neural networks with labelled data.

While radar/lidar/vision-aided communications are quite attractive to reduce the real-time training overhead without consuming the precious communication resources, additional hardware, waveform, and signal processing are usually needed, which increases the cost, size, and complexity of communication systems. To resolve such issues, there are some preliminary research efforts to predict the beam direction using the reflected communication signal from the target UE \cite{3043}, without the need of transmitting separate radar signals. However, this relies on a strong assumption that the dominating LoS link always exists between the BS and UE. Besides, deep learning-based approach for beam alignment based on sensory data requires huge labelled data and costly neural network training. Moreover, vision-aided communications are vulnerable to bad weathers and its sensing range is limited due to the vision LoS requirement. 
By contrast, the CKM-enabled communications can be implemented without such requirements by exploiting the environment awareness. 
Besides, the additional measurements from radar/lidar/vision can also be exploited by CKM for more accurate prediction.

\subsubsection{Fingerprinting-based localization and channel charting} The key idea of CKM is to construct site-specific databases and infer location-specific channel knowledge by using device location information. This makes CKM conceptually the inverse process of the extensively studied fingerprinting-based localization \cite{2073,3019,3020}, whose objective is to estimate the device location based on radio measurements, such as RSS and CSI. In fact, both techniques are built upon the fact that for a given environment, wireless channels are critically dependent on device locations. However, a subtle difference between the two is that for fingerprinting-based localization,  unambiguous localization would require that the fingerprints and set of anchor nodes be deliberately chosen so that different locations would lead to distinct fingerprints, similar to the bijective location-to-CSI mapping assumption for channel-to-channel mapping technique.  By contrast, this requirement is not needed for CKM-based channel inference. Therefore, fingerprinting-based localization was mainly used for indoor scenarios, not just because conventional techniques like \ac{gnss} fails in such scenarios, but also indoor environment provides much richer multi-path fingerprints. By contrast, CKM is expected to work well for both indoor and outdoor due to  the relaxation of bijective mapping criterion. It is worth noting that with the growing interest of ILSAC, future wireless networks are expected to regularly serve both UEs requiring communication services and those requiring localization/sensing services. As such, constructing CKMs would achieve the benefit of ``killing two birds with one stone", since the same CKM can serve dual purposes: used in a forward manner to facilitate CSI acquisition for communication UEs and in a reverse manner as fingerprinting-based positioning for localization UEs. This hopefully would provide more incentives for various stakeholders to construct CKMs.

Recently, another localization technique called {\it channel charting} was proposed \cite{2007}, which is an unsupervised technique without requiring the user location information for offline training or online inference. The key idea is to construct a low-dimensional channel chart from the high-dimensional CSI that locally preserves the spatial geometry, i.e., UEs that are close in  physical space will also be close in the channel chart. Compared to fingerprinting-based localization, channel charting does not require extensive offline measurement, but it can only infer the relative positions between UEs, rather than absolute locations. Table~\ref{table:channelInference} provides a summarized comparison between the CKM versus the above relevant channel inference techniques.

{Finally, it is worth remarking on the relation between CKM and ISAC techniques. In  a nutshell, CKM and ISAC are two different concepts that were  developed  for addressing different challenges in wireless systems. Specifically, ISAC mainly aims to integrate wireless communication and radio sensing functionalities that were independently developed over the past few decades, so as to more efficiently utilize the wireless infrastructure, hardware, signal processing modules, and radio resources. On the other hand, CKM was proposed mainly to enable environment-aware wireless systems, so as to address the challenging issue of  CSI acquisition in future networks. %With CKM, various a priori channel knowledge can be obtained based on devices' physical or virtual locations, so that real-time CSI can be more efficiently acquired. 

On the other hand, ISAC and CKM are closely related in various aspects. For example, during the CKM construction phase, ISAC may provide a way to collect location-tagged data that reflects the local wireless environment. Besides, during the CKM utilization phase, ISAC can be an efficient way to sense or localize the device and/or scatterer, so that the corresponding {\it a priori} channel knowledge can be efficiently learned in CKM. It is worth emphasizing that ISAC is not the only method for serving the aforementioned purposes for CKM construction and utilization, since the same purposes can also be achieved by alternative techniques such as separate sensing and communication, ray tracing-generated data, and GPS. On the other hand, CKM can also be used to further enhance the performance of ISAC. For example, if a high-quality CKM has been built, it may help avoid unnecessary repeated sensing in ISAC. Besides, CKM can also be useful to enhance sensing performance in terms of efficient clutter rejection, anchor node selection, sensing resource allocation, etc. Some preliminary results toward this direction are discussed in Section~\ref{sec:utilization:6}.}

\begin{table*} \footnotesize
\caption{Comparison of CKM and relevant channel inference techniques.}\label{table:channelInference}
\centering
\begin{tabular}{|p{2cm}|p{4.5cm}|p{4.5cm}|p{5cm}|}
\hline
{\bf Channel Inference Techniques} & {\bf     ~~~~~Main characteristics} & {\bf     ~~~~~Pros} & {\bf     ~~~~~Cons}\\
\hline
\vspace{0.1in} Time-domain channel prediction\cite{3015,3016}
& \begin{itemize}[leftmargin=1em]
\item Forecast future CSI based on past observations
\item   Avoid channel aging caused by channel training and feedback delays\end{itemize}
 &  \begin{itemize}[leftmargin=1em]
 \item  Historical channel observations are readily available \end{itemize}
 &  \begin{itemize}[leftmargin=1em]
 \item Time-domain intra-device channel inference only
\item Frequent channel measurements still needed
\end{itemize}\\
\hline
\vspace{0.1in} Channel inference using out-of-band measurements\cite{3005,3006,3008,3007,3009,3010,3013,3014,2124}
& \begin{itemize}[leftmargin=1em]
\item Infer channel using measurements at a different band
\item Use case 1: FDD downlink channel extrapolation using uplink measurements
\item Use case 2: mmWave inference using sub-6 GHz measurements\end{itemize}
& \begin{itemize}[leftmargin=1em]
\item Utilize the fact that multi-path signals of different bands interact with the same physical objects \end{itemize}
& \begin{itemize}[leftmargin=1em]
\item Frequency-domain intra-device channel inference only
\item Quality of inference degrades with large frequency separation
\end{itemize}\\
\hline
\vspace{0.1in} Radar/lidar/vision based channel inference\cite{3003,3002,2112,2119,2116,2118,2129,2131,2114,2117,2120,2121,126}
& \begin{itemize}[leftmargin=1em]
\item Use sensory outputs of radar/lidar/camera for channel inference\end{itemize}
& \begin{itemize}[leftmargin=1em]
\item Do not consume precious communication resource \end{itemize}
& \begin{itemize}[leftmargin=1em]
\item Additional infrastructure for radar/lidar/vision sensing required
\item Mainly rely on LoS condition
\item Vulnerable to bad weathers
\end{itemize}\\
\hline
\vspace{0.1in} Channel-to-channel mapping\cite{2073,3019,3020,2007}
& \begin{itemize}[leftmargin=1em]
\item Use deep learning to predict channels across frequency/space\end{itemize}
& \begin{itemize}[leftmargin=1em]
\item Channel inference across both frequency and space domains
\item Do not explicitly localize user \end{itemize}
& \begin{itemize}[leftmargin=1em]
\item Relies on bijective mapping between user location and CSI
\item Heavy real-time CSI measurement still needed
\item Typically relies on ¡°blackbox¡± implementation via neural networks
\item Do not utilize the almost ¡°free¡± a-prior location information such as GPS
\end{itemize}\\
\hline
\vspace{0.1in} Channel knowledge map (CKM) &
\begin{itemize}[leftmargin=1em]
\item Channel knowledge inference from site-specific database based on (virtual) location
 \end{itemize} &
 \begin{itemize}[leftmargin=1em]
 \item Inter-device channel inference across time/frequency/space
 \item Utilize the almost ¡°free¡± a-prior location information such as GPS
 \item Dual usage for forward channel inference and backward localization
\end{itemize}&
\begin{itemize}[leftmargin=1em]
\item Requires offline CKM construction
\item CKM in dynamic environment
\end{itemize}\\
\hline
\end{tabular}
\end{table*}

\subsection{Summary}
In summary, CKM is a promising new technique that utilizes the location-specific channel data, either from  the historical measurements or offline ray tracing, to enable environment-aware communications, by providing more accurate {\it a priori} knowledge about the local wireless environment than the conventional environment-unaware communication. As such, it may significantly reduce the real-time channel acquisition overhead and enhance the channel estimation quality, which thus provides a promising solution to the challenging channel acquisition problem in 6G with dense  wireless links and large channel dimensions. One appealing advantage of CKM over alternative channel inference techniques lies in that it can utilize the almost ``free'' (virtual) location information of mobile devices that can be obtained by a multitude of positioning or sensing techniques. Besides, another unique feature of CKM is its inter-device channel inference capability across time, frequency, and space domains. {It is worth mentioning that CKM is by no means exclusive to existing training overhead reduction or channel inference techniques. In fact, combining CKM with such existing techniques may maximally exploit their respective advantages. For example, time-domain and cross band channel inference may help to reduce the required measurement data for CKM construction.} Therefore, CKM-enabled environment-aware communication may avoid the unnecessary repeated channel estimation or environment sensing, and hence is expected to significantly enhance the communication, sensing, and environment reconstruction performance in 6G.

\section{Constructing Channel Knowledge Maps}\label{sec:construc}
In order to realize CKM-enabled communications, we need to first construct the CKM in an efficient manner. {Note that CKM can be most effectively constructed for static or semi-static environment, where only a small portion of scatterers are time-varying. Such an environment can be viewed as the static background of the mobile device, and the CKM aims at capturing the channel information over such background.} 

\begin{figure}
\centering
\includegraphics[scale=0.38]{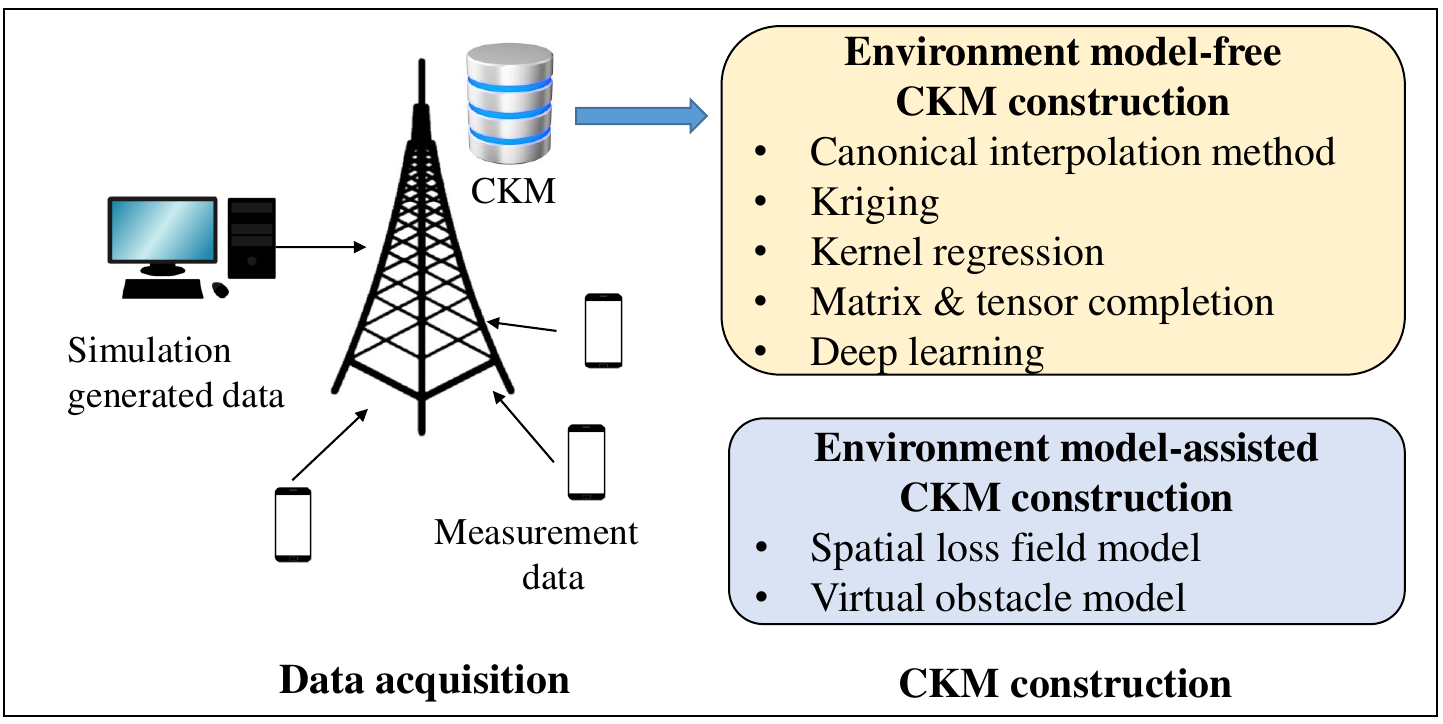}
\caption{An illustration of data acquisition and CKM construction methods.}\label{F:CKMconstruction}
\end{figure}

Fundamentally, the CKM construction corresponds to interpolating the channel knowledge of all interested locations based on the data acquired
at a few locations. In this section, we first give an overview
on the CKM data acquisition, and then introduce {environment model-free} CKM
construction approaches and {environment model-assisted} approaches respectively, as illustrated in Fig.~\ref{F:CKMconstruction}. Finally, we
provide a performance evaluation approach for CKM construction.

\subsection{Data Acquisition}

An essential issue in CKM construction is the collection of location-specific
data for the actual radio environment. %The data acquisition is different from the conventional channel measurements in wireless systems in the following aspects. First, conventional channel measurements during the pilot phase of a transmission aim for link resource allocation and link adaptation. As a result, the measurements include both large timescale quantities, such as reference signal received power (RSRP), and small timescale quantities, such as rank indicator (RI) and precoding matrix index (PMI). By contrast, most CKM focuses on predicting the radio propagation features in both small and large timescale, such as path loss, shadowing, \ac{aoa}/\ac{aod}, and so on. Second, conventional channel measurements usually do not require the location information, because radio resource allocation and link adaptation are based on the channel measurement rather than the location of the device. In fact, traditionally, the location information of a wireless device is usually not directly available to the network due to privacy concern. However, in CKM, the location information, either true location or virtual location, is an essential quantity to be acquired because the CKM construction and exploitation are at least implicitly based on the location of the device. Finally, in a conventional system, all the measurements are instantaneous and the data is mostly kept locally except for very limited signaling among devices. As a result, the issues of channel data storage and data privacy were not considered. However, in CKM, message passing for CKM measurements, data storage, and privacy are important issues to be addressed.
% Due to the above challenges, data collection for CKM construction deserves further studies. 
The two main streams of data acquisition
methods are simulation-based and measurement-based, respectively.

\subsubsection{Simulation-generated Data}

Physical environment maps or city maps may be used to simulate wireless channels at
different transceiver locations. The models used for simulating wireless
channels range from a fully stochastic model to a fine-grained
deterministic ray-tracing model depending on the information granularity
of the city maps, the precision requirement and
the geographical scale of the radio map \cite{110,111}.

Stochastic channel models are widely adopted for simulating wireless
channels for terminals scattered in a large area, such as in a city.
Typical application scenarios include constructing city-level
radio maps for coverage analysis and BS planning. Physical environment maps or city maps play a
role in determining an appropriate propagation \emph{scenario} for
the configuration of model parameters. A typical set of scenarios
includes sub-urban, urban, urban rooftop-to-street, etc., which
can be inferred from a city map, or more generally, a geographic database.
A typical stochastic channel model constructs the channel gain from
three parts, including path loss, shadowing, and small-scale fading. First,
the path loss component computes the signal attenuation as a function
of distance. 
%, where the parameters are dependent on the propagation scenarios. For instance, signal typically has a stronger attenuation over distance in a dense urban environment than in a suburban environment. 
Second, the shadowing component is usually modeled as a spatial random process with a configured drift and covariance depending on the chosen scenario. % If a mobile device locates indoor, the shadowing with respect to all outdoor BS are strongly correlated; for an outdoor user, the shadowing with respect to BSs at different directions are less correlated. 
Finally, the computation of the small-scale fading
component requires to specify the distribution of the fading based on some priori information such as the probability of the existence of an LoS Link. % If a device is in a rich scattering environment without LoS link,  a Rayleigh distribution is a proper choice to model the small-scale fading. If there is a strong reflection path or LoS path, then adopting a Rician distribution may better model the fading component. T
%o summarize, even for a classical fully stochastic
%channel model, the geographical information and the user locations
%are essential to determining the key parameters for the model.

In contrast to the stochastic channel models, ray-tracing approaches require
fine-grained details of the environment for simulating the wireless
channel. The environment information includes, but not limited to,
the location, the geometry model, and the material of the surrounding
objects, the electromagnetic characteristics of the scatterers, and
the dielectric parameters of the atmosphere and the terrain. As a
result, while a ray-tracing approach may generate channel data with
finer details and higher accuracy, it requires much more geographical
information and a substantially higher computational complexity than
the stochastic channel models. An example workflow
is as follows. First, environment models are built, where the surrounding
can be modeled as a collection of geometry objects. Then, based on
the locations of the transmitter and receiver, propagation rays are
simulated. Specifically, for each planar surface, the angles of an
incident ray and an emergent ray can be calculated. Such a process
can be very complicated because the number of specular reflective
rays substantially increases if the number of permitted reflections
increases; and in addition, if scattering is simulated, then every
planar surface can launch a ray. The computational complexity tremendously
grows even by slightly increasing the numbers of simulated reflective,
diffractive, and scattering rays. As a result, system engineers need
to carefully specify these parameters to balance the complexity and
accuracy trade-off. Third, for each ray, a propagation model is applied
to calculate the signal attenuation, where the model parameters depend
on the material of the scatterers and the parameters of the atmosphere.
Finally, the multi-paths need to be integrated while antenna configurations
are taken into account.

Hybrid models can take advantages of both stochastic models and ray-tracing
based models, and thus, balance the complexity and accuracy. There
could be several levels of integration, depending on the generation
of scatterers and channel parameters \cite{110}. In a high-tower
BS urban user scenario, a \emph{map-based stochastic channel model}
corresponds to generating the large-scale component from ray-tracing
approaches, and small-scale component due to local scatterers from
stochastic models. A \emph{map-based hybrid model} corresponds to
generating the large-scale component from ray-tracing, but generating
the small-scale component based on a merged deterministic and stochastic
scattering environment, for instance, with half of the local scatterers
analyzed from ray-tracing based on the map data, and the other half
of the local scatterers modeled as random. Furthermore, real-world
measurements can be integrated to refine the model parameters.

\subsubsection{Measurement data}

There are two major practices to collect real data from communication
networks: dedicated offline measurement and online measurement. In
offline measurements, dedicated devices are sent to collect measurement
data from the network. The measurement campaign involves a carefully
planned measurement route, controlled mobility speed, configured communication
scenarios, and dedicated data cleaning. While the process requires
intensive labor effort, the human-in-the-loop measurement campaign
provides high-quality datasets for performance evaluation and network
optimization. For online measurement, data is collected and stored
during the conventional channel measurement, estimation, and communication
processes. Instead of discarding them immediately after the communication
session, the data can be kept for CKM construction.

To enhance the measurement capability, minimization of drive tests
(MDT) is a feature that has been introduced since 3GPP Release 9 for
the collection of radio measurements from UE while reducing
the traditional drive tests for the operator to assess the network
performance \cite{85,87}. % MDT has been standardized for \ac{lte} and is built with two main features: 1) periodic reporting of GPS locations of the UE, if GPS is enabled, and 2) periodic reporting of Layer 3 measurements if the UE is at idle mode and Layer 2 measurements if the UE is at a connected mode. Layer 3 measurements include GPS locations, RSRP and \ac{rsrq} of both the serving cell and neighboring cells, and Layer 2 measurements include wide-band \ac{cqi}, rank indicator, and modulation related parameters. The UE keeps the measurement logs and may report to the network. When the network wants to retrieve the logged message indicated by the UE, it may trigger a process to acquire the measurement reports.
While MDT provides a powerful tool for the network operator to build
CKMs, its features may not be fully implemented
due to commercial reasons or may be partially disabled by the UE due
to battery or privacy concerns. Therefore, a key issue here is how
to build a full CKM while only partial measurement data is available
to the network \cite{85}. This is to be discussed in the following
subsections.

\subsection{{Environment} Model-free CKM Construction}

{For environment model-free approaches, their applicable scenarios are usually lack of parametric models to characterize the signal propagation as a function of the geometry structure of the geographic environment. The CKM construction is essentially an interpolation or extrapolation problem in the spatial domain, without explicitly exploiting the hidden correlation of the measurement data due to the geometry structure of the environment.}

A classical spatial interpolation problem in 2D area is to construct a function
$f(\mathbf{q})$ for each location $\mathbf{q}$ in a bounded area
$\mathcal{D}\subseteq\mathbb{R}^{2}$ based on $N$ scattered measurements
$\{z_{i}\}$ taken at locations $\{\mathbf{q}_{i}\}$. For CKM construction,
the methods commonly adopted in recent literature include Kriging \cite{14,36},
kernel regression, and matrix completion \cite{42}. Kriging yields the best unbiased linear estimator under some stationary assumption on the spatial process $f(\mathbf{q})$. The kernel approach models $f(\mathbf{q})$ as a linear combination of a set of kernel functions, where the weights are computed from the data. The matrix completion approach views the radio map as a matrix sampled from a grid topology, and it completes the matrix from sparse observations by exploiting the low rank assumption on the matrix. There are also other
related methods, including $K$-nearest-neighbor (KNN) \cite{58},
inverse-distance-weighted (IDW) \cite{ChiDel:B09}, and polynomial
regression \cite{FanGij:B18,SunChe:C22}, which appeared in earlier literature and are thus regarded as canonical interpolation methods. In the following, we briefly elaborate and compare the canonical interpolation methods, Kriging, kernel regression, and matrix completion methods for CKM construction. 

\subsubsection{Canonical Interpolation Methods}

The IDW method was first proposed in the literature on geostatistics and REM
construction. The method constructs $f(\mathbf{q})$ at $\mathbf{q}_{0}$
by the weighted average of the measurements, $f(\mathbf{q}_{0})=\sum_{i}w_{i}z_{i}$,
where the $i$th weight $w_{i}$ is proportional to the reciprocal
of the distance $\|\mathbf{q}_{0}-\mathbf{q}_{i}\|$ to the $i$th
measurement at $\mathbf{q}_{i}$. More generally, the weights can
be chosen as $w_{i}=c\|\mathbf{q}_{0}-\mathbf{q}_{i}\|^{-\alpha}$,
where $c$ is a normalizing constant such that the weights sum to
$1$. It is known  that a proper choice of $\alpha$ can lead to a good
estimate. However, it is in general very difficult
to specify a good parameter $\alpha$, because the propagation field
could be very complicated and there is usually a lack of prior information
of the propagation environment.

KNN method constructs $f(\mathbf{q})$ at $\mathbf{q}_{0}$ as the
weighted average of the $K$ measurements $z_{i}$ selected from the
nearest neighbors \cite{58}. Specifically, one computes $f(\mathbf{q}_{0})=\sum_{k\in\mathcal{N}(\mathbf{q}_{0})}w_{k}z_{k}$,
where $\mathcal{N}(\mathbf{q}_{0})$ is a set that collects the $K$
measurements that yield the smallest distance $\|\mathbf{q}_{0}-\mathbf{q}_{i}\|$,
and $w_{k}$ are the weights that can be specified by the IDW rule.
In addition, a kernel function $w_{k}=w(\mathbf{q}_{0},\mathbf{q}_{k})$
can be used to compute the weight. Some simple choices of kernel functions
include Gaussian kernel $w(\mathbf{q}_{0},\mathbf{q}_{k})=c\cdot\mbox{exp}(-||\mathbf{q}_{0}-\mathbf{q}\|^{2}/\sigma)$
and Laplacian kernel $w(\mathbf{q}_{0},\mathbf{q}_{k})=c\cdot\mbox{exp}(-||\mathbf{q}_{0}-\mathbf{q}\|/\sigma)$.

Local polynomial regression is a powerful tool for many interpolation
problems. To begin with, a first-order \emph{global} polynomial model can be written as $\hat{f}(\mathbf{q};\mathbf{a})=a_0+a_1(\mathbf{q}-\mathbf{q}_0)$, where the coefficients $\bm{a}=[a_{0},a_{1}]$ are obtained
by solving a least-squares problem to minimize $\sum_{i=1}^{N}(z_{i}-\hat{f}(\mathbf{q}_{i};\bm{a}))^{2}$. The least-squares problem is convex and there exists a closed-form
solution. However, it is known that the estimate $\hat{f}(\mathbf{q};\bm{a})$
may be accurate just around the chosen reference location
$\mathbf{q}_{0}$, and therefore, a global model has very limited representability of the data. The \emph{local} polynomial regression adopts the same model
but it adapts the reference point $\mathbf{q}_{0}$ to each location
$\mathbf{q}$ to be interpolated, and determines the coefficients $\bm{a}$
using data only in the neighborhood of $\mathbf{q}_0$. As such, the least-squares problem is modified into
a weighted least-squares $\sum_{i=1}^{N}(z_{i}-\hat{f}(\mathbf{q}_{i};\bm{a}))^{2}w(\mathbf{q}_{0},\mathbf{q}_{i})$,
where $w(\mathbf{q}_{0},\mathbf{q}_{i})$ can be a kernel function
that assigns a large weight to a small distance between the reference
location $\mathbf{q}_{0}$ and the $i$th measurement location $\mathbf{q}_{i}$,
a small weight to a large distance, and a zero weight for measurements
beyond a distance from $\mathbf{q}_{0}$. In other words, there are
individual polynomial models $\hat{f}(\mathbf{q};\bm{a}(\mathbf{q}_{0}))$
for each location $\mathbf{q}_{0}$, and the polynomial coefficients
$\bm{a}(\mathbf{q}_{0})$ are estimated from the weighted least-squares
problem mostly based on the measurements near $\mathbf{q}_{0}$.

In the  local polynomial regression, the weighting function $w(\mathbf{q}_{0},\mathbf{q}_{i})$ is an important
component that affects the regression performance.
Consider a parametric form of the weighting function $w(\mathbf{q}_{0},\mathbf{q}_{i};\sigma)$
that depends on a parameter $\sigma$. The parameter $\sigma$ may
have a strong impact on the performance of the interpolation performance.
Nevertheless, there exists a systematic method to analyze the MSE
$\mathcal{E}(\sigma)$ of the expected interpolation error, and consequently,
a cross-validation approach may be adopted to pick the best parameter
$\hat{\sigma}$ for the least interpolation MSE. As a result, the
local polynomial regression will be less sensitive to the choice of
the weighting function $w(\cdot)$. Furthermore, the choice of order
$p$ is more an engineering problem than a mathematical tractable
one, while some theoretical analysis is given in \cite{FanGij:B18}.
A rule of thumb for the choice of $p$ is to consider both the volume
of the measurement data and how much variation that the CKM $f(\mathbf{q})$ is believed to have.

\subsubsection{Kriging Interpolation}

Kriging is a powerful interpolation tool in geostatistics for the
reconstruction of a geostatistics map over an area based on a limited
number of observations. It has been widely used for map construction. The general philosophy is to first
analyze the data modeled as spatial statistics, and then a predictor
is designed with some statistical property enforced. The methodology
is developed within the scope of a second-order statistical model
involving only the mean $m(\mathbf{q})$ and the covariance $C(\mathbf{q},\mathbf{q}')$
assumed known. 

There are several variants of Kriging. \emph{Simple Kriging} assumes that
the mean $m(\mathbf{q})=m$ is constant and known. \emph{Ordinary
Kriging} assumes that the mean is constant but unknown, and therefore,
the mean needs to be estimated first. \emph{Universal Kriging} assumes that
the mean $m(\mathbf{q})$ is varying and unknown, and hence, an estimator
$\hat{m}(\mathbf{q})$ needs to be developed to track the mean.

Denote $\{\mathbf{q}_{i},z_{i}=f(\mathbf{q}_{i})\}$ as the set of measurements 
taken at locations $\mathbf{q}_{i}$. The Kriging method aims at constructing an unbiased estimator of the form
\begin{equation}
\hat{f}(\mathbf{q})=\sum_{i=1}^{N}\lambda_{i}z_{i}+\lambda_{0} \label{eq:kriging-estimator}
\end{equation}
where the weights $\lambda_{i}$ are selected to minimize the MSE
$\mathbb{E}\{(\hat{f}(\mathbf{q})-f(\mathbf{q}))^{2}\}$. 

Specifically,
$\lambda_{0}$ is chosen to cancel the bias. For a stationary random
function with zero mean $\mathbb{E}\{f(\mathbf{q})\}=0$, we have $\lambda_{0}=0$. In such a case, the MSE coincides with the variance
of $\sum_{i}\lambda_{i}f(\mathbf{q}_{i})-f(\mathbf{q})$, which can be shown as
 $\mathbb{V}\{\sum_{i}\lambda_{i}f(\mathbf{q}_{i})-f(\mathbf{q})\} =\sum_{i}2\lambda_{i}\gamma(\mathbf{q}_{i}-\mathbf{q})-\sum_{i}\sum_{j}\lambda_{i}\lambda_{j}\gamma(\mathbf{q}_{i}-\mathbf{q}_{j})$, where  
the function $\gamma(\mathbf{u})=\frac{1}{2}\mathbb{V}\{f(\mathbf{q}+\mathbf{u})-f(\mathbf{q})\}$ is termed as the semivariogram of $f(\mathbf{q})$ which has been mathematically considered as a stationary and isotropic spatial process. 

The minimizer $\bm{\lambda}^{*}$ of the above variance can be obtained by taking its derivative
to zero, leading to the following simple Kriging system $\bm{\Gamma}\bm{\lambda}=\bm{\gamma}_{0}$, 
where $\bm{\Gamma}$ is a matrix with the $(i,j)$th entry given by
$\gamma(\mathbf{q}_{i}-\mathbf{q}_{j})$ and $\bm{\gamma}_{0}$
is a vector with the $i$th element given by $\gamma(\mathbf{q}_{i}-\mathbf{q})$.
The simple Kriging system is solved by enforcing the equality constraint
$\sum_{i}\lambda_{i}=1$ for the unbiasedness of the estimator $\hat{f}(\mathbf{q})$.

The solution $\bm{\lambda}^{*}$ applied to (\ref{eq:kriging-estimator}) leads to the best unbiased
prediction under the stationary assumption with a constant and known
mean $m(\mathbf{q})$. Furthermore, the estimation variance $\sigma_{\mbox{\scriptsize SK}}^{2}$,
called the \emph{Kriging variance}, can also be found in closed form
\cite{ChiDel:B09}.

In general, to apply Kriging for CKM construction, one needs to first
construct an estimated semivariogram  $\hat{\gamma}(h)$ that approximates $\gamma(h)$ based on the collected data,
and then solve for the Kriging system $\bm{\Gamma}\bm{\lambda}=\bm{\gamma}_{0}$ with the parameters
$\bm{\Gamma}$ and $\bm{\gamma}_{0}$ obtained from the semivariogram.
Finally, the CKM can be obtained from (\ref{eq:kriging-estimator}).
However, there are additional challenges in applying Kriging for CKM construction in specific
communication scenarios. Several examples and tentative solutions
are discussed below.

% --
%

\subsubsection{Kriging for static CKM construction}

CKM construction requires smart
engineering and a systematic design with careful consideration on
data acquisition, message passing, information processing, and resource
allocation and control strategies. There have been studies on the
construction and exploitation of interference
maps \cite{15,96}, signal-to-interference-plus-noise ratio (SINR) maps \cite{31}, and shadowing maps \cite{100}.

\emph{Interference maps} and \emph{SINR maps} can be constructed as
a direct application of the Kriging principle. Several issues need to be discussed. First, the network architecture needs to be
considered according to the types of CKMs. In \cite{15}, an Interference Cartography Manager entity is introduced
that collects terminal measurements and builds a Kriging interpolated
interference cartography, which is stored in a database. Second, for SINR map construction, the heterogeneous structure of
practical cellular networks should be taken into account.  In \cite{31}, a two-tier Kriging interpolation
model is studied, where two separate spatial processes are constructed
for the Kriging interpolated SINR maps, $S_{\text{M}}(\mathbf{q})$
and $S_{\text{m}}(\mathbf{q})$, due to the transmission of the macro
\acpl{bs} and small \acpl{bs}, respectively. In addition, based
on the two-tier SINR maps, a spatial cross-tier correlation function
can be constructed as $R(\mathbf{u})=\mathbb{E}\{S_{\text{M}}(\mathbf{q})S_{\text{m}}(\mathbf{q}+\mathbf{u})\}$
to perform range expansion and coverage optimization for the heterogeneous
network.

Constructing \emph{Shadowing maps} require some modeling efforts before adopting the Kriging techniques. The concept of shadowing map appeared in \emph{ dynamic spectrum
access} in a \emph{cognitive radio network}, where a \ac{su} may
utilize the licensed spectrum without disrupting the communication
of the \acpl{pu}. Such a \ac{lsa} feature has been supported in
\ac{lte}. However, the fundamental challenge of \ac{lsa} is to estimate
the potential interference at the \ac{pu} due to the possible transmission
and power usage at the \ac{su}, where building a shadowing map to
express the radio environment for communication power control could
be of great help.

The basic approach is to first use Kriging interpolation to estimate the
large scale shadowing based on the \ac{rss} of the \ac{pu} measured
by the sensors, and then, to compute the interference at the \acpl{pu}
based on the shadowing map and the power allocation of the \ac{su}.
Specifically, consider a grid model of $N$ points of the target area.
Denote $\bm{p}_{\text{t}}\in\mathbb{R}^{N}$ as a power emission vector,
where the $j$th element of $\bm{p}_{\text{t}}$ represents the effective
isotropic radiated power at the $j$th grid point. Similarly, let
$\bm{p}_{\text{r}}\in\mathbb{R}^{N}$ be a vector that captures the
\ac{rss} at the corresponding grid points. Let $\bm{G}\in\mathbb{R}^{N\times N}$
be a path gain matrix, where the $(i,j)$th entry $G_{ij}$ captures
the large-scale path gain from the $i$th grid point to the $j$th
grid point that can be calculated by a propagation model with known
parameters. Similarly, let $\bm{S}\in\mathbb{R}^{N\times N}$ capture
the shadowing from each grid point $i$ to another grid point $j$.
As a result, a radio propagation model can be given by $\bm{p}_{\text{r}}=(\bm{G}\circ\bm{S})\bm{p}_{\text{t}}+\bm{n}$,
where ``$\circ$'' represents an element-wise multiplication, and
the vector $\bm{n}$ represents the noise. It is clear that if the
shadowing map $\bm{S}$ is known, then the \ac{rss} map can be dynamically
computed in terms of the power allocation $\bm{p}_{\text{t}}$.

To construct the shadowing map $\bm{S}$ based on the measurements
from $M$ sensors, one can construct a sensing matrix $\bm{\Phi}\in\mathbb{R}^{M\times N}$
to capture the location of the sensors, where the $(m,j)$th element
$\Phi_{m,j}=1$ if the $m$th sensor locates at the $j$th grid point,
and $\Phi_{m,j}=0$, otherwise. In addition, let $\bm{p}_{\text{s}}\in\mathbb{R}^{M}$
be a vector of the measured \ac{rss} from the $M$ sensors. As a
result, the measurement model can be constructed as $\bm{p}_{\text{s}}=\bm{\Phi}(\bm{G}\circ\bm{S})\bm{p}_{\text{t}}+\bm{\Phi}\bm{n}$.
Thus, partial entries of $\bm{S}$ can be estimated by solving the
measurement equation, forming a set $\mathcal{S}=\{\bm{c}_{ij},\hat{S}_{ij}:(i,j)\in\Omega\}$,
where $\bm{c}_{ij}\in\mathbb{R}^{4}$ represents the location pairs
from the $i$th grid point to the $j$th grid point, $\hat{S}_{ij}$
is the estimated entry of $\bm{S}$, and $\Omega$ is the set of entries
$(i,j)$ that can be determined from solving the equation. Finally, Kriging
interpolation can be used to construct the full shadowing map based
on the set of limited observations $\mathcal{S}$. An application
of the shadowing map to power allocation for cognitive transmission
is further discussed in \cite{100}.

% --
%

\subsubsection{Kriged Kalman filtering for time-varying CKM} 

Radio propagation
environment is inherently time-varying, and so are CKMs. To construct
and track the possibly time-varying CKMs, the spatio-temporal channel
correlation structure can be exploited using a \emph{Kriged Kalman
filtering} approach \cite{14}. Consider to construct a time-varying
shadowing map $f(\mathbf{q},t)$. Measurements
$z_{i}(t)=f(\mathbf{q}_{i},t)+\epsilon_{i}(t)$ at $N$ locations
$\mathbf{q}_{i}$ across $T$ time slots $t=1,2,\dots,T$ are needed
for the construction of $f(\mathbf{q},t)$, where $\epsilon_{i}(t)$
represents the measurement noise. Following the framework of Kalman filtering, the spatio-temporal Kriged Kalman
filter consists of two key steps. 

First, a state-space model is needed
$f(\mathbf{q},t) =\bar{f}(\mathbf{q},t)+\nu(\mathbf{q},t)$ and 
$\bar{f}(\mathbf{q},t)  =\int w(\mathbf{q},\mathbf{u})\bar{f}(\mathbf{u},t-1)d\mathbf{u}+\eta(\mathbf{q},t)$, where $w(\mathbf{q},\mathbf{u})$ describes the spatial correlation
from location $\mathbf{u}$ in the previous time slot $t-1$ to location
$\mathbf{q}$. The processes $\nu(\mathbf{q},t)$ and $\eta(\mathbf{q},t)$
are spatially correlated yet temporally white zero-mean Gaussian stationary
random processes, with covariance structure $\text{cov}\{\nu(\mathbf{q},t)\nu(\mathbf{u},\tau)\}=C_{\nu}(\mathbf{q}-\mathbf{u})\delta(t-\tau)$
and $\text{cov}\{\eta(\mathbf{q},t)\eta(\mathbf{u},\tau)\}=C_{\eta}(\mathbf{q}-\mathbf{u})\delta(t-\tau)$,
where $\delta(\cdot)$ denotes the Dirac delta function. In addition,
$\nu(\mathbf{q},t)$ and $\eta(\mathbf{q},t)$ are uncorrelated.

Second, a finite representation is needed for $\bar{f}(\mathbf{q},t)$
and $w(\mathbf{q},\mathbf{u})$ to reduce the dimensionality of the
state space. Note that the functions $\bar{f}(\mathbf{q},t)$ and
$w(\mathbf{q},\mathbf{u})$ are continuous in space and time, and the
set of measurement data is finite. One approach is to use a basis-expansion
representation based on a set of orthonormal basis $\psi_{k}(\mathbf{x})$,
$k=1,2,\dots,K$. Then, the two functions $\bar{f}(\mathbf{q},t)$
and $w(\mathbf{q},\mathbf{u})$ can be expressed as $\bar{f}(\mathbf{q},t) =\sum_{k=1}^{K}a_{k}(t)\psi_{k}(\mathbf{q})$, and 
$w(\mathbf{q},\mathbf{u}) =\sum_{k=1}^{K}b_{k}(\mathbf{q})\psi_{k}(\mathbf{u})$,
where $\mathbf{a}(t)=[a_{1}(t),a_{2}(t),\dots,a_{K}(t)]^{\text{T}}$
and $\mathbf{b}(\mathbf{q})=[b_{1}(\mathbf{q}),b_{2}(\mathbf{q}),\dots,b_{K}(\mathbf{q})]^{\text{T}}$
are the coefficients. Upon defining $\bm{\psi}(\mathbf{q})=[\psi_{1}(\mathbf{q}),\psi_{2}(\mathbf{q}),\dots,\psi_{K}(\mathbf{q})]^{\text{T}}$,
the state evolution can be written
as $\bm{\psi}(\mathbf{q})^{\text{T}}\mathbf{a}(t)=\mathbf{b}^{\text{T}}\mathbf{a}(t-1)+\eta(\mathbf{q},t)$.
Denote $\mathbf{B}$ as an $N\times K$ matrix with the $i$th column
given by $\mathbf{b}(\mathbf{q}_{i})$, $\bm{\Psi}\in\mathbb{R}^{N\times K}$
with the $i$th column given by $\bm{\psi}(\mathbf{q}_{i})$, and
$\bm{\eta}(t)$ as a vector with the $i$th element given by $\eta(\mathbf{q}_{i},t)$.
Then, the state equation can be expressed as
\begin{equation}
\mathbf{a}(t)=(\bm{\Psi}^{\text{T}}\bm{\Psi})^{-1}\bm{\Psi}^{\text{T}}\mathbf{B}\mathbf{a}(t-1)+(\bm{\Psi}^{\text{T}}\bm{\Psi})^{-1}\bm{\Psi}^{\text{T}}\bm{\eta}(t)\label{eq:kkf-state-equation}
\end{equation}
and the measurement equation can be formulated as
\begin{equation}
\mathbf{y}(t)=\bm{\Psi}\mathbf{a}(t)+\bm{\nu}(t)+\bm{\epsilon}(t)\label{eq:kkf-measurement-equation}
\end{equation}
where $\mathbf{y}(t)$ is the measurement vector with the $i$th element
given by $f(\mathbf{q}_{i},t)+\epsilon_{i}(t)$, in which $\epsilon_{i}(t)$
is the measurement noise, $\bm{\nu}(t)=[\nu(\mathbf{q}_{1},t),\nu(\mathbf{q}_{2},t),\dots,\nu(\mathbf{q}_{N},t)]^{\text{T}}$,
and $\bm{\epsilon}(t)$ is the vector form of the noise.

Based on (\ref{eq:kkf-state-equation}) and (\ref{eq:kkf-measurement-equation}),
a space-time Kalman filter can be derived via an ordinary Kalman filtering
approach. This yields the Kalman-filter-based trend estimation supported
by Kriging interpolation.

\begin{table*} \footnotesize
{
\caption{Summary and comparison of major CKM construction techniques.}\label{table:ckm-construction}
\centering
\begin{tabular}{|p{2cm}|p{4.5cm}|p{4.25cm}|p{5.25cm}|}
\hline
{\bf Technique} & {\bf     ~~~~~Main characteristics} & {\bf     ~~~~~Pros} & {\bf     ~~~~~Cons}\\
\hline
\vspace{0.1in} Canonical interpolation methods
& \begin{itemize}[leftmargin=1em]
\item Construct each point of the CKM as a weighted average of the measurement samples in the neighborhood
\item Typical weighting methods include IDW, KNN and local polynomial regression
\end{itemize}
 &  \begin{itemize}[leftmargin=1em]
 \item Simple to implement and usually has low computational complexity
 \item Can be fully distributed
 \end{itemize}
 &  \begin{itemize}[leftmargin=1em]
 \item Usually has low accuracy 
\item Cannot learn the structure of the environment
\end{itemize}\\
\hline
\vspace{0.1in} Kriging
& \begin{itemize}[leftmargin=1em]
\item The best unbiased linear estimator if the CKM is a stationary spatial process
\item Interpolate each point as a weighted sum of all the samples in the dataset
\end{itemize}
& \begin{itemize}[leftmargin=1em]
\item Usually has a higher accuracy
\item Very few hyper-parameters to choose
 \end{itemize}
& \begin{itemize}[leftmargin=1em]
\item Required to learn some global statistics of the CKM, such as the semi-variogram.
\item Higher computational complexity for a large dataset
\item Cannot learn the structure of the environment
\item May be difficult to extend to a spatially high-dimensional case due to the curse of dimensionality
\end{itemize}\\
\hline
\vspace{0.1in} Kernel regression
& \begin{itemize}[leftmargin=1em]
\item Interpolate each point as a weighted sum of all the samples in the dataset, where the weights are constructed using a kernel function
\item Identical to Kriging if the kernel function is identical to the spatial correlation function of the CKM process in a zero-mean stationary random field
\end{itemize}
& \begin{itemize}[leftmargin=1em]
\item Usually has a higher accuracy  \end{itemize}
& \begin{itemize}[leftmargin=1em]
\item The performance depends on the choice of the kernel function and some other hyper-parameters
\item Higher computational complexity for a large dataset 
\item Cannot learn the structure of the environment
\item May be difficult to extend to a spatially high-dimensional case due to the curse of dimensionality
\end{itemize}\\
\hline
\vspace{0.1in} Matrix and tensor completion
& \begin{itemize}[leftmargin=1em]
\item Spatially discretize the area of interest into a set of grid points and structure the CKM as a matrix or tensor
\item Matrix or tensor completion techniques applied to CKM construction
\end{itemize}
& \begin{itemize}[leftmargin=1em]
\item Can exploit the global structure of the CKM, and thus, require substantially less samples for dense but spatially non-uniform sampling 
\end{itemize}
& \begin{itemize}[leftmargin=1em]
\item May suffer from discretization error where the measurement samples do not locate at the grid center 
\item May suffer from identifiability issue where the amount of samples and the sampling pattern need to satisfy some condition
\end{itemize}\\
\hline
\vspace{0.1in} Deep learning &
\begin{itemize}[leftmargin=1em]
\item Treat the CKM has a collection of images and migrate the deep learning techniques for image processing to CKM construction
 \end{itemize} &
 \begin{itemize}[leftmargin=1em]
 \item Has the flexibility to be generalized and fuse multimodal data, such as satellite image and city map, to enhance the CKM construction and extend the potential application of CKM
\end{itemize}&
\begin{itemize}[leftmargin=1em]
\item Lack of a clear model for the relationship between the channel information and the environment
\item Require significant effort on data preprocessing and hyper-parameter tuning
\end{itemize}\\
\hline
\vspace{0.1in} Environment model-assisted approaches &
\begin{itemize}[leftmargin=1em]
\item Based on a model that characterizes the signal propagation as a function of the geometry of the environment
\item May need an integration with other environment model-free methods for CKM construction
 \end{itemize} &
 \begin{itemize}[leftmargin=1em]
 \item Can learn the geometry structure of the environment
\item Relatively easy to extend to a spatially high-dimensional case where both TX and RX vary their locations
\end{itemize}&
\begin{itemize}[leftmargin=1em]
\item Computational complexity may be high 
\item It is hard to construct an accurate model to describe the channel from the geometry of the environment, and hence, the accuracy could be low as compared to traditional ray-tracing
\end{itemize}\\
\hline
\end{tabular}
}
\end{table*}

\subsubsection{Kernel Regression}

Kernel-based function estimation provides a framework for nonparametric
regression \cite{19,21,23}. \Ac{rkhs} can be related to Gaussian
processes when defining their covariances via kernels. In addition,
\ac{rkhs} and Kriging are equivalent in the context of field estimation
when the pertinent covariance matrix equals the kernel \emph{Gram}
matrix \cite{Cressie:B15}.

In the context of \ac{rkhs}, consider an estimation of the map $f(\mathbf{q})$
via a set of $N$ measurement samples $z_{i}=f(\mathbf{q}_{i})+\epsilon_{i}$.
A kernel function $w$, defined to be symmetric and positive definite,
specifies a linear space of interpolating functions $f(\mathbf{q})$
given by $\mathcal{H}=\{f(\mathbf{q})=\sum_{i=1}^{\infty}a_{i}w(\mathbf{q}_{i},\mathbf{q}):a_{i}\in\mathbb{R},i\in\mathbb{N}\}$. The space $\mathcal{H}$ becomes a Hilbert space if an inner product
$\langle f,f'\rangle_{\mathcal{H}}=\sum_{i,j}a_{i}a_{j}'w(\mathbf{q}_{i},\mathbf{q}_{j})$
is defined, and the associated norm is $\|f\|_{\mathcal{H}}=\langle f,f\rangle_{\mathcal{H}}$.
The choice of kernels yields a family of functions that obey some
structural property. For example, the spline kernel generates low-curvature
functions.

The \emph{representation theorem }\cite{23} in \ac{rkhs} interpolation
asserts that, based on the $N$ data samples $\{(\mathbf{q}_{i},z_{i})\}$,
the optimal interpolator $f$ in $\mathcal{H}$ that
minimizes the regularized least-squares $\sum_{i=1}^{N}(z_{i}-f(\mathbf{q}_{i}))^{2}+\mu\|f\|_{\mathcal{H}}^{2}$ admits a finite-dimension representation
\begin{equation}
\hat{f}(\mathbf{q})=\sum_{i=1}^{N}a_{i}w(\mathbf{q}_{i},\mathbf{q}).\label{eq:rkhs-representation}
\end{equation}

Denote $\mathbf{z}=[z_{1},z_{2},\dots,z_{N}]^{\text{T}}$ and the
kernel-dependent Gram matrix as $\mathbf{W}$ with the $(i,j)$th
entry given by $W_{ij}=w(\mathbf{q}_{i},\mathbf{q}_{j})$. The optimal
coefficient $\mathbf{a}=[a_{1},a_{2},\dots,a_{N}]^{\text{T}}$ in
(\ref{eq:rkhs-representation}) as the minimizer of the regularized
least-squares cost is given by $\mathbf{a}=(\mathbf{W}+\mu\mathbf{I})^{-1}\mathbf{z}$.

The \ac{rkhs} estimator relates to the Kriging estimator $\hat{f}(\mathbf{q})=\mathbf{z}^{\text{T}}\bm{\lambda}$
in (\ref{eq:kriging-estimator}) as follows. In simple Kriging under
zero-mean stationary random field, the optimal coefficient $\bm{\lambda}$
minimizes the MSE $\mathbb{E}\{(f(\mathbf{q})-\mathbf{z}^{\text{T}}\bm{\lambda})^{2}\}$,
which leads to the solution $\hat{\bm{\lambda}}=\mathbf{R}_{zz}^{-1}\mathbf{r}_{z}$ to the Kriging system, where $\mathbf{R}_{zz}=\mathbb{E}\{\mathbf{z}\mathbf{z}^{\text{T}}\}$
and $\mathbf{r}_{z}=\mathbb{E}\{\mathbf{z}f(\mathbf{q})\}$. Now,
consider a noisy measurement $z_{i}=f(\mathbf{q}_{i})+\epsilon_{i}$,
where $\epsilon_{i}\sim\mathcal{N}(0,\sigma^{2})$. Then, $\mathbf{R}_{zz}=\mathbf{R}_{ff}+\sigma^{2}\mathbf{I}$,
where $\mathbf{R}_{ff}=\mathbb{E}\{\mathbf{f}\mathbf{f}^{\text{T}}\}$
with $\mathbf{f}=[f(\mathbf{q}_{1}),f(\mathbf{q}_{2}),\dots,f(\mathbf{q}_{N})]^{\text{T}}$,
and $\mathbf{r}_{z}$ becomes $\mathbf{r}_{z}=\mathbb{E}\{\mathbf{f}f(\mathbf{q})\}$.
Hence, the simple Kriging estimator $\hat{f}(\mathbf{q})$ takes the form $\mathbf{z}^{\text{T}}(\mathbf{R}_{ff}+\sigma^{2}\mathbf{I})^{-1}\mathbf{r}_{z}=\sum_{i=1}^{N}a_{i}w(\mathbf{q}_{i},\mathbf{q})$, where $\mathbf{a}^{\text{T}}=\mathbf{z}^{\text{T}}(\mathbf{R}_{ff}+\sigma^{2}\mathbf{I})^{-1}$
and the kernel function can be chosen as $w(\mathbf{q}_{i},\mathbf{q})=\mathbb{E}\{f(\mathbf{q})f(\mathbf{q}_{i})\}$.
The Kriging estimator thus matches with (\ref{eq:rkhs-representation}).

\subsubsection{Kernel Approach for CKM Construction} In practice, CKM may
consist of a complicated data structure, and could be of high dimension.
Therefore, for complexity considerations, one needs to take into account
two important issues.

First, an essential step is to look for low dimension representation
of CKMs. For instance, consider to construct a power spectrum map
$\Gamma(\mathbf{q},\omega)$ over a bounded area $\mathcal{R}$ in
$\mathbb{R}^{D}$, which describes the \ac{psd} measured at each
location $\mathbf{q}\in\mathcal{R}$. It is observed that $\Gamma$
is $(D+1)$-dimensional. A narrowband approximation may help reduce
the dimension of the frequency domain. Specifically, consider the
model $\Gamma(\mathbf{q},\omega)=\sum_{m=1}^{M}f_{m}(\mathbf{q})\phi_{m}(\omega)+n$,
where $f_{m}(\mathbf{q})$ is the path loss from the $m$th transmitter
to the receiver at location $\mathbf{q}$, $\phi_{m}(\omega)$ is
the \ac{psd} of the $m$th transmitter, and $n$ is the noise. Thus,
$\{\phi_{m}(\omega)\}$ are common terms that are independent of the
location $\mathbf{q}$. Consequently, if one can estimate the common
terms $\{\phi_{m}(\omega)\}$ or acquire some prior information on
the \ac{psd} of the transmitters, then the problem degenerates to
constructing a 2D power map $\hat{f}(\mathbf{q})=\sum_{m=1}^{M}c_{m}\hat{f}_{m}(\mathbf{q})$,
and the problem can be reformulated following the kernel regression approach.

Second, quantization needs to be considered. As the number of sensors
increases, real-time monitoring of the spectrum environment yields
a significant burden on the sensor network for sensing data transmission.
In practice, measurements are quantized before transmitting to a fusion
center, and as a result, the impact of quantization needs to be considered
for CKM construction. A way of quantization-aware construction is
to redesign the cost function \cite{20}. Specifically, instead of
penalizing the squared cost $l(x)=x^{2}$, one only penalizes the
difference beyond the quantization error. For example, consider a
uniform quantization with quantization error falling in $(-\epsilon,\epsilon)$.
Then, an $L_{1}$ $\epsilon$-insensitive loss function can be designed
as $l_{1}(x)=\max\{0,|x|-\epsilon\}$, and the $L_{2}$ version of
the $\epsilon$-insensitive loss can be $l_{2}(x)=\max\{0,x^{2}-\epsilon\}$.
Thus, the cost function can be specified as $\sum_{i=1}^{N}l(z_{i}-f(\mathbf{q}_{i}))+\mu\|f\|_{\mathcal{H}}^{2}$.

Finally, wireless channel
models are usually constructed from multi-scale components, typically,
a large-scale component due to propagation distance, a medium-scale
component due to shadowing related to the local environment, and a
small-scale component due to multi-paths. While CKMs usually focus
on the large-scale channel knowledge, a dual kernel approach to track
the path loss and shadowing may be applied \cite{95}, where there
is a DC-component kernel, slowly changing over distance, and a fast-varying-component
kernel. Altogether, a dual kernel function can be designed as $w(\mathbf{q},\mathbf{q}')=\exp(-\|\mathbf{q}-\mathbf{q}'\|/\theta_{\text{\scriptsize DC}})+\exp(-\|\mathbf{q}-\mathbf{q}'\|/\theta_{\text{\scriptsize var}})$,
where the parameters are chosen as $\theta_{\text{\scriptsize DC}}\gg\theta_{\text{\scriptsize var}}$.
It is found that the dual-kernel approach benefits the CKM construction,
with an increased accuracy compared with both the single-kernel approach
and multi-kernel approach \cite{95}. 

% The multi-kernel approach defines different candidate kernels $w_{1},w_{2},\dots,w_{P}$, and the estimator $\hat{f}$ can be constructed from a convex combination of the kernels $w=\sum_{i=1}^{P}a_{i}w_{i}$, where $a_{i}\geq0$, and $\sum_{i=1}^{P}a_{i}=1$. The parameters $a_{i}$ can be incorporated as variables for estimating $\hat{f}$ via the least-squares approaches. Specifically, one can compute an outer loop to select ``the best'' kernel that minimizes the least-squares cost  over $k\in\{1,2,\dots,P\}$ and $f\in\mathcal{H}^{k}$, with $\mathcal{H}^{k}$ denoting the \ac{rkhs} corresponding to the kernel $w_{k}$. Alternatively, one can construct $\hat{f}=\sum_{i=1}^{P}a_{i}f_{i}$, with $f_{i}\in\mathcal{H}^{i}$, and optimize for the weights $a_{i}$.

\subsubsection{Matrix and Tensor Completion}

The CKM $f(\mathbf{q})$ can be discretized into a matrix or tensor
form over a bounded area $\mathbf{q}\in\mathcal{R}$, and thus, CKMs
can be constructed via matrix and tensor completion techniques \cite{107,SunChe:J22,ZhaFuWanZha:J20}.
\Ac{wlog}, consider a rectangle area that covers $\mathcal{R}$ in
2D. Discretize the area using $M\times M$ grid points, such that
the grid points are aligned in rows and columns. Specifically, a uniform
grid formation is usually considered. However, if the measurements
$\{\mathbf{q}_{i},z_{i}\}$ are non-uniformly distributed in $\mathcal{R}$,
an optimized non-uniform grid topology may provide a better identifiability
for the matrix completion problem \cite{SunChe:C21}. Denote $\mathbf{c}_{jk}$
as the location of the $(j,k)$th grid point, and $\mathbf{H}\in\mathbb{R}^{M\times M}$
as the matrix that represents the discretized CKM $f(\mathbf{q})$
sampled over the grid points $\{\mathbf{c}_{jk}\}$, \emph{i.e.},
the $(j,k)$th element of $\mathbf{H}$ is defined as $H_{j,k}=f(\mathbf{c}_{jk})$.
From the measurements $\{\mathbf{q}_{i},z_{i}\}$, a sparse observation
matrix $\hat{\mathbf{H}}$ can be formed, where $\hat{H}_{jk}=z_{i}$,
if the measurement location $\mathbf{q}_{i}$ is within the $(j,k$)th
grid cell.

One property of the model $H_{j,k}=f(\mathbf{c}_{jk})$ is that, the
matrix $\mathbf{H}$ is likely low rank. A special example is for
the Gaussian propagation field $f(\mathbf{q})=\alpha\exp(-\gamma\|\mathbf{q}-\mathbf{q}_{\text{s}}\|^{2})$,
where $\mathbf{q}_{\text{s}}$ is the location of an energy source,
the corresponding matrix representation $\mathbf{H}$ is provably
rank-1 \cite{107}. More numerical and experimental examples for the
low rank property of $\mathbf{H}$ under different propagation field
models are reported in \cite{SunChe:J22,ZhaFuWanZha:J20,42}. Given
that $\mathbf{H}$ is low rank, it can be completed from the sparse
observation $\hat{\mathbf{H}}$ via solving a nuclear norm minimization
problem:
\begin{align}
\underset{\mathbf{X}}{\text{minimize}} & \qquad\|\mathbf{X}\|_{*}\label{eq:matrix-completion}\\
\text{subject to} & \qquad X_{jk}=\hat{H}_{jk},\quad\forall(j,k) \in\Omega,\nonumber
\end{align}
where $\Omega$ is the set of observed entries in $\hat{\mathbf{H}}$.
It has been studied in the literature that if the measurement locations
are uniformly random and there are sufficient observations $\mathcal{O}(M\log^{2}M)$
in $\hat{\mathbf{H}}$ \cite{CanRec:J12}, or if the deterministic
observation pattern in $\hat{\mathbf{H}}$ satisfies some identifiability
conditions \cite{PimBosNow:J16}, the full $\mathbf{H}$ can be reconstructed
as the solution to (\ref{eq:matrix-completion}) with high probability.
These conditions require that the resolution $M^{2}$ for $\mathbf{H}$
cannot be too large in order to guarantee the identifiability for
the matrix completion problem (\ref{eq:matrix-completion}).

It is natural to extend the matrix model to
a tensor when constructing CKMs in higher dimension. Extension
can also be applied to other domains, such as the frequency domain
that captures the spectrum information and the domain that represents
the measurement modality for multimodal data fusion.

In spectrum cartography, a multi-emitter model can be considered,
which naturally leads to a rank-($ L_r, L_r,1$) (LL1) tensor model. Specifically, let $x_{jkl}$
be the \ac{psd} at location coordinates $(j,k)$ in a 2D geographical
area and the $l$th frequency band. Then, the \ac{psd} can be \emph{approximately}
given by the power aggregation from the $R$ emitters as $x_{jkl}=\sum_{r=1}^{R}\mathbf{S}_{r}(j,k)c_{r,l}$,
where $\mathbf{S}_{r}\in\mathbb{R}^{2}$ is the propagation map for
the $r$th emitter and $c_{r,l}$ is \ac{psd} of the $r$th emitter
at the $l$th frequency band. This model has assumed that $\mathbf{S}_{r}$
depends only on the environment and the location of the emitter, but
not on the frequency bands. Hence, the model essentially captures the
path loss and shadowing; it may also capture the small-scale fading
for narrowband signals.

Such a model has the following tensor expression: $\bm{\mathcal{X}}=\sum_{r=1}^{R}\mathbf{S}_{r}\circ\mathbf{c}_{r}$,
where $\mathbf{c}_{r}$ is a vector that captures the \ac{psd} of
the $r$th emitter with the $r$th element given by $c_{r,l}$. The
tensor $\bm{\mathcal{X}}\in\mathbb{R}^{M\times M\times R}$ thus denotes
spectrum map. Applying the same low-rank argument on $\mathbf{S}_{r}$
as in the matrix model, one may construct a decomposition model as
$\mathbf{S}_{r}=\mathbf{A}_{r}\mathbf{B}_{r}^{\text{T}}$, resulting
in
\begin{equation}
\bm{\mathcal{X}}=\sum_{r=1}^{R}(\mathbf{A}_{r}\mathbf{B}_{r}^{\text{T}})\circ\mathbf{c}_{r},\label{eq:tensor-model-LL1}
\end{equation}
where $\mathbf{A}_{r},\mathbf{B}_{r}\in\mathbb{R}^{M\times L}$ are
rank-$L$ matrices.

A classical alternating least-squares approach can be applied to construct the tensor under such LL1 structure from sparse measurements \cite{ZhaFuWanZha:J20}.

% ---
\subsubsection{Issues and Remedies for Matrix-based CKM Construction}

The first important issue is the observation noise. Note that the observation $\hat{\mathbf{H}}$ contains
two types of errors. One is the measurement noise\emph{ }from the
measurement model $\hat{H}_{jk}=f(\mathbf{q}_{i})+\epsilon_{i}$,
and the other is the discretization noise due to the fact that $\hat{H}_{jk}=z_{i}$
is usually not sampled at the grid center $\mathbf{c}_{jk}$, \emph{i.e.},
$\mathbf{q}_{i}$ may deviate from $\mathbf{c}_{ij}$ in practice.
To address the observation noise, one may consider a least-squares
objective $\|\mathbf{W}\odot(\mathbf{X}-\hat{\mathbf{H}})\|_{F}^{2}+\mu\|\mathbf{X}\|_{*}$
\cite{19}, where $\odot$ denotes the Hadamard product, \emph{i.e.},
$\mathbf{W}\odot\mathbf{X}$ is an $M\times M$ matrix computed
entry-by-entry with $[\mathbf{W}\odot\mathbf{X}]_{jk}=W_{jk}X_{jk}$,
and $\mathbf{W}\in\mathbb{R}^{M\times M}$ is an indicator matrix
with $W_{jk}=1$ if the $(j,k)$th entry in $\hat{\mathbf{H}}$ is
observed, and $W_{jk}=0$ otherwise. Such a formulation needs to
specify a parameter $\mu$ to trade-off the least-squares cost and
the nuclear norm. Another formulation is to modify the observation
constraint if the knowledge of the observation error is available:
\begin{align}
\underset{\mathbf{X}}{\text{minimize}} & \qquad\|\mathbf{X}\|_{*}\label{eq:matrix-completion-with-confident-interval}\\
\text{subject to} & \qquad X_{jk}-\hat{H}_{jk}\in\mathcal{I}_{ij},\quad\forall(j,k)\in\Omega,\nonumber
\end{align}
where $\mathcal{I}_{ij}$ is the confident interval of the $(j,k)$th
observation, which can be derived using a local Kriging, or a local
polynomial regression approach \cite{SunChe:J22}.

Another important issue
for a classical matrix completion approach in CKM construction is
the trade-off between the resolution $M^{2}$ and the identifiability
of the matrix completion problem. Given a measurement set of $N$
samples, the number of observed entries in $\hat{\mathbf{H}}$ will
be upper bounded by $N$. Consequently, the matrix dimension $M$
cannot be too large, and otherwise, $\hat{\mathbf{H}}$ will be too
sparse. Although there is a guideline in the literature that advices
$N$ should scale with $M$ according to $\mathcal{O}(M\log^{2}M)$
\cite{CanRec:J12}, there is no accurate formula to specify a maximum
$M$ that guarantees the identifiability. In practice, a safe choice
is usually to choose a small $M$ with an \emph{empirically} sufficient
margin.

A remedy to the identifiability issue is to modify the way that forms
the sparse observation matrix $\hat{\mathbf{H}}$ using an interpolation-assisted
approach \cite{SunChe:J22}. Specifically, one may want to first \emph{locally}
interpolate a subset of entries $\hat{H}_{jk}$ where there are sufficient
measurements \emph{near} $\mathbf{c}_{jk}$. Typical interpolation
methods may include Kriging and local polynomial regression. Then,
based on the interpolated, yet sparse, matrix $\hat{\mathbf{H}}$,
a nuclear norm minimization problem (\ref{eq:matrix-completion-with-confident-interval})
is solved to obtain the full CKM. The key of the integrated interpolation
and matrix completion approach is the determination of the radius
$b$ of the observation window, which affects the observation pattern
$\Omega$ of $\hat{\mathbf{H}}$ and the estimation of $\hat{H}_{jk}$
from the local interpolation. There are two typical approaches to
determine the window size $b$. (i) {\it Uniform sampling}: Form the observation
set $\Omega$ by sampling $CM\log^{2}M$ grid points uniformly at
random to form $\hat{\mathbf{H}}$, where the parameter $C$ can be
chosen to guarantee a sufficient number of observations for the $N\times N$
matrix $\hat{\mathbf{H}}$. Then, for each entry $(j,k)$ in $\Omega$,
choose a specific parameter $b_{jk}$ to minimize the estimation error
of $\hat{H}_{jk}$, which is estimated from local polynomial regression
or local Kriging. (ii) {\it Measurement-aware sampling}: Form an observation
set of grid points where there are sufficient measurements nearby.
Specifically, the observation set is constructed as $\Omega=\{(j,k):\sum_{i=1}^{N}\mathbb{I}\{\|\mathbf{q}_{i}-\mathbf{c}_{jk}\|_{2}<b\}\geq N_{0}\}$,
where $\mathbb{I}\{A\}=1$ if condition $A$ is satisfied, and $\mathbb{I}\{A\}=0$
otherwise. The parameter $b$ can be chosen to minimize the average
estimation error of $\hat{H}_{jk}$ for all $(j,k)\in\Omega$. Note
that, there exists analytical forms for the estimation error of $\hat{H}_{jk}$
using the local polynomial regression approach \cite{SunChe:J22}.

The above interpolation-assisted approach addresses both the resolution
issue and the identifiability issue via the choice of the window size
$b$. For a small enough $b$, the scheme degenerates to a conventional
matrix completion approach; for a large enough $b$, the scheme converges
to interpolating all the entries of $\hat{\mathbf{H}}$ (via a large
enough window size $b$). However, it is found that a moderate and
adaptive $b$ yields the best performance, because it is not optimal
to interpolate all the entries. For a local area that the measurements
are very sparse, interpolation method may locally yield a very poor
performance. In this case, matrix completion can fill in the missing
values based on the global information of $\hat{\mathbf{H}}$. Thus,
the interpolation-assisted matrix completion approach is expected
to combine the benefit from both interpolation, that exploits the
local correlation of the data, and matrix completion, that exploits
the global structure of the data.

% ---

\subsubsection{Deep Learning with Neural Network}

Deep learning can also play a role in CKM construction. In early
attempts, neural networks were applied to replace propagation models
that are difficult to construct in complicated environments \cite{30}. Some attempt exploits the geometric information of the multipaths together with the \ac{cnn} architecture to extrapolate the angle and delay for constructing radio maps with AoA information in an indoor localization application \cite{li2023automatic}. 
Some recent works consider to treat the radio map as a collection
of images, and develops neural networks for an end-to-end construction
of radio maps. A typical deep learning structure is the Autoencoder,
which takes a sparsely observed radio map as input, and completes
the radio map in the output \cite{108,ShrFuHon:J22}. Specifically, an Autoencoder consists an encoder module that compresses
a full radio map into a few feature vectors with much smaller dimensions
via a sequence of convolutional layers and pooling layers. This process
is similar to downsampling with a purpose to extract the \emph{spatial
structure} of radio maps. A possible mathematical interpretation of
the possible spatial structure of a radio map is the low-rankness
under the matrix model and the representation theorem under kernel
learning. Then, a decoder module reverses the process of the encoder
and upsamples the feature vectors into a full dimensional radio map
via a sequence of convolutional layers and pooling layers. A typical
cost function used for training the Autoencoder is to compare the
difference between the input and output.

Many recent studies have been working on fusing the environment information
with radio map construction. One possible direction is to fuse the
environment information using satellite images and a three-module
neural network structure. Intuitively, satellite images can be used
to identify different propagation scenarios, such as urban, suburban,
commercial districts, and residential areas; then, different propagation
models can be applied. A typical three-module structure consists of
a \ac{cnn} module to extract relevant features from satellite images,
a neural network module to process the parametric features such as
transceiver locations and propagation distance, and a neural network
module to fuse the features from the satellite images and the features
from the parametric inputs \cite{ThrZibChr:J20}. Another direction
is to use the city map as an input to learn the shadowing status between
the transmitter and the receiver by encoding various input features
into images. Based on the idea of ray-tracing, the key features for
computing the propagation channel quality include the environment
map and the positions of the transmitters and receivers. A city map
can be encoded into a binary image, with ``1'' indicating a pixel
occupied by a building, and ``0'' indicating an empty area. The
transceiver locations can be encoded into an almost blank image with
only two pixels, where the pixel locations indicate the transmitter
and receiver locations. These feature images can be fed into a neural
network using a U-net structure \cite{2008}.

\subsection{{Environment} Model-assisted CKM Construction}

{Environment} model-assisted CKM construction tries to reconstruct the propagation
environment as an intermediate step to construct the CKM. {Environment-aware models can be established to describe the signal propagation due to the geometry structure of the environment. As a result, the hidden correlation of the data, which is collected from a shared environment, can be discovered and exploited. Once the geometry of the propagation environment is constructed as a byproduct, a CKM with full spatial degrees of freedom, i.e., for arbitrary locations of the transmitter and receiver, can be efficiently constructed.} 

{A classical environment model-assisted CKM construction technique is the ray-tracing channel modeling. However, classical ray-tracing models are complicated, and it is very challenging to solve the corresponding inverse problem that recovers the environment from the channel measurement. Thus, recent literature focuses on simplified models that may jointly reconstruct the environment and the CKM based on little or no prior information on the environment.} For example, \emph{spatial loss
field} \cite{HamMaBaxMat:J13,43,39} establishes a 2D model for the
propagation environment, and a CKM can be constructed for any transmitter
and receiver locations in 2D. Joint 3D environment and radio map construction
approaches can construct a 3D propagation environment based on a {\em virtual obstacle model} that only consider the direct propagation path, and consequently, a CKM for
any transmitter and receiver locations in 3D can be constructed \cite{CheEsrGesMit:C17,ZhaChe:C20,106}.

\subsubsection{Spatial Loss Field Model}

The spatial loss field $g(\mathbf{q})$ models an incremental shadowing
loss when the signal propagates through location $\mathbf{q}$. Based
on the spatial loss field, one may construct a linear model to describe
the shadowing loss between two locations $\mathbf{q}_{j}$ and $\mathbf{q}_{k}$ \cite{2063}:
\begin{equation}
s(\mathbf{q}_{j},\mathbf{q}_{k})=\int g(\mathbf{q})b(\mathbf{q}_{j},\mathbf{q}_{k},\mathbf{q})d\mathbf{q}\label{eq:spatial-loss-field-for-shadowing}
\end{equation}
where the function $b(\mathbf{q}_{j},\mathbf{q}_{k},\mathbf{q})$
describes the weighted area that are covered by all possible propagation
paths from locations $\mathbf{q}_{j}$ to $\mathbf{q}_{k}$, and the
integral is taken over the entire area of interest to construct the
CKM.

The simplest model for $b(\mathbf{q}_{j},\mathbf{q}_{k},\mathbf{q})$
can be the direct path model, i.e., $b(\mathbf{q}_{j},\mathbf{q}_{k},\mathbf{q})$ equals to a distance-scaled Dirac delta function for $\mathbf{q}$ locating on the line segment joining $\mathbf{q}_j$ and $\mathbf{q}_k$. Besides, there could be more possibility for $b(\mathbf{q}_{j},\mathbf{q}_{k},\mathbf{q})$.
For example, in a \emph{normalized ellipse model}, it assumes that
propagation paths may travel in an ellipse area with $\mathbf{q}_{j}$
and $\mathbf{q}_{k}$ being the foci. Thus, the function $b(\mathbf{q}_{j},\mathbf{q}_{k},\mathbf{q})$
describes an ellipse area, with a zero weight outside the ellipse
area. In an \emph{inverse area elliptical model}, it also assumes
an ellipse propagation area using $\mathbf{q}_{j}$ and $\mathbf{q}_{k}$
as the foci, and in addition, it allocates higher weights for locations
closer to the direct path $\mathcal{L}$ from $\mathbf{q}_{j}$ to
$\mathbf{q}_{k}$, and the weights decrease when moving closer to
the edge of the propagation ellipse.

In (\ref{eq:spatial-loss-field-for-shadowing}), while $b(\mathbf{q}_{j},\mathbf{q}_{k},\mathbf{q})$
is completely defined by the propagation path model and the transmitter
and receiver locations, the spatial loss field $g(\mathbf{q})$ needs
to be estimated from data. This can be done by discretization of the
area of interest. Let $\mathbf{g}$ be a discrete approximation of the continuous spatial
loss field $g(\mathbf{q})$. Let $s_{j,k}=s(\mathbf{q}_{j},\mathbf{q}_{k})$, where $\mathbf{s}$ denotes a vector that collects all measurements.
The model (\ref{eq:spatial-loss-field-for-shadowing}) can be approximated
by the vector-vector multiplication form $s_{j,k}=\mathbf{b}_{j,k}^{\text{T}}\mathbf{g}$,
where $\mathbf{b}_{j,k}^{\text{T}}$ is a row vector. Let $\mathbf{B}$
be a matrix that collects the row vectors $\mathbf{b}_{j,k}^{\text{T}}$
for all measurement pairs $\{(\mathbf{q}_{j},\mathbf{q}_{k})\}$.
Then, the measurement model (\ref{eq:spatial-loss-field-for-shadowing})
can be arranged into a matrix form as $\mathbf{s}=\mathbf{B}\mathbf{g}$.
The discretized spatial loss field can be estimated as $\hat{\mathbf{g}}=\mathbf{B}^{-1}\mathbf{s}$.

\subsubsection{Virtual Obstacle Model}

\begin{figure}[htb]
	\centering
	\begin{subfigure}[b]{0.45\textwidth}
		\centering
		\includegraphics[width=\textwidth]{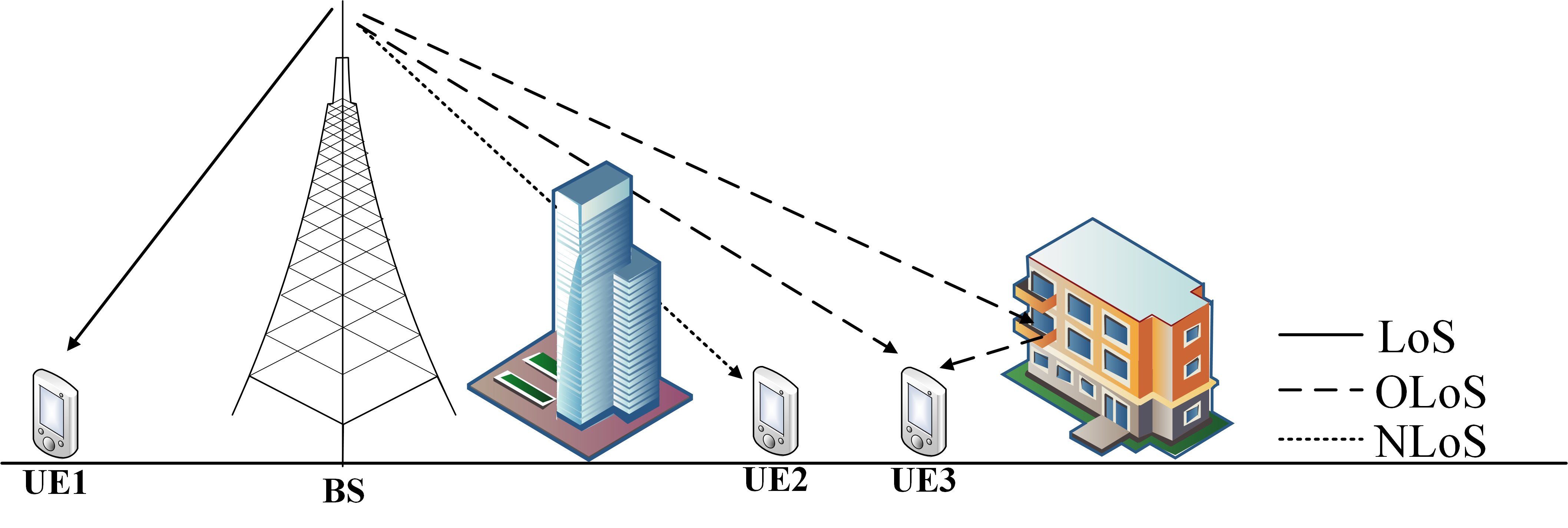}
		\caption{Real propagation environment.}
	\end{subfigure}
	\hfill
	\begin{subfigure}[b]{0.45\textwidth}
		\centering
		\includegraphics[width=\textwidth]{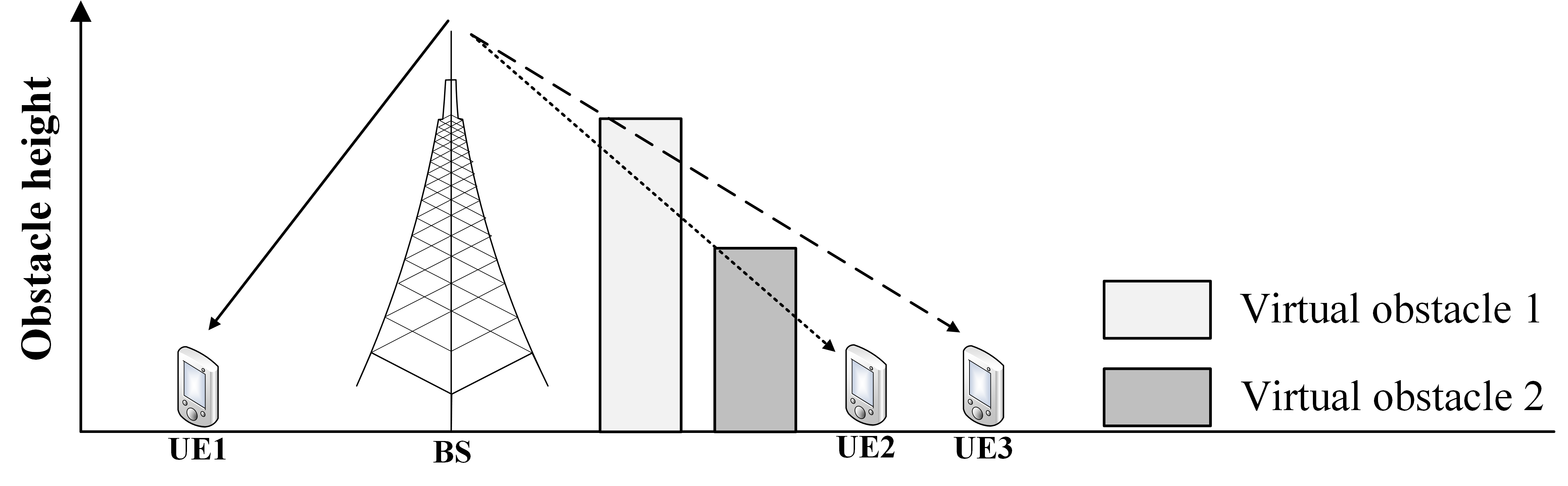}
		\caption{Virtual obstacle model.}
	\end{subfigure}
	\caption{The virtual obstacle model in (b) approximates the propagation over all scattered paths in (a) by an equivalent direct path that passes through effective virtual obstacles.}
	\label{fig:Virtual _obstacle}
\end{figure}

It is also possible to reconstruct the environment with geometric
meaning using a\emph{ multi-class virtual obstacle model} \cite{CheEsrGesMit:C17,ZhaChe:C20,106}. The general idea is
to employ a certain class of equivalent virtual obstacle at a certain
location with appropriate shape and dimension to represent the equivalent
signal attenuation due to the geometry relation between the transceivers
and the environment. As a result, the fundamental purpose of the virtual
obstacle model is not to reconstruct the usual \emph{visual} environment,
but to reconstruct a radio environment where the structures are automatically
labeled with some electromagnetic meaning, as shown in Fig.~\ref{fig:Virtual _obstacle}.
For example, if the propagation between locations $\mathbf{q}_{j}$
and $\mathbf{q}_{k}$ is in deep shadow, there should be a solid virtual
obstacle that blocks the direct path form $\mathbf{q}_{j}$ to $\mathbf{q}_{k}$ as the case for UE 2;
if the propagation is only slightly attenuated compared to the \ac{los}
case, there should be a light virtual obstacle in place as the case for UE 3; if there is an LoS condition, there should not be any virtual obstacle blocking the propagation as the case for UE 1.

A mathematical description of a $L$-class virtual obstacle environment
can be established as follows. Define $L+1$ classes of propagation
regions $\{\mathcal{D}_{l}\}$, where $\mathcal{D}_{0}\subseteq\mathbb{R}^{6}$
represents the region where the direct path between the location pair
$(\mathbf{q},\mathbf{q}')$ is in \ac{los}, and $\mathcal{D}_{l}$
represents the region where the direct path $(\mathbf{q},\mathbf{q}')$
is blocked by a class-$l$ virtual obstacle. Specifically, $(\mathbf{q},\mathbf{q}')\in\mathcal{D}_{0}$,
if no obstacle blocks the direct path of $(\mathbf{q},\mathbf{q}')$,
and $(\mathbf{q},\mathbf{q}')\in\mathcal{D}_{l}$, if a class-$l$
obstacle blocks the direct path and no class-$n$ obstacle, with $n>l$,
blocks the direct path. It follows that $\{\mathcal{D}_{l}\}$ are
disjoint.

A \emph{segmented propagation model }can be used to describe the channel
based on the propagation regions $\{\mathcal{D}_{l}\}$:
\begin{equation}
f(\tilde{\mathbf{q}};\bm{\theta},\{\mathcal{D}_{l}\})=\sum_{l=0}^{L}g_{l}(\tilde{\mathbf{q}};\bm{\theta})w_{l}(\tilde{\mathbf{q}};\{\mathcal{D}_{l}\})+\xi(\tilde{\mathbf{q}}),\label{eq:segmented-channel-model}
\end{equation}
where $\tilde{\mathbf{q}}=(\mathbf{q},\mathbf{q}')$ denotes the location
pair of a transmitter at $\mathbf{q}$ and a receiver at $\mathbf{q}'$;
$g_{l}(\tilde{\mathbf{q}};\bm{\theta})$ describes the path loss for
the $l$th propagation region, a typical model for $g_{l}(\cdot)$
being $g_{l}(\tilde{\mathbf{q}};\bm{\theta})=\beta_{l}+\alpha_{l}\log_{10}(\|\mathbf{q}-\mathbf{q}'\|_{2})$,
in which, $\bm{\theta}=\{\alpha_{l},\beta_{l}\}$ is a collection
of parameters; $w_{l}(\cdot)$ describes the likelihood of $\tilde{\mathbf{q}}\in\mathcal{D}_{l}$,
a typical choice of $w_{l}(\cdot)$ being $w_{l}(\tilde{\mathbf{q}};\{\mathcal{D}_{l}\})=\mathbb{I}\{\tilde{\mathbf{q}}\in\mathcal{D}_{l}\}$;
and $\xi(\tilde{\mathbf{q}})$ describes the residual shadowing.

Note that, without a geometry model for the virtual obstacle environment,
these propagation regions $\{\mathcal{D}_{l}\}$ can still be constructed
from {an environment} model-free approach, using Kriging or subspace clustering \cite{2093}
and the segmented channel model (\ref{eq:segmented-channel-model}).
On the other hand, it is not difficult to describe $\{\mathcal{D}_{l}\}$
based on the geometry of the environment using a simplified ray-tracing
method.

Consider a discretized area of $M$ grid cells. Denote $\mathbf{H}\in\mathbb{R}^{M\times L}$
as a matrix with the $m$th entry of the $l$th column $\mathbf{h}_{l}$
of $\mathbf{H}$ representing the height of the class-$l$ virtual
obstacle located at the $m$th grid cell. Denote $\mathcal{B}(\tilde{\mathbf{q}})$
as the set of grid points that are covered by the direct path of $\tilde{\mathbf{q}}$.
For each grid cell $m\in\mathcal{B}(\tilde{\mathbf{q}})$, denote
$z_{m}(\tilde{\mathbf{q}})$ as the altitude when the direct path
of $\tilde{\mathbf{q}}$ passes over the grid cell $m$. Then, it
follows that $\tilde{\mathbf{q}}\in\mathcal{D}_{0}$ if $h_{m,l}<z_{m}(\tilde{\mathbf{q}})$
for all $m\in\mathcal{B}(\tilde{\mathbf{q}})$ and all $1\leq l\leq L$.
Likewise, $\tilde{\mathbf{q}}\in\mathcal{D}_{l}$ if $h_{m,l}>z_{m}(\tilde{\mathbf{q}})$
for at least one $m\in\mathcal{B}(\tilde{\mathbf{q}})$, and $h_{m,n}<z_{m}(\tilde{\mathbf{q}})$
for all $m\in\mathcal{B}(\tilde{\mathbf{q}})$ and all $l<n\leq L$.
These conditions can be mathematically written in the following form:
\begin{equation}
\mathbb{I}\{\tilde{\mathbf{q}}\in\mathcal{D}_{l}(\mathbf{H})\}=\prod_{m\in\mathcal{B}(\tilde{\mathbf{q}})}\prod_{n=1}^{L}\mathbb{I}\{h_{m,n}<z_{m}(\tilde{\mathbf{q}})\}\label{eq:vitural-obstacle-model-1}
\end{equation}
for $l=0$, and
\begin{align}
\mathbb{I}\{\tilde{\mathbf{q}}\in\mathcal{D}_{l}(\mathbf{H})\} & =\Big(1-{\displaystyle \prod_{m\in\mathcal{B}(\tilde{\mathbf{q}})}}(1-\mathbb{I}\{h_{m,l}\geq z_{m}(\tilde{\mathbf{q}})\})\Big)\nonumber \\
 & \qquad\times{\displaystyle \prod_{m\in\mathcal{B}(\tilde{\mathbf{q}})}\prod_{n=l+1}^{L}}\mathbb{I}\{h_{m,n}<z_{m}(\tilde{\mathbf{q}})\}\label{eq:vitural-obstacle-model-2}
\end{align}
for $l\geq1$.

With the environment modeling, the segmented channel model in (\ref{eq:segmented-channel-model})
can be fully characterized as $f(\tilde{\mathbf{q}};\bm{\theta},\mathbf{H})$
by the propagation parameter $\bm{\theta}$ and the environment parameter
$\mathbf{H}$, which can be estimated from a least-squares approach
using a set of measurement data $\{y_{i},\tilde{\mathbf{q}}_{i}\}$
and the cost function can be typically formulated as $\frac{1}{N}\sum_{i}(y_{i}-f(\tilde{\mathbf{q}}_{i};\bm{\theta},\mathbf{H}))^{2}$. Once the parameters $\bm{\theta}$ and $\mathbf{H}$ are obtained,
the CKM can be fully reconstructed from the segmented propagation
model (\ref{eq:segmented-channel-model}).

The estimation of the environment parameter $\mathbf{H}$
in the virtual obstacle model (\ref{eq:vitural-obstacle-model-1})
and (\ref{eq:vitural-obstacle-model-2}) can be very challenging because
the problem can be non-convex and even ill-posed due to the indicator
functions in (\ref{eq:vitural-obstacle-model-1}) and (\ref{eq:vitural-obstacle-model-2}).
Apart from the likelihood-based approach developed in \cite{LiuChe:J23},
deep learning based approaches \cite{ZenChe:C22} can also help the
estimation of $\mathbf{H}$. The general idea is to parameterize the
expression of $\mathbf{H}$ and use the \ac{cnn} and \ac{dnn} structure to express the virtual
obstacle model (\ref{eq:vitural-obstacle-model-1}) and (\ref{eq:vitural-obstacle-model-2}),
where the coefficients of the CNN model are tuned by off-the-shelf
deep learning optimizers. 

\subsection{Performance Evaluation}

Evaluating radio construction and utilization relies on a large amount of data collected or simulated over real geographic environments. Table \ref{table:dataset} summarizes several public radio map datasets reported in the literature. 

The most prevailing performance evaluation approach is to calculate
the pixel-by-pixel \ac{mse} or \ac{mae} of the constructed CKM.
%However, such an approach faces difficulties in practice. First, it is challenging to define the ground truth. Wireless channels suffer from small-scale fading, which yields up to 10 dB channel gain variation within 1 second or 1 meter movement due to the multipath effect and the time-variation of the propagation environment. As a result, it seems not practical to capture the small-scale fading for CKM, but more reasonable to characterize only the large-scale fading that includes the path loss and the shadowing of the propagation environment. However, it is still difficult to collect the ground truth of the CKM, because this requires the measurements to be averaged over both time and spatial domain in a small neighborhood of each sampled location. 
% Second, the pixel-centric evaluation philosophy lacks the connection with the utilization of CKM. For example, if the CKM is used for localizing an unknown transmitter, then calculating the \ac{mse} of all pixels is obviously not an ideal metric for CKM construction. In another example, when the CKM is used for location-based communication optimization \cite{CheMitGes:J21}, then it may be more important to characterize the structure of the \ac{los} region rather than pursuing the pixel-by-pixel quality of the CKM. However, it is unknown yet how to evaluate the CKM quality when there is a target CKM application.
In the following, we evaluate the performance of several {environment} model-free
and {environment} model-assisted approaches for CKM construction.

\begin{table*} \footnotesize
\caption{Some public data sets useful for radio and channel knowledge maps.}\label{table:dataset}
\centering
\begin{tabular}{|p{2.0cm}|p{2.0cm}|p{7.5cm}|p{4.5cm}|}
\hline
{\bf Data set} & {\bf     ~~Source} & {\bf     ~~~~~Feature} & {\bf     ~~~~~Application}\\
\hline
		%%%%%%%%%%%%%%%%%%%%%%%%%%%%%%%%%%%%%%%%%%%%%%%%%%%%%%%%%%%%%%%
		\vspace{0.1in} High resolution indoor office REM \cite{BurLiBiz:D22}
		& 	\vspace{0.1in} Measurement
		&  \begin{itemize}[leftmargin=1em]
			\item  Two measured high-resolution REMs at 3.75GHz
			\item  A 92cm wide and over 18m long radio channel scan is recorded
			\item Spatial resolution 10 mm in x-direction and 0.8 mm in y-direction
		\end{itemize}
		&  \begin{itemize}[leftmargin=1em]
			\item Channel characterization
			\item REM interpolation application
		\end{itemize}\\
		\hline
		
		%%%%%%%%%%%%%%%%%%%%%%%%%%%%%%%%%%%%%%%%%%%%%%%%%%%%%%%%%%%%%%%
		\vspace{0.1in} DroneRFa  \cite{YuMaoZhoSun:D23}
		& \vspace{0.1in} Measurement
		& \begin{itemize}[leftmargin=1em]
			\item RF signals of UAV controller detected by USRP
			\item 25 categories of signals: indoor (\#9), outdoor (\#15), background reference signal (\#1)
			\item 3 frequency bands: 915 MHz (902$\sim$928 MHz), 2.4 GHz (2400$\sim$2484.5 MHz) and 5.8 GHz (5725$\sim$5850 MHz)
		\end{itemize}
		& \begin{itemize}[leftmargin=1em]
			\item Detection (identification) of UAV RF signals
		\end{itemize}\\
		\hline
		
		%%%%%%%%%%%%%%%%%%%%%%%%%%%%%%%%%%%%%%%%%%%%%%%%%%%%%%%%%%%%%%%
		\vspace{0.1in} A2G RadioMap \cite{WanWen:D23}
		& \vspace{0.1in} Simulation (Wireless Insite)
		& \begin{itemize}[leftmargin=1em]
			\item Two outdoor radio maps with more than 1 million channel measurements
			\item Three bands: 2.5GHz, 5.9GHz, 28GHz
			\item Each map: 640 $\times$ 640 m$^2$ with 10$~$m-resolution
		\end{itemize}
		& \begin{itemize}[leftmargin=1em]
			\item 3D radio map construction
			\item Source localization
		\end{itemize}\\
		
		%%%%%%%%%%%%%%%%%%%%%%%%%%%%%%%%%%%%%%%%%%%%%%%%%%%%%%%%%%%%%%%
		\hline
		\vspace{0.1in} RadioMapSeer \cite{YapLevKut:D22}
		& \vspace{0.1in} Simulation  (WinProp)
		& \begin{itemize}[leftmargin=1em]
			\item 701 city maps and 80 transmitters per map
			\item Two types of simulations: with and without cars
			\item Each channel pathloss map: 256 $\times$ 256 m$^2$ with 1 m-resolution
		\end{itemize}
		& \begin{itemize}[leftmargin=1em]
			\item Pathloss radio map estimation
			\item Localization and target sensing
		\end{itemize}\\
		
		%%%%%%%%%%%%%%%%%%%%%%%%%%%%%%%%%%%%%%%%%%%%%%%%%%%%%%%%%%%%%%%
		\hline
		\vspace{0.1in} Sensiverse \cite{LuoZhouYan:D23}
		& \vspace{0.1in} Simulation
		& \begin{itemize}[leftmargin=1em]
			\item 25 city and 100+ scenes
			\item Four bands: 3.5 GHz, 10 GHz, 26 GHz and 100 GHz.
			\item Three components: 3D maps, channel data and examples.
		\end{itemize}
		& \begin{itemize}[leftmargin=1em]
			\item Moving target detection
			\item 3D environment imaging reconstruction
			\item Integrated sensing and communication solution evaluation
		\end{itemize}\\
		
		%%%%%%%%%%%%%%%%%%%%%%%%%%%%%%%%%%%%%%%%%%%%%%%%%%%%%%%%%%%%%%%
		\hline
		\vspace{0.1in} DeepREM \cite{ChaAndVit:D22}
		& \vspace{0.1in} Simulation  (WinProp)
		& \begin{itemize}[leftmargin=1em]
			\item RSRP maps for BS coverage
			\item 200 maps and 4 transmitters on each coverage area
			\item Each RSRP map: $2290 \times 3670$ m$^2$ to $3810 \times 5160$ m$^2$ with 10 m-resolution
		\end{itemize}
		& \begin{itemize}[leftmargin=1em]
			\item Radio map and coverage estimation
			\item Localization and target sensing
		\end{itemize}\\
		%%%%%%%%%%%%%%%%%%%%%%%%%%%%%%%%%%%%%%%%%%%%%%%%%%%%%%%%%%%%%%%

\hline
\end{tabular}
\end{table*}

\subsubsection{{Environment} model-free approaches for CKM construction}

We consider to reconstruct an indoor CKM (storing RSS information) and evaluate the
performance using real data \cite{KinKopHae:Data08}. In this dataset,
there are $M=166$ RSS measurements for each emitter. The measurements
are taken in a $14\times34\ \text{m}^{2}$ indoor area. The area is
divided into $1\times1\ \text{m}^{2}$ grid cells such that the $166$
measurements are in the grid center. Among the $166$ measurements,
$M'=20$ to $ 140 $ measurements are randomly selected for
reconstructing the prorogation field. We denote $M'/M$ as the sampling
rate.

The evaluation criterion is \ac{mse} which is calculated through
$\|\bm{X}-\hat{\bm{X}}\|_F^2/(N_1N_2)$ where $\bm{X}\in\mathbb{R}^{N_{1}\times N_{2}}$,
$N_{1}=14$, $N_{2}=34$, a matrix representation of the CKM $f(\bm{\mathrm{\bm{q}}})$
of the RSS evaluated on grids. Here, we evaluate seven representative
methods:
\begin{itemize}
\item{{\bf $k$-LP} \cite{VerFunRaj:J16}: $k$-nearest neighbor local polynomial interpolation
(k-LP), a first order local polynomial regression
method is used to estimate $f(\bm{\mathrm{\bm{q}}})$ using $k$ nearest
measurements, where $k$ is chosen through cross validation.}
\item{{\bf Kriging} \cite{96}: Ordinary Kriging method, where the Kriging weights $\lambda_{i}$ are optimized to minimize the expected error of the estimator using a universal exponential semivariogram model. }
\item{{\bf TPS} \cite{JuaGonGeo:J11}: A thin plate
spline (TPS) method is used to construct $f(\bm{\mathrm{\bm{q}}})$.}
\item{{\bf NNM-t} \cite{SunChe:J22}: Nuclear norm minimization with trust region constraints (NNM-t) for matrix completion, where the trust region is obtained from a local polynomial interpolation model.}
\item{{\bf GPR} \cite{SunXueYu:J18}: Gaussian process regression (GPR) 
with a squared exponential kernel.}
\item{{\bf KNN} \cite{58}: a K-nearest neighbors algorithm with $K = 5$.}
\item{{\bf ALS} \cite{TanNeh:J11}: Alternating least
square for matrix completion (ALS), where a matrix
$\bm{X}$ is completed through alternatively minimizing $||\bm{y}-\bm{A}\mbox{vec}(\bm{X)}||$
with $\bm{X}=\bm{LR}$, $\bm{A}$ is sensing matrix.}
\end{itemize}

\begin{figure}
	\includegraphics[width=0.5\textwidth]{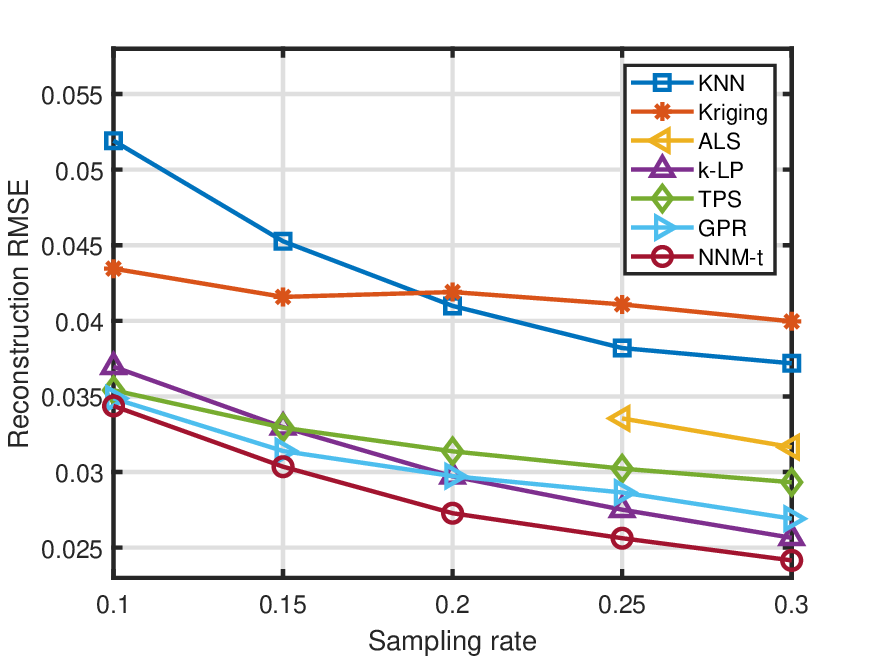}
	\caption{\label{fig:Model-free-CKM simulation}{Environment} model-free CKM construction
		versus sampling rate.}
\end{figure}

\
Fig.~\ref{fig:Model-free-CKM simulation} shows the numerical results
of the CKM reconstruction \ac{mse} versus different sampling ratio.
In general, k-LP, TPS, GPR, and NNM-t methods yield satisfactory performance,
and, in particular, the NNM-t scheme performs the best among all the
7 methods we tested. KNN exhibits poor performance as it
is  not robust for measurements that are non-uniformly distributed.
The performance of Ordinary Kriging is also unsatisfactory in this
setting, since Ordinary Kriging assumes a constant mean throughout
the entire area of interest, limiting its capability to accurately
capture the energy drift of the CKM. Traditional matrix completion
approach ALS here probably encounters recoverability issue for a sampling
rate below $0.2$, resulting in a large MSE, and thus its performance
under a sampling rate below $0.2$ is ignored. While TPS and
GPR yield a good performance, they may encounter some difficulties
in selecting a good kernel for the algorithm. The NNM-t addresses
the recoverability issue in matrix completion through local polynomial
regression, and it performs better than both the pure matrix-based schemes, i.e., ALS, and pure interpolation-based schemes, i.e., $k$-LP, Kriging, TPS, GPR, and KNN.

\subsubsection{{Environment} model-assisted approaches for environment and radio map construction}

Here, we consider joint 3D environment and radio map construction
based on simulated RSS measurements over an urban area in central
Shanghai, China. There are around a dozen buildings with heights ranging
from 10 to 130 meters. Some buildings are circular and there are also
some other oddly-shaped buildings and obstacles in this area. The
material of all structures is considered to be concrete. The receivers
are placed on the street randomly and the transmitters are uniformly
distributed in the sky with altitude ranging from 50 to 120 meters.
The measurements are generated using Remcom Wireless Insite, with ray-tracing
configured for up to 6 reflections and 1 diffraction. The waveforms
are chosen as narrowband sinusoidal signals at frequency 2.5 GHz.

\begin{figure}[!t]
	\centering \includegraphics[width=0.5\textwidth]{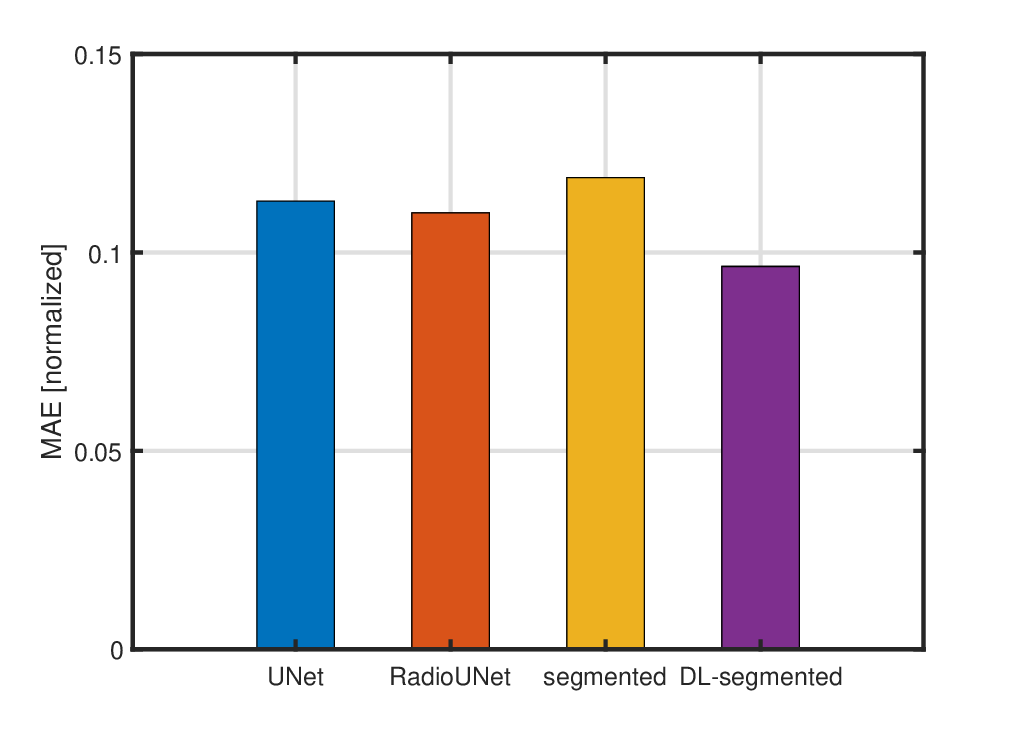}
	\caption{The prediction MAE of the radio map construction using different methods.}
	\label{fig:RadioMap_MAE}
\end{figure}

We first present the performance of radio map construction using the
following methods, U-net \cite{RonFisBro:J15}, RadioUNet \cite{2008},
segmented model \cite{61} and DL-based segmented model \cite{ZenChe:C22}.
Among them, UNet and RadioUNet are both based on the vanilla UNet
architecture with multiple encoder layers connected with decoders.
They require geographical city maps as input and predict path loss
at every point in a target area in a single inference. Segmented model
and DL-based segmented model take the position pair of the transmitter
and receiver as input to predict the corresponding path loss and reconstruct
the virtual 3D radio environment. The prediction MAE of the radio
map construction using different methods are presented in Fig.~\ref{fig:RadioMap_MAE}.
The DL-based segmented model outperforms the others in terms of radio
map construction accuracy. In addition, the UNet and RadioUNet require
the information of geographic environment.

The true city map and the virtual obstacle maps generated by the segmented
model \cite{61} and DL-based segmented model \cite{ZenChe:C22} are
also plotted in Fig. \ref{fig:virtual_obsmap}. It can be seen that
the geometry and distribution of the virtual obstacles quite coincide
with the true city map which shows that the hidden geographic spatial
information can be uncovered from the RSS measurements of radio signals.
Compared with the DL-based method, the geometry of virtual obstacle
map generated by the segmented model is closer to the true city map
while the DL-based one is vague. Yet, it is noted that the virtual
obstacle map, which serves as a geometry interpretation of the radio
propagation environment, is not necessarily expected to be identical
to the true city map.

\begin{center}
	\begin{figure*}[!t]
		\includegraphics[width=0.8\paperwidth]{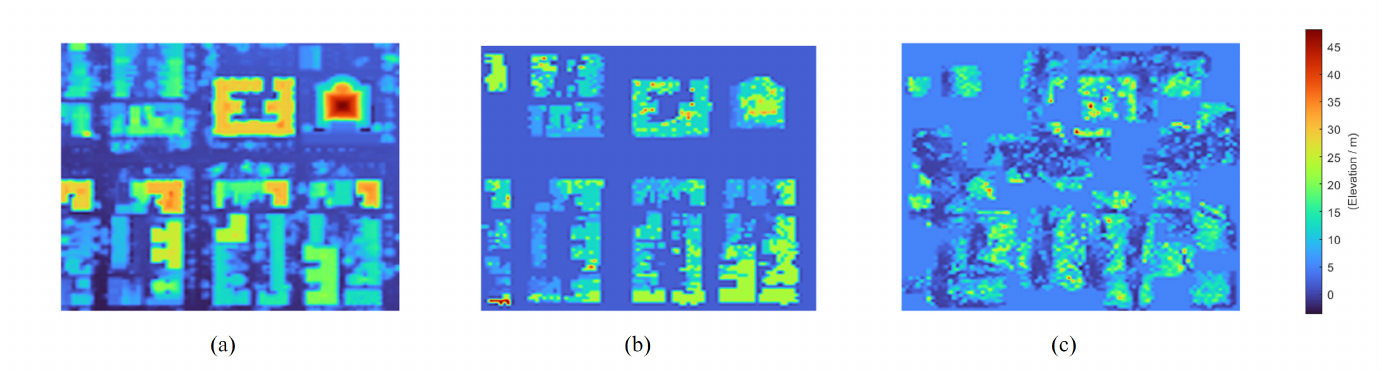}
		\caption{(a) The true city map. (b) The virtual obstacle generated by segmented
			model. (c) The virtual obstacle generated by DL-based segmented model.}
		\label{fig:virtual_obsmap}
	\end{figure*}
	\par\end{center}

\subsection{Summary}
To summarize for the CKM construction techniques, classical {environment} model-free approaches, such as Kriging and kernel regression, are computationally efficient when the data is sparse, but their scalability is poor for a massive amount of high-dimensional data over a large map. Tensor-based methods need to take care of the identifiability issue, where a finer grid topology usually requires much more data to satisfy the identifiability condition for tensor completion. {Environment} model-assisted approaches generally require more data to learn the model parameters and are more computationally heavy in the training phase; however, these methods are easier to scale up to a geographically large scenario for constructing a large CKMs from a massive amount of high-dimensional data. 

\section{Utilizing Channel Knowledge Maps}\label{sec:utilizing}
Building upon the constructed CKMs in Section~\ref{sec:construc}, this section discusses how to utilize them to enable environment-aware communications efficiently. %First, Section \ref{sec:utilization:1} presents a general framework for utilizing CKM to optimize wireless communications networks. Next, based on this framework, the remaining subsections provide several CKM-utilization use cases, in which the CKMs are utilized to enable training-free or light-training communications, predictive communications for yet-to-reach locations, resource management for non-cooperative nodes and dense networks, as well as localization and sensing.

\subsection{Optimization Framework for CKM Utilization}\label{sec:utilization:1}

\begin{figure}
\centering
\includegraphics[scale=0.8]{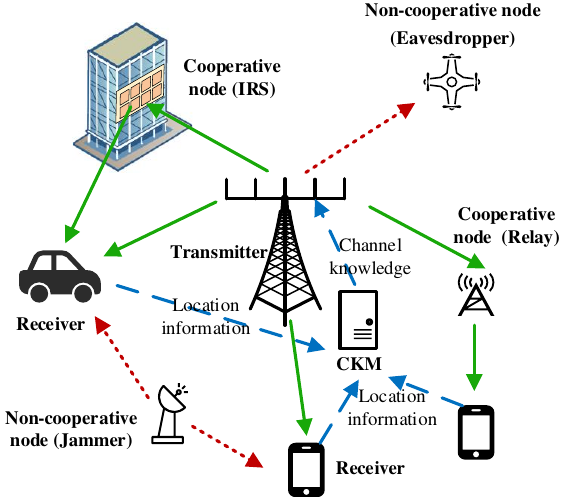}
\caption{A generic wireless system for illustrating the optimization framework with CKM.}
\label{F:UtilizationModel}
\end{figure}

This subsection first presents an optimization framework for CKM utilization. Without loss of generality, we consider a generic wireless system as shown in Fig. \ref{F:UtilizationModel}, which consists of $K_1 \ge 1$ transmitters, $K_2 \ge 1$ receivers, $K_3 \ge 0$ cooperative nodes, and $K_4\ge 0$ non-cooperative nodes. Let $\mathcal K_1$, $\mathcal K_2$, $\mathcal K_3$, and $\mathcal K_4$ denote their corresponding sets, and $\mathcal K = \bigcup_{i=1}^4 \mathcal K_i$ denote the set of all communication nodes. 
%of transmitters, receivers, cooperative nodes, and non-cooperative nodes, respectively, 
In this system, the transmitters aim to deliver information to the respective receivers. The cooperative nodes (such as relays and IRSs) try to help their communications (e.g., via relaying), and the non-cooperative nodes (such as interferers, eavesdroppers, and jammers) may harm their communications (e.g., by jamming or eavesdropping). 

%point-to-point channel (i.e., $K_1=K_2=1$),  broadcast channel (i.e., $K_1 = 1$ BS transmitter sends individual messages to $K_2>1$ UE receivers), multicast channel (i.e., $K_1 = 1$ BS transmitter sends common messages to $K_2>1$ UE receivers), multiple access channels (i.e., $K_1>1$ UE transmitter sends individual messages to $K_2=1$ BS receiver), and interference channel (i.e., the $K_1 > 1$ transmitters each send individual messages to the corresponding $K_2=K_1$ receivers) as special cases.

%the set of which is denoted by $\mathcal{K} = \{1,\ldots, K\}$. Suppose that each communication node $k\in\mathcal K$ is equipped with $M_k$ antenna elements. In general, the communication nodes can be classified into four categories including source nodes (or transmitters), destination nodes (or receivers), cooperative nodes, and non-cooperative nodes.

We focus on the wireless communications over $T$ channel coherence blocks, with each block consisting of $S$ symbol durations. Let $\mathcal{T} = \{1,\ldots, T\}$ denote the set of channel blocks, and $\mathcal{S} = \{1,\ldots, S\}$ denote the set of symbols in each block. We consider the block fading channel model, where the node locations and wireless channels remain unchanged over each block, but may change over different blocks. At block $t\in\mathcal T$, let $\boldsymbol{q}_i[t]$ denote the position of node $i\in\mathcal K$, and $\boldsymbol{H}_{i,j}[t]$ denote the channel matrix from node $i \in \mathcal K$ to node $j \in \mathcal K, i\neq j$. In general, the dimension of each channel matrix $\boldsymbol{H}_{i,j}[t]$ depends on the antenna configuration at nodes $i$ and $j$, and the number of subcarriers.

%For ease of exposition, we consider frequency non-selective fading channels, though our developed framework can be easily extended to the case with frequency-selective fading channels.

%In this case, let  denote the channel matrix from  at channel coherence block $t\in\mathcal T$. In the narrowband transmission, $\boldsymbol{H}_{i,j}[t]$ corresponds to an $M_j \times M_i$ complex-valued matrix with each element denoting the channel coefficient from one transmit to one receive antenna. In the wideband transmission with $N$ subcarriers, $\boldsymbol{H}_{i,j}[t]$ can be expressed as an $M_j \times M_i \times N$ complex-valued tensor with each element denoting the channel coefficient from one transmit to one receive antenna over one subcarrier. For the ease of exposition, we focus on the narrowband transmission and our developed framework is easily extendible to the wideband transmission.

In general, our objective is to optimize certain decision variables at the transmitters,  receivers, and cooperative nodes for maximizing the system utility over the whole $T$ blocks. %As the non-cooperative nodes operate without coordination with the transmitters and the receivers, we cannot control or optimize their transmission/reception. 
Without loss of generality, let $\boldsymbol{\mathcal{A}}_i[s,t]$ denote the decision variables by each node $i \in \bigcup_{i=1}^3 \mathcal K_i$ at symbol $s \in \mathcal S$ of block $t\in \mathcal T$.\footnote{In practice, the decision variables can be the transmit power allocations, the bandwidth allocation, the transmit/receive/reflective beamforming vectors or precoding/combining matrices, and even the placement locations $\boldsymbol{q}_i[t]$'s (e.g., for mobile robots).} 
%\footnote{Note that the wireless channels remain unchanged over each block. Therefore, if the CSI is known {\it a priori} at the system, then each node $i$ may adopt the same transmission/reception strategy over one block by setting $\boldsymbol{\mathcal{A}}_i[1,t] = \cdots = \boldsymbol{\mathcal{A}}_i[S,t]$. In practice, however, the CSI may not be available at the beginning of each block, and as a result, $\boldsymbol{\mathcal{A}}_i[s,t]$'s may vary over different symbol durations $s$'s in each block, e.g., for channel training and data transmission, respectively.} 
Based on the channel matrices $\{\boldsymbol{H}_{i,j}[t]\}$ and the design variables $\{\boldsymbol{\mathcal{A}}_i[s,t]\}$, the system utility function at block $t\in\mathcal T$ is expressed as $U_t(\{\boldsymbol{H}_{i,j}[t]\} , \{\boldsymbol{\mathcal{A}}_i[s,t]\})$, and the total system utility over the whole $T$ blocks is expressed as $U(\{\boldsymbol{H}_{i,j}[t]\} , \{\boldsymbol{\mathcal{A}}_i[s,t]\}) = \sum_{t\in\mathcal T} U_t(\{\boldsymbol{H}_{i,j}[t]\} , \{\boldsymbol{\mathcal{A}}_i[s,t]\})$. As a result, the system utility maximization problem is mathematically formulated as
\begin{align}\label{eqn:utilization:1}
\max_{\{\boldsymbol{\mathcal{A}}_i[s,t]\}}&U(\{\boldsymbol{H}_{i,j}[t]\} , \{\boldsymbol{\mathcal{A}}_i[s,t]\}) \\
\mathrm{s.t.}~& \{\boldsymbol{\mathcal{A}}_i[s,t]\} \in \mathcal{O}_i, \forall i\in \cup_{i=1}^3 \mathcal K_i, \nonumber
\end{align}
where $\mathcal{O}_i$ denotes the feasible set of $\{\boldsymbol{\mathcal{A}}_i[s,t]\}$ for each node $i$ characterized by practical constraints.% on, e.g., maximum bandwidth and maximum transmit power.

However, the optimization of $\{\boldsymbol{\mathcal{A}}_i[s,t]\}$ in the system utility maximization problem \eqref{eqn:utilization:1} is challenging. This is due to the fact that the evaluation of the system utility function $U_t(\{\boldsymbol{H}_{i,j}[t]\} , \{\boldsymbol{\mathcal{A}}_i[s,t]\})$ highly relies on the current and future CSI $\{\boldsymbol{H}_{i,j}[t]\}$, % of the four types of communication nodes, 
but obtaining such CSI is generally difficult, costly, or even impossible, as specified in the following.
\begin{itemize}
  \item First, obtaining the current CSI among the transmitters, receivers, and cooperative nodes at a particular block $t$ requires sophisticated channel training and feedback, which may consume a large portion of time-frequency resources and compromise the communication performance, especially when the number of  nodes and the associated channel dimensions become large.
  \item Next, predicting the future CSI $\{\boldsymbol{H}_{i,j}[\tau]\}$ of blocks $\tau > t$ relies on the channel correlations over time, which may become less accurate when the nodes are highly mobile and the wireless channels are fast-changing.
  \item Furthermore, it is difficult if not impossible for the system to acquire the CSI associated with non-cooperative nodes, as the conventional channel training is infeasible due to the nodes' non-cooperative nature.
\end{itemize}

To resolve the above issues, CKM is a viable new solution, which can support the acquisition of the cooperating nodes' current CSI in a light-training or even training-free manner, enable the efficient prediction of their future CSI, and help obtain the non-cooperative nodes' CSI, thus outperforming conventional channel training or environment-unaware based designs. In the following, we first discuss the conventional designs without CKM in Sections \ref{sec:conventional1} and \ref{sec:conventional2}, and then present the new CKM-enabled environment-aware solution in Section \ref{sec:CKM:solution}.
%, thus supporting the environment-aware communications. 

%it is  to  of  in practice. First, in order to optimize $\{\boldsymbol{\mathcal{A}}_i[s,t]\}$ in one particular block $t\in\mathcal T$ highly relies on the availability of CSI $\{\boldsymbol{H}_{i,j}[t]\}$ at that block, and designing $\{\boldsymbol{A}_i[s,t]\}$ over different blocks $t$'s further requires the prediction of CSI $\{\boldsymbol{H}_{i,j}[\tau]\}$ at future blocks $\tau > t$.

\subsubsection{Conventional channel training based design}\label{sec:conventional1}

First, we focus on one particular communication pair between transmitter $i$ and receiver $j$. They implement channel training to obtain the CSI of $\boldsymbol{H}_{i,j}[t]$ at each block $t$, which consists of $M_{\mathrm{BS}} M_{\mathrm{UE}} N_{\mathrm{SC}}$ elements to be estimated. In order to obtain the CSI at receiver side, transmitter $i$ sends $M_{\mathrm{BS}} N_{\mathrm{SC}}$ predetermined training or pilot symbols at each block $t$,\footnote{The required training signals may become less if the channel correlation and sparsity across different subcarriers and different antennas are exploited.} such that receiver $j$ can successfully estimate  $\boldsymbol{H}_{i,j}[t]$. Furthermore, to obtain the CSI at transmitter side, receiver $j$ further implements limited feedback to send its estimated CSI back to transmitter $i$, or perform the reverse-link channel training if the channel reciprocity holds (e.g., TDD systems). As such, the signaling overhead for channel training is proportional to $M_{\mathrm{BS}} N_{\mathrm{SC}}$ in general, which is quite significant, especially when $M_{\mathrm{BS}}$ and  $N_{\mathrm{SC}}$ become large.

The signaling overhead issue for channel training may become more severe when there are more than one transmitter or receiver ($K_1>1$ and/or $K_2>1$), as the number of channel matrices to be estimated and accordingly the signalling overhead under orthogonal pilot transmission become $K_1K_2$ times of that with only one communication pair. This is unacceptable when $K_1K_2$ becomes sufficiently large for future dense networks with massive devices. In this case, non-orthogonal pilot transmission can be applied, which, however, results in the so-called pilot contamination issue \cite{470} that may degrade the communication performance. On the other hand, the conventional channel training design may become difficult or infeasible, when additional cooperative nodes (e.g., IRSs) and non-cooperative nodes (e.g., interferers) are involved in this system. For instance, it is difficult for the transmitters and receivers to obtain the CSI associated with the cooperative IRS nodes (e.g., the CSI of transmitter-to-IRS and IRS-to-receiver links), as the IRSs are passive nodes without signal transmission or processing capabilities for pilot signals transmission or channel estimation. It is also difficult for the transmitters and receivers to obtain the CSI associated with the non-cooperative nodes, as they operate without the coordination with the transmitters and receivers. Although there have been various studies investigating new channel training protocols for IRSs (e.g., \cite{3052}) and new channel learning methods for non-cooperative nodes (e.g., \cite{6884811,6578560}), these approaches require much higher signaling overheads and implementation complexity.% as compared to the case with cooperative nodes.

\subsubsection{Location-aware design without CKM}\label{sec:conventional2}

Besides channel training, the nodes' location information provided by the GPS and sensing systems can be exploited to enable location-aware design for saving the training overhead. The basic idea of such location-aware designs is to use the simplified channel models (e.g., LoS channel models, location-dependent probabilistic-LoS and Rician channel models) to approximate the corresponding wireless channels. This can help avoid the channel training or reduce the training overhead in scenarios with strong LoS connections between communication nodes. However, such design does not apply for the scenarios with rich scatterers and obstructions in the environments. %This thus calls for new designs by jointly exploiting the location information and wireless environment information.

%and also provide a viable method to roughly predict the future CSI based on the UE trajectory. For instance, based on the UE location, a massive MIMO BS transmitter can design the transmit beamforming based on the AoD/AoA associated with the LoS link. In UAV communications, the UAV can decide its flight altitude and trajectory to optimize its communication performance along the flight path (e.g., \cite{1095}). 

\subsubsection{Environment-aware design with CKM}\label{sec:CKM:solution}

%Different from conventional channel training and the location-aware designs, 

Differently, the exploitation of CKM enables both location- and environment-aware wireless communications without the need of heavy channel training. With CKM at hand, the network controller can obtain the corresponding channel knowledge based on the nodes' current locations, and can also infer the future channel information based on the predicted nodes' trajectories. Accordingly, such channel knowledge can be directly used to optimize the wireless communication designs by solving problem \eqref{eqn:utilization:1}, without or with only limited channel training needed.

For the purpose of illustration, consider one particular time block $t$. Let $ \mathcal{M} $ denote the constructed CKM, and suppose that the network controller is able to obtain the nodes' locations $\boldsymbol{q}_i[t]$'s, $\forall i\in\mathcal K$, by using proper localization and/or sensing techniques. By using the CKM $ \mathcal{M} $ together with  $\boldsymbol{q}_i[t]$'s, it follows from (\ref{eq:mapping}) that the network controller is able to infer the channel knowledge between nodes $i$ and $j$, given by
\begin{align}
\boldsymbol{z}_{i,j}[t] = f(\boldsymbol{q}_i[t],\boldsymbol{q}_j[t]).
\end{align}
The structures of $\boldsymbol{z}_{i,j}[t]$'s and their dimension $J$ depend on the specific CKM types employed. If the CMM is employed, then $\boldsymbol{z}_{i,j}[t]$ corresponds to an estimated version of the channel matrix $\boldsymbol{H}_{i,j}[t]$; if CPM is considered, then $\boldsymbol{z}_{i,j}[t]$ contains the delays/Dopplers/AoAs/AoDs of multi-paths, which can be used to construct an estimate of $\boldsymbol{H}_{i,j}[t]$. Let $\hat{\boldsymbol{H}}_{i,j}[t]$ denote the constructed or estimated channel matrix based on the CKM.

Similarly, the network controller can further predict the nodes' trajectory $\boldsymbol{q}_i[\tau]$'s, based on which the estimate of future channel matrices can be inferred as $\bar{\boldsymbol{H}}_{i,j}[t]$'s. As such, the network utility function $U_t(\{{\boldsymbol{H}}_{i,j}[t]\} , \{\boldsymbol{\mathcal{A}}_i[s,t]\})$ at the current block $t$ is approximated as $U_t(\{\hat{\boldsymbol{H}}_{i,j}[t]\} , \{\boldsymbol{\mathcal{A}}_i[s,t]\})$, and  $U_\tau(\{{\boldsymbol{H}}_{i,j}[\tau]\} , \{\boldsymbol{\mathcal{A}}_i[s,\tau]\})$ at future blocks $\tau$ is approximated as $U_\tau(\{\bar{\boldsymbol{H}}_{i,j}[\tau]\} , \{\boldsymbol{\mathcal{A}}_i[s,\tau]\})$. As a result, at time block $t \ge 1$, we obtain an approximate version of problem \eqref{eqn:utilization:1} as
\begin{align}
\max_{\{\boldsymbol{\mathcal{A}}_i[s,t]\}} & U_t(\{\hat{\boldsymbol{H}}_{i,j}[t]\} , \{\boldsymbol{\mathcal{A}}_i[s,t]\}) \nonumber\\
&+ \sum_{ \tau\in \mathcal{T}, \tau > t} U_{\tau}(\{\bar{\boldsymbol{H}}_{i,j}[{\tau}]\} , \{\boldsymbol{\mathcal{A}}_i[s,{\tau}]\}) \nonumber\\
\mathrm{s.t.}~& \{\boldsymbol{\mathcal{A}}_i[s,t]\} \in \mathcal{O}_i, \forall i\in \cup_{i=1}^3 \mathcal K_i.\label{eqn:utilization:3}
\end{align}
It is clear that the network utility objective function in problem \eqref{eqn:utilization:3} can be successfully evaluated without any channel training implemented. Therefore, proper optimization techniques can be applied to design $\{\boldsymbol{\mathcal{A}}_i[s,t]\}$, for solving problem \eqref{eqn:utilization:3}. This thus leads to the training-free communications.

The training-free system design in \eqref{eqn:utilization:3}, however, may suffer from performance degradation, as the constructed CKM $ \mathcal{M} $ and the obtained nodes locations $\{\boldsymbol{q}_i[t]\}$ may deviate from the ground-truth values due to errors induced during the CKM construction and localization processes. Therefore, proper online channel training can be employed to refine the estimated channel matrices. Different from the conventional channel training designs, the online training process here can exploit the coarse estimates from CKM to reduce the pilot overhead, thus achieving light-training communications.

Furthermore, besides the training-free and light-training communications, the utilization of CKM in problem \eqref{eqn:utilization:3} can enable predictive communication with yet-to-reach locations by properly planning $\{\boldsymbol{\mathcal{A}}_i[s,t]\}$ over blocks, and support interference management with non-cooperative nodes by acquiring their channel knowledge without sophisticated channel learning. Furthermore, CKM can also be exploited to facilitate network planning and link scheduling as well as localization and sensing. We will focus on these applications by discussing their representative use cases in detail next.

\subsection{Training-Free and Light-Training Communications}\label{sec:utilization:2}

%Conventional wireless communications rely on channel training to obtain the CSI for adapting the transmit and receive strategies (e.g., power allocation, beamforming). However, the channel training may become costly or difficult when the channel dimension becomes too large or when the transmitters and/or receivers have hardware limitations. CKM provides a new solution to resolve such issues to enable 

This subsection presents the CKM-enabled training-free or light-training communications. In the following, we first consider the hybrid beamforming for massive MIMO systems as a representative example to show the benefit of CKM-enabled designs, and then discuss the extension to other scenarios.

\subsubsection{Hybrid beamforming for mmWave massive MIMO}
We consider the hybrid analog-digital beamforming for a point-to-point mmWave massive MIMO communication system with one transmitter and one receiver \cite{CKM_Utilization_Yong_Hybrid_BF}. The transmitter has a hybrid analog-digital architecture with $M_t$ transmit antennas and $M_t^{\text{RF}}$ RF chains, with $M_t^{\text{RF}}<M_t$. The receiver is equipped with $M_r$ receive antennas and $M_r^{\text{RF}}$ RF chains, with $ M_r^{\text{RF}}<M_r $. The transmitter sends $M_s$ parallel data streams. Over one particular time block $t$, the design variables $\{\boldsymbol{\mathcal{A}}_1[s,t]\}$ at the transmitter include the transmit analog and digital beamformers, i.e., $\boldsymbol{F}_{\text{RF}} \in \mathbb{C}^{M_t \times M_t^{\text{RF}}}$ and $\boldsymbol{F}_{\text{BB}} \in \mathbb{C}^{ M_t^{\text{RF}} \times M_s}$, and the design variables $\{\boldsymbol{\mathcal{A}}_2[s,t]\}$  at the receiver include the receive analog and digital beamformers, i.e., $\boldsymbol{W}_{\text{RF}} \in \mathbb{C}^{M_r \times M_r^{\text{RF}}}$ and $\boldsymbol{W}_{\text{BB}} \in \mathbb{C}^{ M_r^{\text{RF}} \times M_s}$. Suppose that the MIMO channel matrix during this block is given by $\boldsymbol{H}[t]$. Consider that the optimal receive digital beamformers are employed at the receiver. The achievable rate over $ T $ coherent blocks is used as the system utility, which is given by \cite{CKM_Utilization_Yong_Hybrid_BF}
\begin{align}\label{eqn:Hybrid:BF:1}
R = \sum_{t=1}^T\log\det\left(\boldsymbol{I}_{M_r^{\text{RF}}} + \frac{1}{\sigma^2}  \boldsymbol{H}_e[t] \boldsymbol{F}_{\text{BB}}[t] \boldsymbol{F}_{\text{BB}} ^H[t]\boldsymbol{H}_e[t]^H \right),
\end{align}
where $\boldsymbol{H}_e[t] = (\boldsymbol{W}_{\text{RF}}^H[t] \boldsymbol{W}_{\text{RF}}[t] )^{-1/2} \boldsymbol{W}_{\text{RF}}[t] \boldsymbol{H}[t] \boldsymbol{F}_{\text{RF}}[t]$ denotes the equivalent MIMO channel with the transmit and receive processing. The design objective is to optimize the hybrid beamforming to maximize the rate or system utility given in \eqref{eqn:Hybrid:BF:1}.

%\begin{figure}[htb]
%\centering
%\includegraphics[scale=0.18]{HybridBeamforming}
%\caption{The structure of the hybrid analog-digital beamforming \cite{CKM_Utilization_Yong_Hybrid_BF}.}\label{F:hybrid}
%\end{figure}

Solving the above hybrid beamforming problem depends on the acquisition of CSI $\boldsymbol{H}[t]$, which, however, is particularly challenging due to the large dimension of massive MIMO channels and the coupling between analog and digital beamformers. Conventionally, training-based channel estimation \cite{7961152} and beam sweeping \cite{9499127} are two widely adopted approaches for hybrid beamforming design. In training-based channel estimation, the complete MIMO channel matrix is first estimated by using the pilots, and then the hybrid beamformers are designed based on the estimated channel matrix. However, the pilot overhead for this scheme increases significantly with respect to the number of antennas, which thus becomes prohibitive for massive MIMO systems. Supposing that each coherent block $ t $ contains $ N $ symbol durations, and $N_{\text{tr}}$ of them are used for channel training. 
Thus, the actual achievable data rate becomes
\begin{equation}\label{eqn:Hybrid:BF:2}
\small
\begin{aligned}
R = \frac{N-N_{\text{tr}}}{N} \sum_{t=1}^{T}\log\det\left(\boldsymbol{I}_{M_r^{\text{RF}}} + \frac{1}{\sigma^2}  \hat{\boldsymbol{H}}_e[t] \boldsymbol{F}_{\text{BB}}[t] \boldsymbol{F}_{\text{BB}} ^H[t]\hat{\boldsymbol{H}}_e[t]^H \right),
\end{aligned}
\end{equation}
where $\hat{\boldsymbol{H}}_e[t]$ denotes the estimated equivalent MIMO channel. It is observed from \eqref{eqn:Hybrid:BF:2} that the achievable rate is compromised if the training duration $T_{\text{tr}}$ becomes large.

On the other hand, in beam sweeping, the transmitter and receiver search over possible training beamformers from pre-defined codebooks, and then choose the best beamforming pairs without explicitly estimating the MIMO channel. Nevertheless, the training overhead for this scheme is proportional to the number of possible beam combinations, which can be quite large in order to find good beamforming pairs.

Alternatively, CKM provides a viable new solution to realize the hybrid beamforming without training or with only light training needed, by exploiting the nodes' location information and the environment information provided by CKM. In particular, the design of training-free and light-training hybrid beamforming depends on the types of CKM utilized. In the following, we discuss the use of CMM and CPM to enable training-free hybrid beamforming, and then consider the \ac{cam} and BIM to support light-training hybrid beamforming. For ease of illustration, we focus on one channel coherent block.

\subsubsection{CMM and CPM enabled training-free hybrid beamforming}

CMM is a straightforward type of CKM, which directly provides the channel matrix based on transmitter and receiver locations. In the mmWave massive MIMO scenario with the BS location fixed, we only need to obtain the location $\hat{\boldsymbol{q}}[t]$, and accordingly obtain an estimate of the channel matrix as $\hat{\boldsymbol{H}}[t]$. As CMM can directly provide the channel matrix without any further computation needed, it has the lowest implementation complexity, but an excessive storage cost for storing channel matrices with large dimensions.

To reduce the high storage cost of CMM, CPM is proposed as another type of CKM that offers location-specific channel path information, including the number of significant paths and their power, phases, and AoA/AoDs, at any possible transmitter and receiver locations. For mmWave massive MIMO of our interest, we can directly obtain these channel path information based on the estimated UE location  $\hat{\boldsymbol{q}}[t]$. At time block $t$, suppose that the number of significant paths is $L[t]$, and the complex gain, AoA, and AoD of the $l$-th path are given by $\alpha_l[t]$, $\theta_l[t]$, and $\phi_l[t]$, respectively, where the AoA and AoD may contain both zenith and azimuth directions, i.e., $\theta_l[t] = (\theta_l^z[t], \theta_l^a[t])$, and $\phi_l[t] = (\phi_l^z[t], \phi_l^a[t])$, with the superscript $z$ and $a$ denoting zenith and azimuth directions, respectively. Accordingly, we can obtain the reconstructed MIMO channel matrix at time block $t$ as
\begin{align}
\hat{\boldsymbol{H}}[t] = \sum_{l=1}^{L[t]} \alpha_l[t] \boldsymbol{a}_r(\theta_l[t])  \boldsymbol{a}_t^H(\phi_l[t]),
\end{align}
where $\boldsymbol{a}_r(\cdot)$ and $\boldsymbol{a}_t(\cdot)$ denote the transmit and receive array response vectors, respectively.

Based on the estimated channel matrix $\hat{\boldsymbol{H}}[t]$ from CMM or CPM, the BS transmitter and the UE receiver can accordingly design the hybrid beamformers. It is clear that no extra channel training is needed for the above two CKMs, and as a result, training-free hybrid beamforming is achieved. Note that the performance of such training-free designs highly depends on the accuracy of the CMM/CPM and the UE location. In practice, the time-variation nature of wireless environment (e.g., due to dynamic scatterers like pedestrians) and the localization errors may affect the accuracy of the resultant channel matrix estimations, especially for the elements that change fast over time (e.g., the complex channel gain $\alpha_l[t]$'s). Therefore, additional but limited training may further benefit the utilization of CKM, as shown next.

\subsubsection{CAM and BIM enabled light-training hybrid beamforming}

\begin{figure}[htb]
\centering
\begin{subfigure}[b]{0.48\textwidth}
\centering
\includegraphics[width=\textwidth]{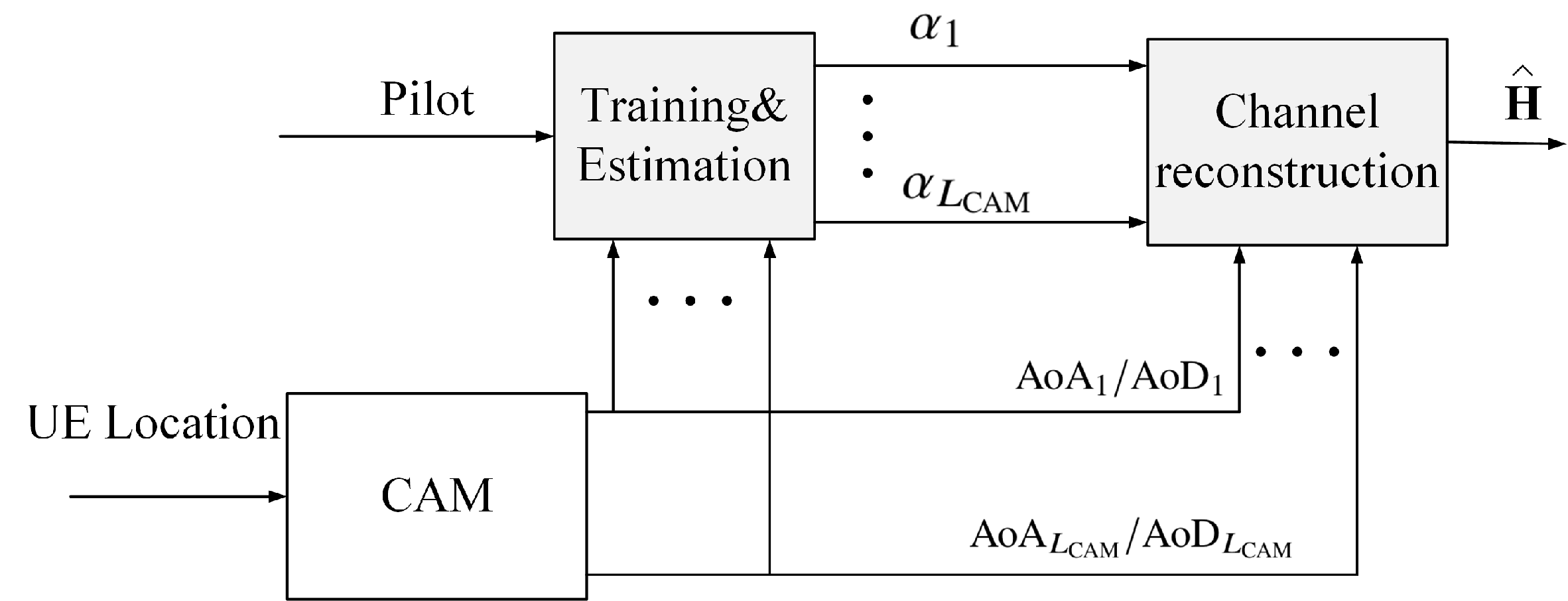}
\caption{}
\end{subfigure}
\hfill
\vspace{0.2cm}
\begin{subfigure}[b]{0.38\textwidth}
\centering
\includegraphics[width=\textwidth]{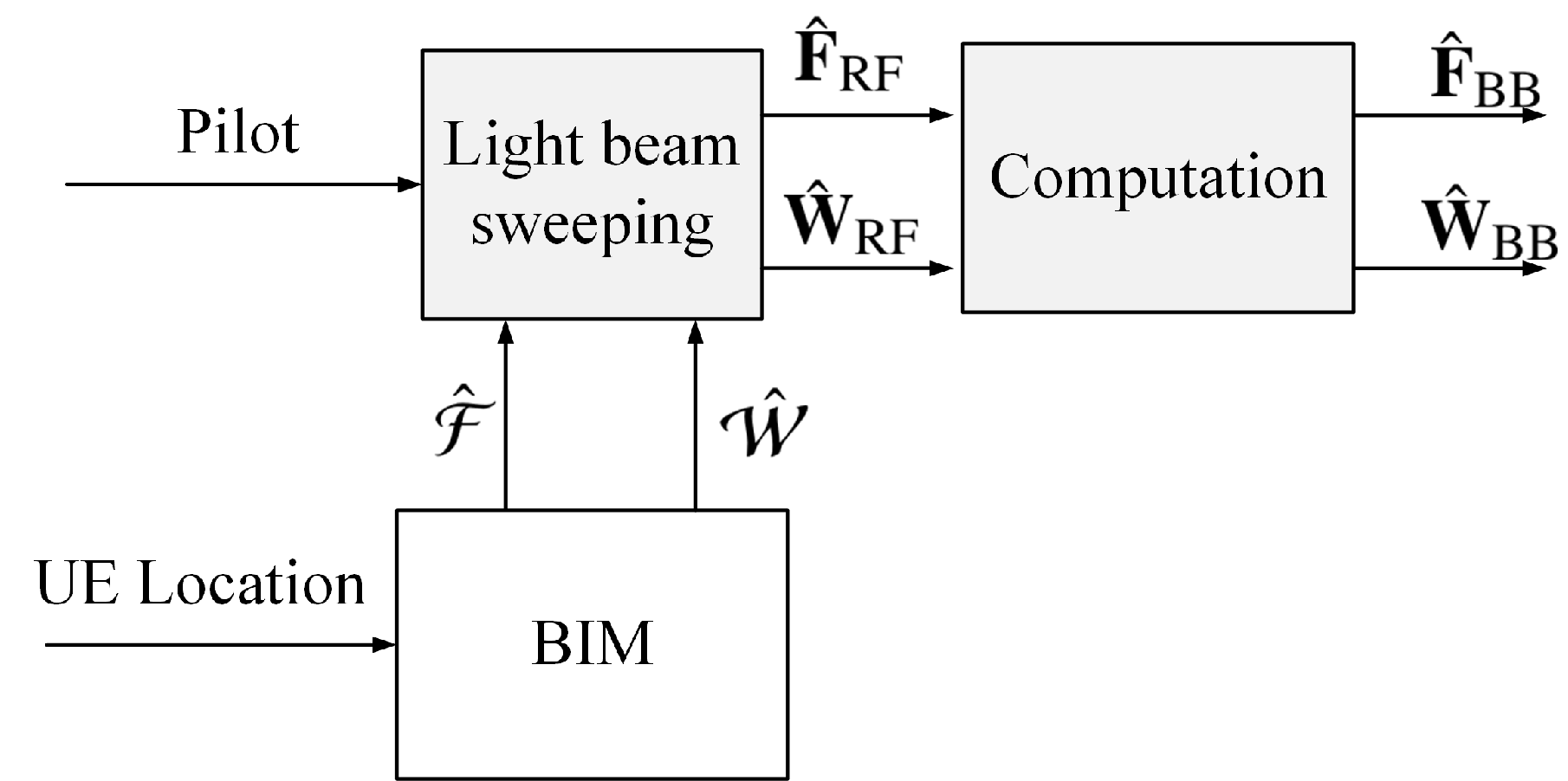}
\caption{}
\end{subfigure}
\caption{Illustration of the light-training based hybrid beamforming design by utilizing CAM (a) and BIM (b), respectively.}
\label{fig:light-training}
\end{figure}

To resolve the potential mismatch issue between the inferred channel matrix and the ground truth, a viable solution is to exploit additional training together with the CKM to obtain refined channel estimates. In particular, we consider two light-training designs based on different CKMs, namely CAM and BIM, respectively, where CAM and BIM focus on providing channel information related to large-scale wireless environment, and the additional light training is used to extract the small-scale channel information that may fluctuate significantly over time. For illustration, Fig. \ref{fig:light-training} shows the diagram of the two light-training hybrid beamforming designs.

%\begin{figure}
%  %\includegraphics[width=\linewidth]{boat.jpg}
%  \caption{Illustration of the light-training hybrid beamforming based on CAM and BIM.}
%  \label{fig:light-training}
%\end{figure}

First, consider the CAM-based light-training design, in which CAM only provides path angle information, i.e., the AoA $\theta_l[t]$'s, and the AoD $\phi_l[t]$'s of the $L[t]$ significant paths, and the additional path gain $\alpha_l[t]$'s are viewed as unknown parameters to be learnt via additional training. As such, we express the channel matrix as follows in a compact form, i.e.,
\begin{align}
\hat{\boldsymbol{H}}[t] = {\boldsymbol{A}}_r(\Theta[t]) \mathrm{diag}(\boldsymbol{\alpha}[t]) {\boldsymbol{A}}_t^H(\Phi[t]),
\end{align}
where $\boldsymbol{\alpha}[t] = [\alpha_1[t],\ldots, \alpha_{\hat{L}[t]}[t]]^T$, $\Theta[t] = \{\theta_l[t], l=1,\ldots,\hat{L}[t]\}$, $\Phi[t] = \{\phi_l[t], l=1,\ldots,\hat{L}[t]\}$, ${\boldsymbol{A}}_r(\Theta[t]) = [{\boldsymbol{a}}_r(\theta_1[t]),\ldots, {\boldsymbol{a}}_r(\theta_{\hat{L}[t]}[t])]$, and ${\boldsymbol{A}}_t(\Phi[t]) = [{\boldsymbol{a}}_t(\phi_1[t]),\ldots, {\boldsymbol{a}}_t(\phi_{\hat{L}[t]}[t])]$. With CAM, we are able to obtain the AoA and AoD, and accordingly acquire ${\boldsymbol{A}}_r(\Theta[t])$ and ${\boldsymbol{A}}_t(\Phi[t])$. Then we only need to estimate $\boldsymbol{\alpha}[t]$, which has $\hat{L}[t]$ parameters. Therefore, the BS transmitter and the UE receiver need to properly design their hybrid beamformers to send pilot signals for estimating the elements in $\boldsymbol{\alpha}[t]$. In general, a total number of $\hat{L}[t]$ symbols are needed for pilot transmission. The overhead is significantly reduced as compared to the conventional training-based design.

Different from CAM-based design that still requires the BS to solve the sophisticated optimization problem for obtaining the hybrid beamforming, BIM-based design is another technique that directly obtains the beamformers, which can be viewed as an advanced beam sweeping technique empowered by CKM. BIM provides a mapping from the transmitter and receiver locations to a set of candidate transmit and receive beamforming vectors. The basic idea of BIM-based design is to first obtain the candidate beamforming vectors based on the UE locations, and then the BS implements the beam sweeping over these candidate beamforming vectors, instead of all possible beamforming vectors for reducing the sweeping overhead. As such, the BIM-based design is expected to achieve a comparable performance as exhaustive beam sweeping, but with significantly reduced overhead. 

\begin{figure}
  \includegraphics[width=\linewidth]{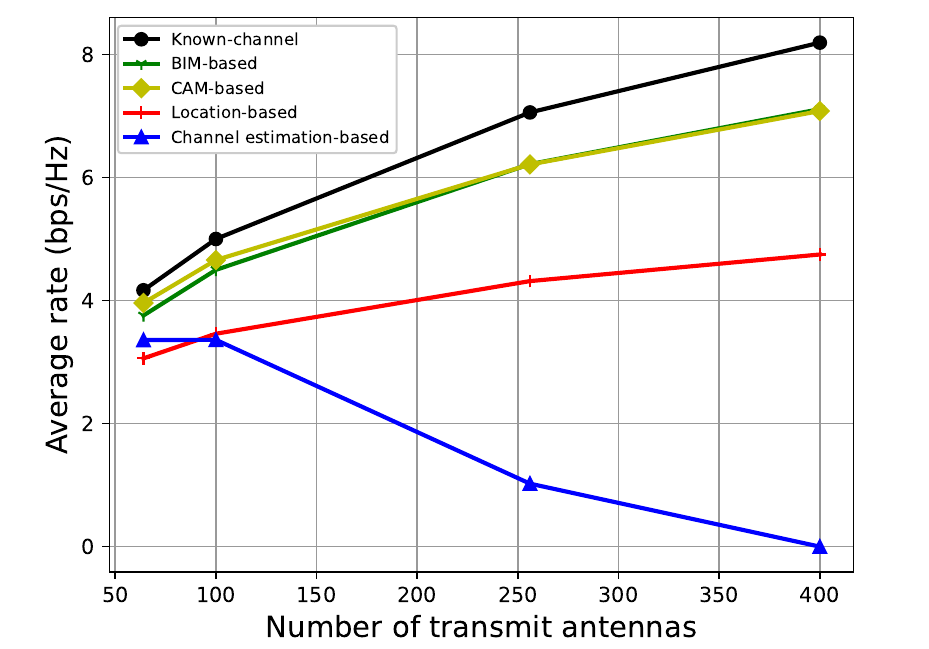}
  \caption{Performance comparison between different CKM-based hybrid beamforming designs versus the conventional channel-estimation based design.}
  \label{fig:performance-Hybrid-BF}
\end{figure}

Fig. \ref{fig:performance-Hybrid-BF} compares the achievable effective communication rate of different approaches. For the benchmark schemes, we consider the least square (LS)-based channel estimation and the location-based beam alignment, respectively\cite{4026,4027}. It is observed from Fig.  \ref{fig:performance-Hybrid-BF}  that the effective communication rate of the channel estimation based design decreases drastically as the number of BS antennas grows, since its training overhead outweighs the resulting beamforming gain for large antenna systems. By contrast, the CAM- and BIM-enabled schemes achieve monotonic rate improvement as BS antennas increase, and they both significantly outperform the location-based scheme, thanks to their drastic reduction of the training overhead with CKM.

\subsubsection{Extensions}\label{sec:utilization:2:1}

Besides hybrid beamforming, the CKM-enabled training-free and light-training communication designs also have abundant applications in various wireless systems. For instance, the authors in \cite{CKM_Utilization_Yong_IRS_BF} proposed to use the CKM to enable the training-free and light-training reflective beamforming for IRS-assisted communications. In particular, the IRS is a passive device with each element being capable of adjusting the signal reflection phases, which can thus form the so-called reflective beamforming for enhancing the reflected signal strength. However, as IRS does not have the signal processing capability due to the lack of signal processing components, it cannot perform the conventional channel training. In this case, we can use the CMM and CPM to enable training-free reflective beamforming designs, and use the CAM and BIM to support light-training designs.

Another possible extension of CKM-enabled training-free and light-training communication is energy beamforming for wireless power transfer (WPT). In energy beamforming, the multi-antenna energy transmitter aims to design the beamformers to efficiently charge a number of energy receivers.
%In WPT, the energy receivers are deployed with rectifiers to charge energy without signal processing capabilities.
In conventional WPT, in order to get the CSI, the energy receiver needs to send pilots in the reverse link and thus the energy transmitter can estimate the channel based on the pilot transmission via exploiting the channel reciprocity, or the energy receiver needs to feed back the received power level such that the energy transmitter can infer the CSI by sending different pilot signals \cite{zhang2018wireless}. As such, the energy receiver needs to consume a large amount of energy to support the CSI acquisition for the energy transmitter, which may compromise the gain provided by energy beamforming. In this case, CKM is expected to be useful to resolve this issue by providing training-free/light-training WPT.

\subsection{Predictive Communication for Yet-to-Reach Locations}\label{sec:utilization:4}

CKM also enables predictive communications, in which wireless channels along the nodes' moving trajectories are predicted before they physically reach there based on CKM, thus supporting more efficient resource management and even mobility control. The CKM-enabled predictive communication is particularly important for B5G/6G wireless networks with machines such as automated guided vehicles (AGVs) in smart factory and UAVs in the sky, whose moving locations or trajectories are controllable and/or highly predictable. Based on the predicted trajectory, we can infer the channel information along the path, which makes it possible to predictively plan the resource allocation and the moving path for optimizing the long-term system utility, instead of only optimizing the one-shot utility in  conventionally designs.% wireless communication systems.

UAV communication is one typical scenario for predictive communication, in which UAVs in the sky are integrated into wireless networks as BSs, relays, or even users, and they can freely control their mobility in the 3D space to communicate with other on-ground or aerial nodes while fulfilling certain tasks. Extensive prior works have been pursued to maximize the average UAV communication performance (e.g., average data rate) over a certain mission period, via joint wireless communication and trajectory design \cite{yan2019comprehensive}. However, these works mainly considered free-space LoS channel models or stochastic wireless channel models (like probabilistic-LoS channels and Rician fading channels), which cannot capture the site- and environment-specific channel characteristics such as the availability of LoS signal paths of the air-to-ground (A2G) links. This thus leads to compromised communication performance. By contrast, CKM provides more accurate channel prediction based on the real environments. In the following, we provide two scenarios with CKM-assisted single-UAV trajectory optimization and multi-UAV placement design, respectively.

\subsubsection{CKM-assisted single-UAV trajectory design}

CKM can facilitate the UAV trajectory design for both scenarios of cellular-connected UAV and UAV-assisted communications \cite{zeng2020uav}. First, consider a cellular-connected single-UAV communication system, in which one single-antenna UAV user flies in the sky to perform certain tasks, and the UAV accesses the terrestrial wireless network to enable remote command and control for efficient and safe operation. Maintaining continuous wireless communications with on-ground BSs is important for the cellular-connected UAV user to safely and successfully accomplish its mission. Towards this end, the UAV user needs to ensure a minimum SINR requirements during the flight. As such, the outage probability can be adopted as a viable utility function in this case, i.e., we have the outage indicator at each block $t$ as
\begin{align}
\mathcal{I}[t] = \left\{
\begin{array}{cc}
  1, & \text{SINR}[t] \ge \Gamma \\
  0, &  \text{SINR}[t] < \Gamma
\end{array}
\right.
\end{align}
where $\text{SINR}[t]$ denotes the SINR of the UAV user at block $t$, and $\Gamma$ denotes the threshold for ensuring the reliable communication. For instance, we may design the UAV trajectory to minimize the outage probability or minimize the mission latency \cite{CKM_Utilization_Shuowen_UAV} over the whole mission period, subject to the initial and final locations of UAV for this mission. In such trajectory design problem, it is important to make decisions based on the channel prediction over the whole flight path. However, conventional designs based on deterministic LoS channels or stochastic channels do not work well as they fail to capture the blockage information of LoS paths. By contrast, based on the CKM idea, the authors in \cite{CKM_Utilization_Shuowen_UAV} first construct an SINR map based on the channel gain map together with the loading factors at potential interfering BSs on the ground, and then adopt the graph and grid-quantization based optimization methods to design the UAV flight trajectories for minimizing the mission latency while ensuring the minimum SINR over the whole path. It is shown that the UAV can fly around the areas with severe LoS path blockage or with strong interference to enhance the communication performance, and conventional designs without CKM work poorly in this case.

Next, consider a UAV-assisted communication system, in which one single UAV BS communicates with multiple on-ground users. One interesting design problem is to optimize the UAV trajectory to maximize the data-rate throughput of the served on-ground users throughput the flight period as the system utility. CKM is also beneficial to enhance the system utility in this case. As the UAV BS needs to serve multiple on-ground users at the same time, it is desired to find a path with good channel qualities with most of these users. As such, the UAV may try to fly over or even hover above locations with strong LoS links with all or most of these users without blockage. This cannot be achieved without CKM, as conventional designs with deterministic/stochastic channels prefer to place the UAV at the location close to all or most of the users, even if the corresponding A2G channels are blocked. How to find the optimal trajectory is still non-trivial due to the complicated relationship between the channels and the nodes' locations, with or without CKM. Some prior works adopted reinforcement learning to find such trajectories for communications \cite{CKM_Utilization_Yuwei_UAV_RL} or for mapping and communication at the same time \cite{2005}.

\subsubsection{CKM-assisted multi-UAV placement design}

CKM is also useful for facilitating the multi-UAV placements or trajectories for CKM-assisted UAV communications. This problem, however, is difficult due to the strong mutual interference among them. In particular, we consider the multi-UAV placement problem, in which multiple UAVs need to optimize their placement locations over the whole communication period for maximizing the network utility (e.g., the weighted sum rate of their A2G links \cite{CKM_Utilization_Haoyun_UAV}). Due to the consideration of CKM, the challenge of this problem is the lack of an analytic function form for such optimization, thus making the conventional convex and non-convex optimization techniques not applicable. To resolve this issue, \cite{CKM_Utilization_Haoyun_UAV} proposes to use the derivative-free optimization to find a close-to-optimal solution to the placement optimization problem.

\begin{figure}[htb]
\centering
\begin{subfigure}[b]{0.5\textwidth}
\centering
\includegraphics[width=\textwidth]{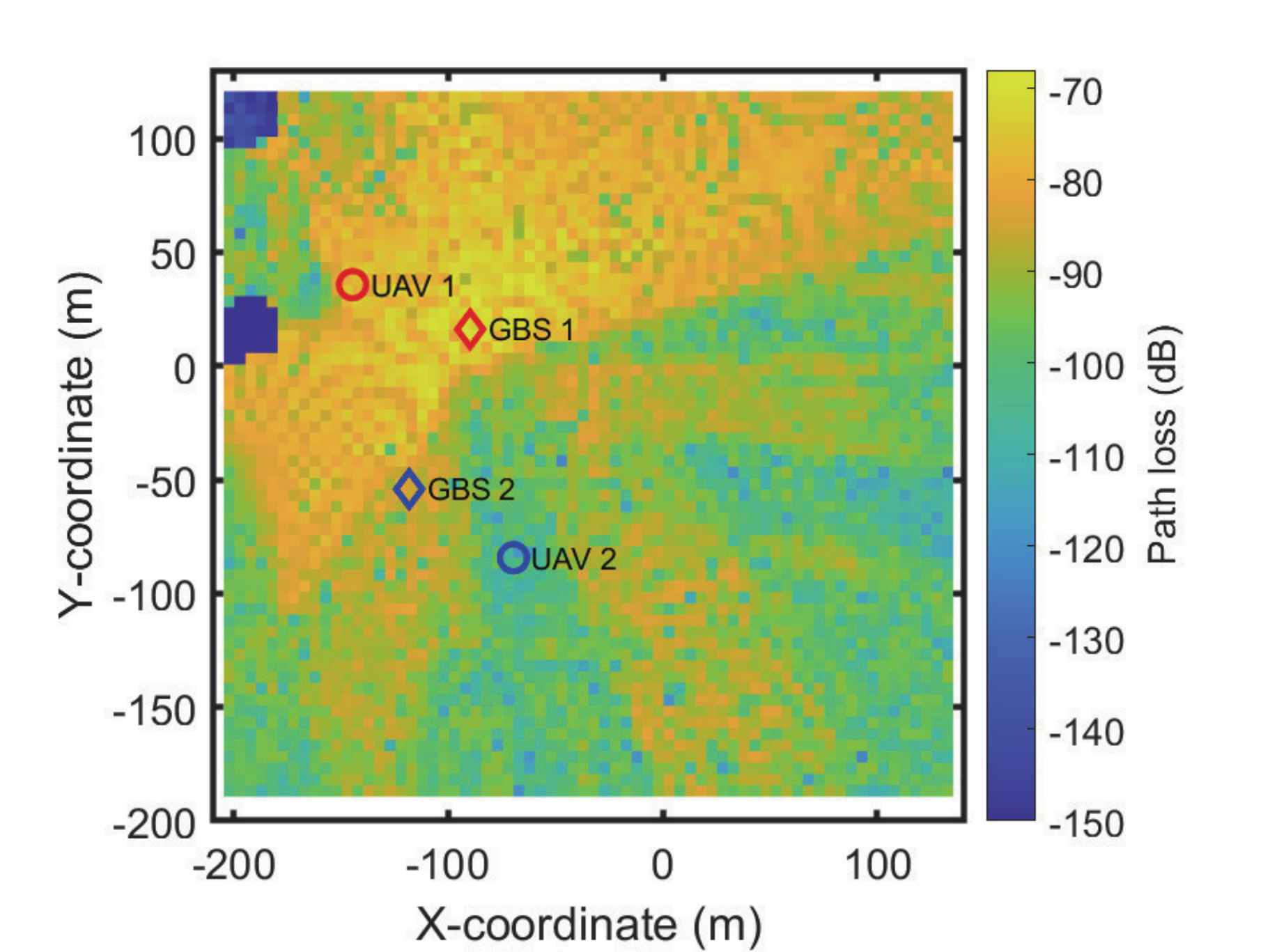}
\caption{CKM of GBS 1.}
\end{subfigure}
\hfill
\begin{subfigure}[b]{0.45\textwidth}
\centering
\includegraphics[width=\textwidth]{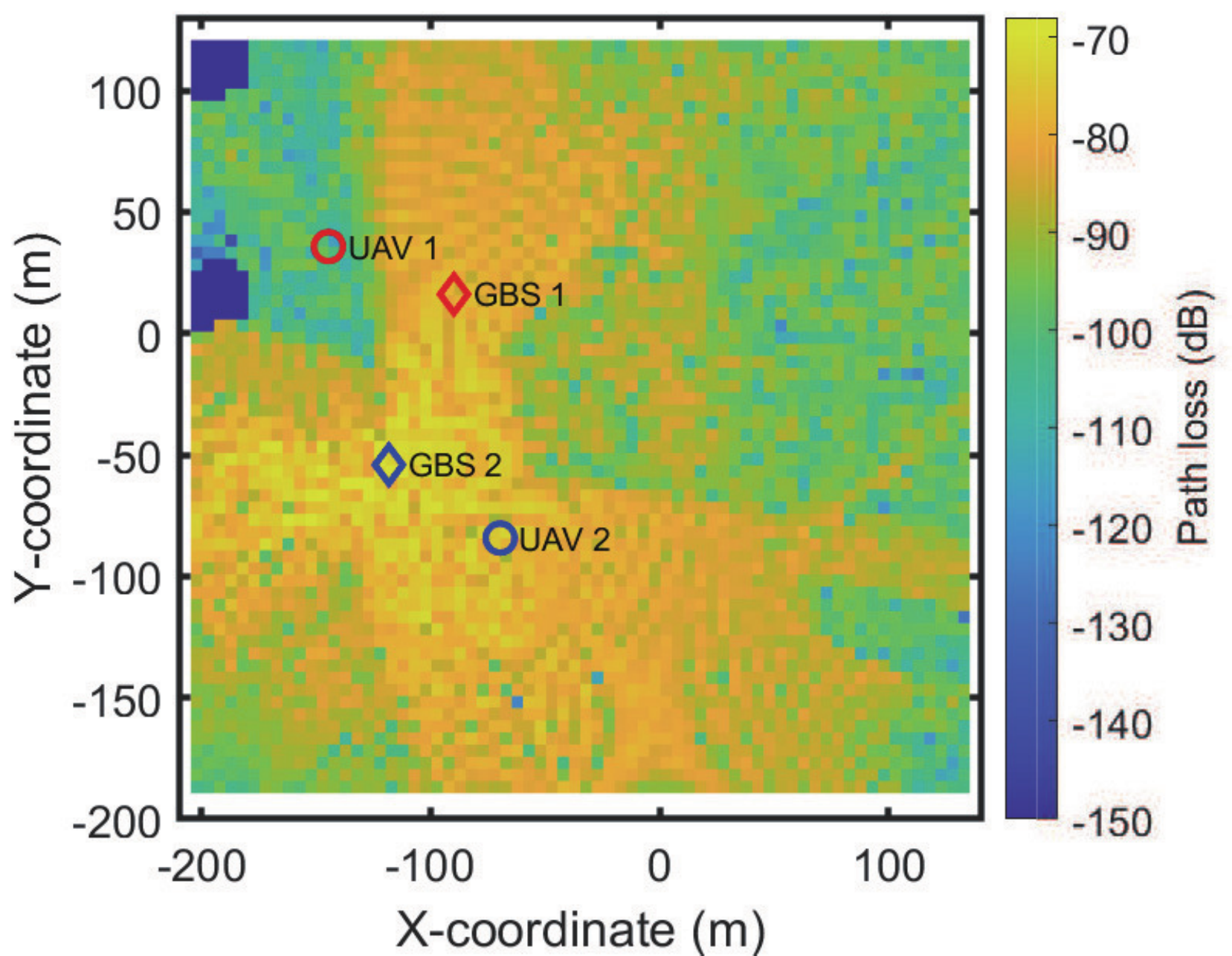}
\caption{CKM of GBS 2.}
\end{subfigure}
\caption{Optimized UAV placement for the case of two UAV users \cite{CKM_Utilization_Haoyun_UAV}.}
\label{fig:UAV:1}
\end{figure}

For illustration, we consider an example with two UAV users sending individual messages to their respectively associated ground BSs (GBSs) over the same frequency band, where each GBS is located at a fixed location. Figs.~\ref{fig:UAV:1}(a) and (b) show the CKM of GBS 1 and GBS 2, respectively, together with the correspondingly optimized locations of UAV 1 and UAV 2 based on the derivative-free optimization \cite{CKM_Utilization_Haoyun_UAV}, where the diamonds indicate GBS locations, the circles indicate the optimized UAV locations. It is observed that UAV 1 (or UAV 2) is placed at a location where its desired link with the correspondingly associated GBS 1 (or GBS 2) enjoys good channel quality, while its interference link with GBS 2 (or GBS 1) is weak, thus mitigating the co-channel interference with the other UAV for enhancing the sum-rate. This cannot be achieved without CKM, thus showing its benefits again.

%%%%%%%%%%%%%%% Add by Jie Xu for revision 2023/11/6
{
\subsubsection{Extensions}
The idea of predictive communications enabled by CKM can also be applied to various ground communication scenarios with, e.g., network-connected ground vehicles and robots. On one hand, if the vehicles/robots move based on predetermined trajectories, then CKM can help provide prior channel information to enable predictive communications. On the other hand, if the mobility of vehicles/robots can be controlled by the network, then such mobility optimization can be exploited as a new degree of freedom (DoF) for performance enhancement. Two new challenges rise for employing predictive communications with CKM for ground vehicles and robots. First, the ground communication channels are usually more complex than A2G channels due to rich scattering environments. In this case, additional light-training might be needed together with CKM to facilitate the channel acquisition and prediction. Next, different from UAVs that can fly freely over the 3D space, vehicles/robots normally have constrained moving paths and trajectories, thus making the trajectory planning more challenging. 
}

%%%%%%%%%%%%%%% Add by Jie Xu for revision 2023/11/6 End here

%It is observed that the horizontal location of UAV 1 converges to $(-144.64\text{m},35.69\text{m})$, and that of UAV 2 converges to $(-69.64\text{m},-84.31\text{m})$ to maximize the sum rate of the two UAVs. The obtained UAV placement locations and the correspondingly achieved sum rate are exactly same as those achieved by the optimal exhaustive search, thus showing the effectiveness of the proposed design. Interestingly

%\begin{figure}
%\centering
%  \subfigure[CKM of GBS 1]{\includegraphics[width=\linewidth]{CKM_GBS1}}
%  \subfigure[CKM of GBS 2]{\includegraphics[width=0.9\linewidth]{CKM_GBS2}}
%  \caption{Optimized UAV horizontal locations for the case with $K = 2$. \cite{CKM_Utilization_Haoyun_UAV}.}
%  \label{fig:UAV:2}
%\end{figure}

\subsection{Resource Management for Non-cooperative Nodes and Dense Networks}\label{sec:utilization:3}

Another use case of CKM-assisted communications is the scenario with non-cooperative nodes such as interferers, eavesdroppers, and jammers. This may correspond to cognitive radio with the cognitive/secondary transmitters and receivers share the spectrum with non-cooperative primary users, and can also correspond to the secrecy communication with the  legitimate transmitters sending confidential information to receivers  in presence of non-cooperative eavesdroppers. In order to design the cognitive communications and the secrecy communications, it is important to acquire the channel information related to the primary users and the eavesdroppers, which, however, is difficult due to their non-cooperative nature. In the literature, there have been various designs that aim at obtaining (part of) their wireless channels by, e.g., monitoring the emitted signals (e.g., \cite{6884811,6578560}). Differently, with the aid of CKM, we can infer the CSI directly based on the location information of non-cooperative nodes, which can be obtained thanks to the recent advancements of device-free localization techniques such as radar and camera sensing.

%need to makes the optimization of such scenarios highly challenging.

%In cognitive radio, the problem becomes ....
%
%
%
%In secrecy communications, the problem can be formulated as ......

CKM is also beneficial to facilitate the link scheduling for networks with dense links, especially for massive D2D links, in which estimating channels over massive communication links is highly costly and even infeasible. In such networks, the D2D user pairs are randomly deployed over a large wireless network, and their transmissions may result in severe interference to degrade the network performance. Proper link scheduling is essential for performance optimization, in which different D2D communication links are allocated with proper time-frequency resources to minimize the co-channel interference. This problem, however, is quite challenging, due to the following two reasons. First, the link scheduling problems normally contain integer optimization variables, which are thus very difficult to be solved from an optimization perspective. Next, similar as UAV communications, we may not be able to express the CKM of communication links as an analytic function of the locations of user pairs, thus making the conventional non-convex optimization inapplicable. In this case, advanced machine learning techniques such as deep learning may be viable solutions. For instance, \cite{8664604} has developed efficient deep learning algorithms for link scheduling. Intuitively, with the CKM at hand, different user pairs can be formed at locations with strong channel conditions but small interference, thus enhancing the network performance.

%\subsection{Network Planning and Link Scheduling for }\label{sec:utilization:5}

%The problem becomes .....
%
%There lacks an
%
%Deep learning may play an important role.

\subsection{CKM-Assisted Localization and Sensing}\label{sec:utilization:6}
While CKM-enabled environment-aware communication is dependent on the availability of nodes' locations, CKM itself can also be inversely utilized to facilitate the localization and sensing. As compared with the conventional localization methods, such as geometric and fingerprinting based localization, the rich location-specific channel information contained in CKM opens a broader design space for wireless localization and sensing in complex environment.

CKM can be utilized as {\it a priori} information to enhance the performance of classical localization methods. For example, with geometric mapping, the target location is estimated based on the intermediate geometric parameters such as distance or direction with respect to the reference points or anchor nodes, and those geometric parameters are usually extracted from the \ac{toa}, \ac{aoa}, and RSS measurements. However, it is known that NLoS anchor nodes cannot contribute to the improvement of localization accuracy, and can even cause interference for localization when no prior information on their NLoS paths is available \cite{Qi2006Loc}. With the aid of a special CKM, i.e., LoS map, the anchor selection procedures can be greatly simplified based on the prior distribution of the target. Besides, it was shown in \cite{Long2022} that localization accuracy, measured by the Bayesian Cram\'{e}r-Rao lower bound (CRLB), can also be significantly improved as compared with the distance-based anchor selection or simply using all anchors.

As compared with geometric mapping, utilizing CKM for fingerprint-based localization seems to be a natural choice, since both CKM and fingerprints are location-tagged and site-specific databases. The traditional fingerprinting mainly uses the measured RSS \cite{3019} to locate the target, and the extension to the channel response is reviewed in \cite{3020}. The use of CAM as fingerprint for localization has been investigated in \cite{2135}. The radio map that contains the path loss information from various BSs to sample locations is stored as a neural network and used for localization in \cite{Yapar2020RealtimeLU}. {Compared with the traditional fingerprints collected by site survey, the CKM enables the flexibility for incorporating heterogenous and diversified channel information, including the channel information from both the BSs and the neighboring devices. Thus, the source of fingerprints is largely expanded and the spatial resolution is en  hanced.} Besides, the channel features extracted in the CKM are usually more stable than the directly measured RSS, and unlike the RSS, the channel information is independent of the device. Thus, CKM-based localization could be more robust to environment dynamics and device variety.

The rich environment information provided by CKM also enables the passive localization and sensing for static obstacles (such as buildings, trees) and dynamic scatterers (such as vehicles and pedestrians) in the ultra-dense network or cell-free massive MIMO. The geometric distribution of the channel features could be used to deduce the location and shape of the obstacles, e.g., the construction of obstacle maps from path loss measurements has been studied in \cite{ZhaChe:C20}. With the movement of the scatterers, the channel information will change accordingly. Hence, the temporal pattern of the CKM or the inconsistence of real-time measurements with CKM, could be used to detect the presence of dynamic scatterers, and even estimate their trajectories. In \cite{ShiQiICC}, the LoS map was utilized to greatly improve the accuracy of Kalman filter for tracing the UAV through the reflected signal and enables the predictive beamforming in complex environment.

%First, consider the classical localization based on geometric mapping.
%}
%\begin{itemize}
%\item{Fingerprint-based indoor localization, RSSI-based, CSI, Channel impulse response.}
%\item{CKM-enabled non-line-of-sight sensing and localization (fingerprinting-based wireless localizations is one simple example and more advanced localization methods can be developed by utilizing more refined CKMs)}
%\item{device-free localization and sensing: localize dynamic scatterer (such as cars, pedestrian) in ultra-dense network (UDN) or cell-free massive MIMO}
%\end{itemize}

\subsection{Summary}\label{sec:utilization:7}
To summarize, the efficient utilization of CKM provides a paradigm shift in wireless systems design from the conventional environment-unaware to the new environment-aware communication, sensing, and localization. In particular, by jointly exploiting the spatial channel knowledge information from CKM and the location information from advanced localization techniques, wireless transceivers can obtain the channel knowledge in a training-free or light-training manner, and accordingly design the resource allocations for performance optimization with low cost. Furthermore, with proper moving path prediction or planning, wireless transceivers can predict or even control the channel conditions in the future, thus enabling a new predictive resource allocation paradigm to further enhance the system performance. Despite these advancements, how to implement the environment-aware communications with outdated and inaccurate CKM is still challenging. Emerging real-time sensing techniques may be employed to update the CKM for supporting robust environment-aware communications.

\section{Open Problems}\label{sec:openProblems}
There are still various open problems that remain to be addressed for CKM-enabled environment-aware communications, which are discussed in the following to inspire future work.

\subsection{How Much Data Is Sufficient?}
The effectiveness of the environment-aware communication based on CKM is dependent on the quality and accessibility of location-specific channel data. Therefore, it is imperative to establish theoretical frameworks to determine how much data is sufficient for constructing accurate CKMs. To this end, spatial statistics may be used to provide theoretical foundations for modelling and predicting the distribution of the interested channel knowledge across a given area, taking into account the spatial dependencies and correlations in the data. {For instance, the required sample density to achieve the desired channel prediction accuracy has been analyzed in \cite{Xu2023arXiv} for CGM construction, which is related to the spatial correlation model and the  sample location distributions.} Besides, the granularity of data samples  also needs to consider the impact of location uncertainty in practical systems. One possible approach is to model the location uncertainty as a probability distribution function (PDF), based on which its impact on channel
prediction can be analyzed.

\subsection{CKM in Highly Dynamic Environment}
In practice, the wireless environment could be highly dynamic, which results in the wireless channels experiencing rapid fluctuations. Therefore, it is crucial to develop CKM-based environment-aware communication techniques that can adapt to the changing environment. One possible approach is to leverage additional sensory information together with machine learning techniques to enhance the robustness of CKM to environment variations. For example,  the integration of sensing and vision technologies such as radar, lidar, and cameras can provide additional information on the dynamic environment, such as the location and movement of people and objects. By leveraging such sensory data together with CKM, effective {\it a priori} channel knowledge can be inferred even in the highly dynamic environment. {Another possible approach is to utilize the physical map, which can be used to aid CKM to distinguish between major and minor environment changes, so that the CKM can be updated accordingly.}

\subsection{CKM Based on Heterogeneous Data Sources}
In practice, the location-specific channel data for CKM construction are typically gathered from diverse data sources (such as smartphones, autonomous vehicles and UAVs), and consolidating them for effective CKM construction poses significant challenges due to the heterogeneity in data format, granularity, quality, and redundancy. Therefore, more studies are needed to develop CKM construction and utilization methods by considering such heterogeneous data sources. 

\subsection{CKM Update with Sequential/Continuous Data Arrival}
CKM needs to be  continuously updated  as new data arrives. The main challenge in online map reconstruction is to update the CKM efficiently  without having to recompute the entire map from scratch every time new data arrives. One potential solution is the incremental learning algorithms \cite{4012}, which enables the CKM to be updated efficiently by incorporating new data into the existing CKM without the need to retrain the entire model. One popular incremental learning algorithm is the Recursive Least Squares (RLS) algorithm \cite{4013}, which implements a recursive update formula to update the CKM with new data, making it a computationally efficient and scalable solution.

%Another approach to online map reconstruction is to use online convex optimization (OCO) algorithms \cite{4014}. OCO algorithms can be used to optimize the CKM using streaming data in real-time by minimizing a convex loss function that captures the prediction error between the CKM and the observed data. OCO algorithms are adaptable to the variations in the wireless environment by continuously updating the CKM as new data arrives.

\subsection{Privacy and Security Issues}
The utilization of user location towards environment-aware communication may bring the critical concern on the user privacy issues. If misused, such data can expose personal and sensitive information concerning the user's daily routine, whereabouts, and social interactions. Therefore, it is of paramount importance to develop privacy-preserving techniques before techniques like CKM can be practically used. Fortunately, this issue can be effectively resolved  by techniques like virtual location and crowdsourcing \cite{2067}.  Specifically, instead of using the devices' true locations, CKM can be constructed by using virtual locations that satisfy three properties: uniqueness, irreversibility, and adjacency invariance. Uniqueness guarantees that any true location can be represented by an
unique virtual location, while irreversibility ensures that it is impossible to infer the user's true location from the virtual location. The adjacency invariance reserves the spatial correlation of channel knowledge. One possible approach to generate such virtual locations is by using hashing algorithms, which are one-way functions that generate a fixed-length output from an input of arbitrary length \cite{2138}. Crowdsourcing is another potential solution to address privacy concerns in environment-aware communication systems \cite{2067}. Crowdsourcing involves aggregating data from multiple users to make it difficult to identify individual users' locations and activities, which can help to preserve the anonymity of individual users while still providing useful information for environment-aware communication.

\subsection{CKM with Device Orientation Information}
Device orientation  is an important parameter in environment-aware communication, especially when UE is equipped with antenna arrays. Therefore, by incorporating device orientation information, more accurate and richer CKM can be constructed.
One way to achieve this is by adding device orientation as a dimension to the CKM. For instance, the CKM can be represented as a tensor with the Cartesian coordinates  and orientation. The CKM can then be computed by measuring the channel response at different locations and orientations, which requires additional measurement efforts. Nevertheless, the resulting map can provide accurate channel inference even when device orientation varies.

\subsection{Prototyping and Experiment Validation}
Before CKM-enabled environment-aware communication can be realized in practical systems, proof-of-concept prototyping design and experimental verification are needed. {A preliminary work toward this end was reported in \cite{4033}, where a prototyping experiment for CKM-enabled training-free mmWave beam alignment was  developed. Specifically, by leveraging BIM, the transmit and receive beam pair of a mmWave massive MIMO system is directly inferred based on  UE's location without requiring channel estimation or beam sweeping.
For both quasi-static and dynamic scenarios considered in \cite{4033}, the experimental results demonstrate that CKM-based training-free mmWave beam alignment achieves comparable performance as exhaustive beam sweeping over 4096 beam pairs. Furthermore, compared with the location-based beam alignment that does not exploit the environment information, the developed CKM-based strategy achieves much higher received power in the considered scenarios. More prototyping and experiment results in the future will help identifying and developing promising use cases for CKM-enabled environment-aware communications.}
%%%%%%%%%%%%%%% Add by Jie Xu for revision 2023/11/6
{
\subsection{CKM for 6G Digital Twin}

CKM is also a promising technology to enable the realization of digital twin of 6G networks. Digital twin is the digital representation of physical entities (e.g., cellular BSs) and systems (e.g., 6G networks of our interest), which can be used to monitor, simulate, and facilitate the operation of physical systems \cite{DigitalTwin_Niyato}. In particular, CKM itself can be viewed as a digital twin of wireless environment, which provides channel information among different 6G network entities such as BSs, IRSs, and mobile devices. This can thus facilitate the operation of digital twin for 6G networks. Furthermore, to ensure accurate simulation and efficient operation of digital twin, we need to construct and update CKM accurately in a timely manner. 

\subsection{CKM with Semantic Communication}

The interplay between CKM and the emerging semantic communication technique is also another interesting topic. While conventional communication design focuses on the transmission of symbols in a technical level, semantic communication considers the exchange of semantics or even the effects of semantics exchange in semantic and effectiveness levels, by exploiting novel joint source and channel coding (JSCC) designs via, e.g., emerging deep learning techniques \cite{SemanticCom1,SemanticCom2}. On one hand, CKM is able to provide semantics of wireless environment for facilitating semantic communication, such that the transceivers can adapt their JSCC design by, e.g., adaptively changing the parameters of deep learning based on the channel semantics\cite{SemanticJie}. On the other hand, the idea of semantic communication can be exploited to facilitate the exchange of CKM among different nodes (e.g., during the distributed training of CKM), in which the essential semantic information of CKM can be extracted to enhance transmission efficiency.}

\section{Conclusion}\label{sec:conclusion}
Environment-aware communication is one of the most promising paradigm shifts towards 6G, which is expected to achieve significant performance gains over the conventional environment-unaware approach. This article provided a comprehensive overview of environment-aware wireless communications enabled by CKM. The basic concept of CKM-enabled environment-aware communications were elaborated, together with the main techniques for its construction and utilization. Several open problems that deserve further investigation are also discussed. It is hoped that such  tutorial discussions would inspire more follow-up works to unlock the full potential of CKM for realizing the paradigm shift from  the conventional environment-unaware networks to future environment-aware networks.

\bibliographystyle{IEEEtran}
\bibliography{IEEEabrv,RadioMap,RadioMap_supp,CKM_Utilization_Xu,IEEEfull}

\begin{thebibliography}{100}
\providecommand{\url}[1]{#1}
\csname url@samestyle\endcsname
\providecommand{\newblock}{\relax}
\providecommand{\bibinfo}[2]{#2}
\providecommand{\BIBentrySTDinterwordspacing}{\spaceskip=0pt\relax}
\providecommand{\BIBentryALTinterwordstretchfactor}{4}
\providecommand{\BIBentryALTinterwordspacing}{\spaceskip=\fontdimen2\font plus
\BIBentryALTinterwordstretchfactor\fontdimen3\font minus
  \fontdimen4\font\relax}
\providecommand{\BIBforeignlanguage}[2]{{%
\expandafter\ifx\csname l@#1\endcsname\relax
\typeout{** WARNING: IEEEtran.bst: No hyphenation pattern has been}%
\typeout{** loaded for the language `#1'. Using the pattern for}%
\typeout{** the default language instead.}%
\else
\language=\csname l@#1\endcsname
\fi
#2}}
\providecommand{\BIBdecl}{\relax}
\BIBdecl

\bibitem{1205}
{6G Research Visions 1, ``Key drivers and research challenges for 6G ubiquitous
  wireless intelligence,'' Matti Latvaaho, Kari Leppanen (eds.), 6G Flagship,
  University of Oulu, Finland,Sep. 2019}.

\bibitem{3021}
{X. H. You, C.-X. Wang, J. Huang, et al.}, ``Towards {6G} wireless
  communication networks: vision, enabling technologies, and new paradigm
  shifts,'' \emph{Sci China Inf Sci}, vol.~64, no.~1, pp. 1--74, Jan. 2021.

\bibitem{3022}
{W. Tong and P. Zhu}, \emph{{6G}: The Next Horizon: From Connected People and
  Things to Connected Intelligence}.\hskip 1em plus 0.5em minus 0.4em\relax New
  York, NY: Cambridge University Press, 2021.

\bibitem{4044}
\BIBentryALTinterwordspacing
``{The ITU-R Framework for IMT-2030}.'' [Online]. Available:
  \url{https://www.itu.int/en/ITU-R/study-groups/rsg5/rwp5d/imt-2030/Documents/IMT-2030%20Framework_WP%205D%20Management%20Team.pdf}
\BIBentrySTDinterwordspacing

\bibitem{3025}
{S. Z. Chen, Y. C. Liang, S. H. Sun, et al.}, ``Vision, requirements, and
  technology trend of {6G}: how to tackle the challenges of system coverage,
  capacity, user data-rate and movement speed,'' \emph{IEEE Wirel. Commun.},
  pp. 218--228, Apr. 2020.

\bibitem{3023}
{Z. Zhang, Y. Xiao, Z. Ma, et al.}, ``{6G} wireless networks: vision,
  requirements, architecture, and key technologies,'' \emph{IEEE Veh. Technol.
  Mag.}, pp. 28--41, Sep. 2019.

\bibitem{3027}
{H. Q. Ngo, A. Ashikhmin, H. Yang, E. G. Larsson, T. L. Marzetta}, ``Cell-free
  massive {MIMO} versus small cells,'' \emph{IEEE Trans. Wirel. Commun.},
  vol.~16, no.~3, pp. 1834--1850, Mar. 2017.

\bibitem{4020}
M.~Ke, Z.~Gao, Y.~Wu, X.~Gao, and K.-K. Wong, ``Massive access in cell-free
  massive {MIMO}-based {Internet of Things}: Cloud computing and edge computing
  paradigms,'' \emph{IEEE J. Sel. Areas Commun.}, vol.~39, no.~3, pp. 756--772,
  Mar. 2020.

\bibitem{4021}
X.~You, D.~Wang, and J.~Wang, \emph{Distributed {MIMO} and Cell-Free Mobile
  Communication}.\hskip 1em plus 0.5em minus 0.4em\relax Springer, 2021.

\bibitem{1095}
{Y. Zeng, Q. Wu, and R. Zhang}, ``Accessing from the sky: a tutorial on {UAV}
  communications for {5G} and beyond,'' \emph{Proc. of the IEEE}, vol. 107,
  no.~12, pp. 2327--2375, Dec. 2019.

\bibitem{569}
T.~S. Rappaport, R.~W. Heath~Jr, R.~C. Daniels, and J.~N. Murdock,
  \emph{{Millimeter Wave Wireless Communications}}.\hskip 1em plus 0.5em minus
  0.4em\relax Prentice Hall, 2014.

\bibitem{3028}
{T. S. Rappaport, Y. Xing, O. Kanhere, et al.}, ``Wireless communications and
  applications above 100 {GHz}: Opportunities and challenges for {6G} and
  beyond,'' \emph{IEEE Access}, vol.~7, pp. 78\,729--78\,757, Jun. 2019.

\bibitem{3029}
{E. Bjornson, L. Sanguinetti, H. Wymeersch, J. Hoydis, and T. L. Marzetta},
  ``Massive {MIMO} is a reality-what is next?: Five promising research
  directions for antenna arrays,'' \emph{Digit. Signal Process.}, vol.~94, pp.
  3--20, 2019.

\bibitem{3030}
{J. Zhang, E. Bjornson, M. Matthaiou, D. W. K. Ng, H. Yang, and D. J. Love},
  ``Prospective multiple antenna technologies for beyond {5G},'' \emph{IEEE J.
  Sel. Areas Commun.}, vol.~38, no.~8, pp. 1637--1660, 2020.

\bibitem{3032}
{H. Lu and Y. Zeng}, ``Communicating with extremely large-scale array/surface:
  Unified modeling and performance analysis,'' \emph{IEEE Trans. Wirel.
  Commun.}, vol.~21, no.~6, pp. 4039--4053, Jun. 2022.

\bibitem{4022}
Y.~Han, S.~Jin, M.~Matthaiou, T.~Q. Quek, and C.-K. Wen, ``Towards extra
  large-scale {MIMO}: New channel properties and low-cost designs,'' \emph{IEEE
  Internet of Things J.}, 2023.

\bibitem{4023}
J.~Wang, C.-X. Wang, J.~Huang, H.~Wang, and X.~Gao, ``{A general 3D
  space-time-frequency non-stationary THz channel model for 6G ultra-massive
  MIMO wireless communication systems},'' \emph{IEEE J. Sel. Areas Commun.},
  vol.~39, no.~6, pp. 1576--1589, Jun. 2021.

\bibitem{4024}
J.~Xu, L.~You, G.~C. Alexandropoulos, X.~Yi, W.~Wang, and X.~Gao, ``{Near-field
  wideband extremely large-scale MIMO transmission with holographic metasurface
  antennas},'' \emph{arXiv preprint arXiv:2205.02533}, 2022.

\bibitem{3033}
{S. Hu, F. Rusek, and O. Edfors}, ``Beyond massive {MIMO}: The potential of
  data transmission with large intelligent surfaces,'' \emph{IEEE Trans. Signal
  Process.}, vol.~66, no.~10, pp. 2746--2758, May 2018.

\bibitem{3034}
T.~L.~M. A.~Pizzo and L.~Sanguinetti, ``Spatially-stationary model for
  holographic {MIMO} small-scale fading,'' \emph{IEEE J. Sel. Areas Commun.},
  vol.~38, no.~9, pp. 1964--1979, Sep. 2020.

\bibitem{3037}
{M. Di Renzo, A. Zappone, M. Debbah, M.-S. Alouini, C. Yuen, J. de Rosny, and
  S. Tretyakov}, ``Smart radio environments empowered by reconfigurable
  intelligent surfaces: How it works, state of research, and the road ahead,''
  \emph{IEEE J. Sel. Areas Commun.}, vol.~38, no.~11, pp. 2450--2525, Nov.
  2020.

\bibitem{3035}
{Q. Wu, S. Zhang, B. Zheng, C. You, and R. Zhang}, ``Intelligent reflecting
  surface aided wireless communications: A tutorial,'' \emph{IEEE Trans.
  Commun.}, vol.~69, no.~5, pp. 3313--3351, May 2021.

\bibitem{3036}
{H. Lu, Y. Zeng, S. Jin, and R. Zhang}, ``Aerial intelligent reflecting
  surface: Joint placement and passive beamforming design with {3D} beam
  flattening,'' \emph{IEEE Trans. Wirel. Commun.}, vol.~20, no.~7, pp.
  4128--4143, Jul. 2021.

\bibitem{4019}
Z.~Wei, W.~Yuan, S.~Li, J.~Yuan, G.~Bharatula, R.~Hadani, and L.~Hanzo,
  ``Orthogonal time-frequency space modulation: A promising next-generation
  waveform,'' \emph{IEEE Wirel. Commun.}, vol.~28, no.~4, pp. 136--144, Aug.
  2021.

\bibitem{3039}
{H. Lu and Y. Zeng}, ``Delay alignment modulation: Enabling equalization-free
  single-carrier communication,'' \emph{IEEE Wirel. Commun. Lett.}, vol.~11,
  no.~9, pp. 1785--1789, Sep. 2022.

\bibitem{4016}
H.~Lu and Y.~Zeng, ``Delay-doppler alignment modulation for spatially sparse
  massive {MIMO} communication,'' \emph{IEEE Trans. Wireless Commun}, 2023.

\bibitem{3040}
{Y.-C. Liang, Q. Zhang, E. G. Larsson and G. Y. Li}, ``Symbiotic radio:
  Cognitive backscattering communications for future wireless networks,''
  \emph{IEEE Trans. Cognit. Commun. Netw.}, vol.~6, no.~4, pp. 1242--1255, Dec.
  2020.

\bibitem{3041}
{Z. Xiao and Y. Zeng}, ``An overview on integrated localization and
  communication towards {6G},'' \emph{Sci. China Inf. Sci.}, vol.~65, no.~3,
  pp. 1--46, Mar. 2022.

\bibitem{4017}
F.~Liu, Y.~Cui, C.~Masouros, J.~Xu, T.~X. Han, Y.~C. Eldar, and S.~Buzzi,
  ``Integrated sensing and communications: Towards dual-functional wireless
  networks for {6G} and beyond,'' \emph{IEEE J. Sel. Area Commun.}, vol.~40,
  no.~6, pp. 1728--1767, Jun. 2022.

\bibitem{4018}
A.~Liu, Z.~Huang, M.~Li, Y.~Wan, W.~Li, T.~X. Han, C.~Liu, R.~Du, D.~K.~P. Tan,
  J.~Lu, S.~Yuan, and C.~Fabiola, ``A survey on fundamental limits of
  integrated sensing and communication,'' \emph{IEEE Communications Surveys \&
  Tutorials}, vol.~24, no.~2, pp. 994--1034, 2022.

\bibitem{3042}
{R. Di Taranto, S. Muppirisetty, R. Raulefs, D. Slock, T. Svensson, and H.
  Wymeersch}, ``Location-aware communications for {5G} networks: How location
  information can improve scalability, latency, and robustness of {5G},''
  \emph{IEEE Signal Process. Mag.}, vol.~31, no.~6, pp. 102--112, Nov. 2014.

\bibitem{3043}
{F. Liu, W. Yuan, C. Masouros, and J. Yuan}, ``Radar-assisted predictive
  beamforming for vehicular links: Communication served by sensing,''
  \emph{IEEE Trans. Wirel. Commun.}, vol.~19, no.~11, pp. 7704--7719, Nov.
  2020.

\bibitem{3045}
{Y. Zhao, X. Xu, Y. Zeng, and F. Liu}, ``Sensing-assisted predictive
  beamforming with {NLoS} identification,'' in \emph{Proc. IEEE Int. Conf.
  Commun. (ICC)}, 2023.

\bibitem{4028}
Y.~Mao, C.~You, J.~Zhang, K.~Huang, and K.~B. Letaief, ``A survey on mobile
  edge computing: The communication perspective,'' \emph{IEEE Commun. Surveys
  Tuts.}, vol.~19, no.~4, pp. 2322--2358, 2017.

\bibitem{4029}
G.~Zhu, Z.~Lyu, X.~Jiao, P.~Liu, M.~Chen, J.~Xu, S.~Cui, and P.~Zhang,
  ``Pushing {AI} to wireless network edge: An overview on integrated sensing,
  communication, and computation towards {6G},'' \emph{Science China
  Information Sciences}, vol.~66, no.~3, p. 130301, Feb. 2023.

\bibitem{3044}
{Z. Yu, C. Feng, Y. Zeng, T. Li, and S. Jin}, ``Wireless communication using
  metal reflectors: Reflection modelling and experimental verification,'' in
  \emph{Proc. IEEE Int. Conf. Commun. (ICC)}, 2023.

\bibitem{3046}
{R. Li, Z. Xiao, and Y. Zeng}, ``Beamforming towards seamless sensing coverage
  for cellular integrated sensing and communication,'' in \emph{Proc. IEEE Int.
  Conf. Commun. (ICC) Workshop}, 2022.

\bibitem{851}
{Y. Zeng and R. Zhang}, ``Cost-effective millimeter wave communications with
  lens antenna array,'' \emph{IEEE Wireless Commun.}, vol.~24, no.~4, pp.
  81--87, Aug. 2017.

\bibitem{574}
{J. Wang, \emph{et al.}}, ``Beam codebook based beamforming protocol for
  multi-{Gbps} millimeter-wave {WPAN} systems,'' \emph{{IEEE} J. Sel. Areas
  Commun.}, vol.~27, no.~8, pp. 1390--1399, Oct. 2009.

\bibitem{575}
F.~Gholam, J.~Via, and I.~Santamaria, ``Beamforming design for simplified
  analog antenna combining architectures,'' \emph{{IEEE} Trans. Veh. Technol.},
  vol.~60, no.~5, pp. 2373--2378, Jun. 2011.

\bibitem{573}
{S. Hur, \emph{et al.}}, ``Millimeter wave beamforming for wireless backhaul
  and access in small cell networks,'' \emph{{IEEE} Trans. Commun.}, vol.~61,
  no.~10, pp. 4391--4403, Oct. 2013.

\bibitem{576}
X.~Zhang, A.~F. Molish, and S.~Y. Kung, ``Variable-phase-shift-based
  {RF}-baseband codesign for {MIMO} antenna selection,'' \emph{{IEEE} Trans.
  Signal Process.}, vol.~53, no.~11, pp. 4091--4103, Nov. 2005.

\bibitem{578}
O.~E. Ayach, S.~Rajagopal, S.~Abu-Surra, Z.~Pi, and R.~W. Heath~Jr, ``Spatially
  sparse precoding in millimeter wave {MIMO} systems,'' \emph{{IEEE} Trans.
  Wireless Commun.}, vol.~13, no.~3, pp. 1499--1513, Mar. 2013.

\bibitem{823}
Y.~Zeng and R.~Zhang, ``Millimeter wave {MIMO} with lens antenna array: a new
  path division multiplexing paradigm,'' \emph{{IEEE} Trans. Commun.}, vol.~64,
  no.~4, pp. 1557--1571, Apr. 2016.

\bibitem{1026}
{Y. Zeng, L. Yang, and R. Zhang}, ``Multi-user millimeter wave {MIMO} with
  full-dimensional lens antenna array,'' \emph{IEEE Trans. Wireless Commun.},
  vol.~17, no.~4, pp. 2800--2814, Apr. 2018.

\bibitem{828}
{J. Singh, O. Dabeer, and U. Madhow}, ``On the limits of communication with
  low-precision analog-to-digital conversion at the receiver,'' \emph{IEEE
  Trans. Commun.}, vol.~57, no.~12, pp. 3629--3639, Dec. 2009.

\bibitem{824}
{R. W. Heath Jr, N. G. Prelcic, S. Rangan, W. Roh, and A. M. Sayeed}, ``An
  overview of signal processing techniques for millimeter wave {MIMO}
  systems,'' \emph{IEEE J. Sel. Topics Signal Process}, vol.~10, no.~3, pp.
  436--453, Apr. 2016.

\bibitem{3047}
{J. Mo and R. W. Heath}, ``Capacity analysis of one-bit quantized {MIMO}
  systems with transmitter channel state information,'' \emph{IEEE Trans.
  Signal Process.}, vol.~63, no.~20, pp. 5498--5512, Oct. 2015.

\bibitem{4025}
C.~Zhang, Z.~Zhou, H.~Wang, and Y.~Zeng, ``Integrated super-resolution sensing
  and communication with {5G NR} waveform: Signal processing with uneven cps
  and experiments,'' \emph{arXiv preprint arXiv:2305.05142}, 2023.

\bibitem{3049}
{Y. Shen and M. Z. Win}, ``Fundamental limits of wideband
  localization��part {I}: A general framework,'' \emph{IEEE Trans. Inf.
  Theory}, vol.~56, no.~10, pp. 4956--4980, Oct. 2010.

\bibitem{3048}
{R. Mendrzik, H. Wymeersch, G. Bauch, and Z. A. Shaban}, ``Harnessing {NLOS}
  components for position and orientation estimation in {5G} millimeter wave
  {MIMO},'' \emph{IEEE Trans. Wirel. Commun.}, vol.~18, no.~1, pp. 93--107,
  Jan. 2019.

\bibitem{4003}
W.~Shi, J.~Cao, Q.~Zhang, Y.~Li, and L.~Xu, ``Edge computing: Vision and
  challenges,'' \emph{IEEE IoT J.}, vol.~3, no.~5, pp. 637--646, Oct. 2016.

\bibitem{4005}
T.~Hewa, G.~G{\"u}r, A.~Kalla, M.~Ylianttila, A.~Bracken, and M.~Liyanage,
  ``The role of blockchain in {6G}: Challenges, opportunities and research
  directions,'' \emph{2020 2nd 6G Wireless Summit (6G SUMMIT)}, pp. 1--5, 2020.

\bibitem{4032}
G.~Zhu, J.~Xu, K.~Huang, and S.~Cui, ``Over-the-air computing for wireless data
  aggregation in massive {IoT},'' \emph{IEEE Wireless Communications}, vol.~28,
  no.~4, pp. 57--65, 2021.

\bibitem{373}
T.~L. Marzetta, ``Noncooperative cellular wireless with unlimited numbers of
  base station antennas,'' \emph{{IEEE} Trans. Wireless Commun.}, vol.~9,
  no.~11, pp. 3590--3600, Nov. 2010.

\bibitem{3050}
{M. Giordani, M. Polese, A. Roy, D. Castor, M. Zorzi}, ``A tutorial on beam
  management for {3GPP NR at mmWave} frequencies,'' \emph{IEEE Commun. Surveys
  \& Tutorials}, vol.~21, no.~1, pp. 173--196, 1st Quarter 2019.

\bibitem{579}
A.~Alkhateeb, O.~E. Ayach, R.~Leus, and R.~W. Heath~Jr, ``Channel estimation
  and hybrid precoding for millimeter wave cellular systems,'' vol.~8, no.~5,
  pp. 831--846, Oct. 2014.

\bibitem{3051}
{Z. Xiao, T. He, P. Xia, and X.-G. Xia}, ``Hierarchical codebook design for
  beamforming training in millimeter-wave communication,'' \emph{IEEE Trans.
  Wirel. Commun.}, vol.~15, no.~5, pp. 3380--3392, May 2016.

\bibitem{470}
J.~Jose, A.~Ashikhmin, T.~L. Marzetta, and S.~Vishwanath, ``Pilot contamination
  and precoding in multi-cell {TDD} systems,'' \emph{{IEEE} Trans. Wireless
  Commun.}, vol.~10, no.~8, pp. 2640--2651, Aug. 2011.

\bibitem{3052}
{B. Zheng, C. You, W. Mei, and R. Zhang}, ``A survey on channel estimation and
  practical passive beamforming design for intelligent reflecting surface aided
  wireless communications,'' \emph{IEEE Commun. Surveys \& Tutorials}, vol.~24,
  no.~2, pp. 1035--1071, 2nd Quarter 2022.

\bibitem{ShadowingModel}
M.~Gudmundson, ``Correlation model for shadow fading in mobile radio systems,''
  \emph{Electron. Lett.}, vol.~27, no.~4, Nov. 1991.

\bibitem{4006}
M.-S. Chen, J.~Han, and P.~S. Yu, ``Data mining: an overview from a database
  perspective,'' \emph{IEEE Trans. on Knowledge and data Engineering}, vol.~8,
  no.~6, pp. 866--883, Dec. 1996.

\bibitem{2101}
Y.~Zeng and X.~Xu, ``Toward environment-aware {6G} communications via channel
  knowledge map,'' \emph{IEEE Wireless Commun.}, vol.~28, no.~3, pp. 84--91,
  2021.

\bibitem{1203}
ICT-317669 METIS Project deliverable D1.4 v 1.0, METIS Channel Models, Feb
  2015.

\bibitem{1063}
{O. Esrafilian, R. Gangula, and D. Gesbert}, ``Learning to communicate in
  {UAV}-aided wireless networks: map-based approaches,'' \emph{IEEE Internet of
  Things Journal}, vol.~6, no.~2, pp. 1791--1802, Apr. 2019.

\bibitem{2132}
M.~S. S. Y. K. C. B.~C. Y.~G.~Lim, Y. J.~Cho and R.~A. Valenzuela, ``Map-based
  millimeter-wave channel models: An overview, data for {B5G} evaluation and
  machine learning,'' \emph{IEEE Wireless Commun.}, vol.~27, no.~4, pp. 54--62,
  2020.

\bibitem{1204}
{S. Y. Seidel and T. S. Rappaport}, ``Site-specific propagation prediction for
  wireless in-building personal communication system design,'' \emph{IEEE
  Trans. Veh. Technol.}, vol.~43, no.~4, pp. 879--891, Nov. 1994.

\bibitem{2008}
G.~K. R.~Levie, C.~Yapar and G.~Caire, ``Radiounet: Fast radio map estimation
  with convolutional neural networks,'' \emph{IEEE Trans. Wireless Commun.},
  vol.~20, no.~6, pp. 4001--4015, 2021.

\bibitem{2100}
B.~L. Y.~Zhao and J.~F. Reed, ``Network support: The radio environment map,''
  \emph{Cognitive Radio Technol.}, pp. 337--363, 2006.

\bibitem{2096}
F.~A. H.~B.~Yilmaz, T.~Tugcu and S.~Bayhan, ``Radio environment map as enabler
  for practical cognitive radio networks,'' \emph{IEEE Commun. Mag.}, vol.~51,
  no.~12, pp. 162--169, Dec. 2013.

\bibitem{2097}
L.~B. A. K. K. K. N. D. R.~K. J.~Perez-Romero, A.~Zalonis, ``On the use of
  radio environment maps for interference management in heterogeneous
  networks,'' \emph{IEEE Commun. Mag.}, vol.~53, no.~8, pp. 184--191, Aug.
  2015.

\bibitem{1067}
{S. Bi, J. Lyu, Z. Ding, and R. Zhang}, ``Engineering radio map for wireless
  resource management,'' \emph{IEEE Wireless Commun.}, vol.~26, no.~2, pp.
  133--141, Apr. 2019.

\bibitem{1210}
{K. Sato and T. Fujii}, ``Kriging-based interference power constraint:
  Integrated design of the radio environment map and transmission power,''
  \emph{IEEE Trans. Cognitive Commun. and Netw.}, vol.~3, no.~1, pp. 13--25,
  Mar. 2017.

\bibitem{2067}
K.~L. Q.~Jiang, Y.~Ma and Z.~Dou, ``A probabilistic radio map construction
  scheme for crowdsourcing-based fingerprinting localization,'' \emph{IEEE
  Sens. J.}, vol.~16, no.~10, pp. 3764--3774, May 2016.

\bibitem{2069}
Y.~Tao and L.~Zhao, ``A novel system for {WiFi} radio map automatic adaptation
  and indoor positioning,'' \emph{IEEE Trans. Veh. Technol.}, vol.~67, no.~11,
  pp. 10\,683--10\,692, Nov. 2018.

\bibitem{2071}
S.~V. S.~Sorour, Y.~Lostanlen and K.~Majeed, ``Joint indoor localization and
  radio map construction with limited deployment load,'' \emph{IEEE Trans.
  Mobile Comput.}, vol.~14, no.~5, pp. 1031--1043, May 2014.

\bibitem{2072}
Z.~Y. C.~Wu and C.~Xiao, ``Automatic radio map adaptation for indoor
  localization using smartphones,'' \emph{IEEE Trans. Mobile Comput.}, vol.~17,
  no.~3, pp. 517--528, Mar. 2017.

\bibitem{1062}
{J. Chen, U. Yatnalli, and D. Gesbert}, ``Learning radio maps for {UAV}-aided
  wireless networks: a segmented regression approach,'' in \emph{Proc. IEEE
  International Conference on Communications (ICC)}, May 2017.

\bibitem{2059}
S.~Zhang and R.~Zhang, ``Radio map based path planning for cellular-connected
  {UAV},'' \emph{IEEE Trans. Wireless Commun.}, vol.~20, no.~3, pp. 1975--1989,
  2020.

\bibitem{2039}
R.~L. G.~C. D.~Schufele and S.~Stanczak, ``Tensor completion for radio map
  reconstruction using low rank and smoothness,'' \emph{IEEE Int. Workshop Sig.
  Proc. Advances Wireless Commun.}, pp. 1--5, 2019.

\bibitem{2005}
{Y. Zeng, X. Xu, S. Jin, and R. Zhang}, ``Simultaneous navigation and radio
  mapping for cellular-connected {UAV} with deep reinforcement learning,''
  \emph{IEEE Trans. Wireless Commun.}, vol.~20, no.~7, pp. 4205--4220, 2021.

\bibitem{2074}
Y.~H. X.~Mo and J.~Xu, ``Radio-map-based robust positioning optimization for
  {UAV}-enabled wireless power transfer,'' \emph{IEEE Wireless Commun. Lett.},
  vol.~9, no.~2, pp. 179--183, Feb. 2019.

\bibitem{2105}
W.~Liu and J.~Chen, ``Geography-aware radio map reconstruction for {UAV}-aided
  communications and localization,'' \emph{IEEE Int. Conf. on Commun.}, pp.
  1--6, May 2021.

\bibitem{3001}
I.~M. C. T. A. E. C.~R. C.~Parera, Q.~Liao and M.~Cesana, ``Transfer learning
  for tilt-dependent radio map prediction,'' \emph{IEEE Trans. Cognitive
  Commun. Networking}, vol.~6, no.~2, pp. 829--843, Jan. 2020.

\bibitem{2107}
Y.~Teganya and D.~Romero, ``Deep completion autoencoders for radio map
  estimation,'' \emph{IEEE Trans. Wireless Commun.}, vol.~21, no.~3, pp.
  1710--1724, 2021.

\bibitem{1206}
{E. D. Anese, S.-J. Kim, and G. B. Giannakis}, ``Channel gain map tracking via
  distributed {Kriging},'' \emph{IEEE Trans. Veh. Technol.}, vol.~60, no.~3,
  pp. 1205--1211, Mar. 2011.

\bibitem{2013}
E.~D. S.~J.~Kim and G.~B. Giannakis, ``Cooperative spectrum sensing for
  cognitive radios using kriged {Kalman} filtering,'' \emph{IEEE J. Sel. Topics
  in Signal Process}, vol.~5, no.~1, pp. 24--36, Feb. 2010.

\bibitem{2038}
D.~L. D.~Romero and G.~B. Giannakis, ``Blind channel gain cartography,''
  \emph{IEEE Globe Conf. Sig. Info. Proc.}, pp. 1110--1115, 2016.

\bibitem{2042}
S.~K. D.~Lee and G.~B. Giannakis, ``Channel gain cartography for cognitive
  radios leveraging low rank and sparsity,'' \emph{IEEE Trans. Wireless
  Commun.}, vol.~16, no.~9, pp. 5953--5966, 2017.

\bibitem{2043}
D.~Lee and G.~B. Giannakis, ``A variational bayes approach to adaptive
  channel-gain cartography,'' \emph{IEEE Int. Conf. Acoustics, Speech and Sig.
  Proc.}, pp. 8434--8438, 2019.

\bibitem{2052}
S.~Z. S.~C. R.~Deng, Z.~Jiang and Z.~Niu, ``A two-step learning and
  interpolation method for location-based channel database construction,''
  \emph{IEEE Globalcom}, pp. 1--6, 2018.

\bibitem{2009}
P.~N. G. R. W. H. A.~K. W.~Zheng, A.~Ali and E.~M. Pari, ``{5G V2X}
  communication at millimeter wave: Rate maps and use cases,'' \emph{IEEE Veh.
  Technol. Conf.}, pp. 1--5, 2020.

\bibitem{2026}
S.~V. S.~S. M.~Kasparick, R. L.~Cavalcante and M.~Yukawa, ``Kernel-based
  adaptive online reconstruction of coverage maps with side information,''
  \emph{IEEE Trans. Veh. Technol.}, vol.~65, no.~7, pp. 5461--5473, Jul. 2015.

\bibitem{2046}
D.~S. C.~Phillips and D.~Grunwald, ``A survey of wireless path loss prediction
  and coverage mapping methods,'' \emph{IEEE Commun. Surveys Tuts.}, vol.~15,
  no.~1, pp. 255--270, 2012.

\bibitem{2014}
B.~S. P.~H. A.~B. H. Alaya-Feki, S. B.~Jemaa and E.~Moulines, ``Informed
  spectrum usage in cognitive radio networks: Interference cartography,''
  \emph{Proc. IEEE Int. Symp. Personal, Indoor Mobile Radio Commun.}, pp. 1--5,
  2008.

\bibitem{2015}
I.~O. G.~F. B.~A.~Jayawickrama, E.~Dutkiewicz and J.~Ding, ``Improved
  performance of spectrum cartography based on compressive sensing in cognitive
  radio networks,'' \emph{IEEE Int. Conf. Commun.}, pp. 5657--5661, Jun. 2013.

\bibitem{2019}
J.~A. Bazerque and G.~B. Giannakis, ``Learning power spectrum maps from
  quantized power measurements,'' \emph{IEEE Trans. Sig. Proc.}, vol.~65,
  no.~10, pp. 2547--2560, May 2017.

\bibitem{2135}
S.~J. D.~Wu, Y.~Zeng and R.~Zhang, ``Environment-aware and training-free beam
  alignment for mmwave massive mimo via channel knowledge map,'' \emph{IEEE
  Int. Conf. Commun. Workshops}, pp. 1--7, 2021.

\bibitem{CKM_Utilization_Yong_Hybrid_BF}
D.~Wu, Y.~Zeng, S.~Jin, and R.~Zhang, ``Environment-aware hybrid beamforming by
  leveraging channel knowledge map,'' 2022, {A}rXiv pre-print cs.IT/2206.08707.
  https://arxiv.org/abs/2206.08707.

\bibitem{CKM_Utilization_Yong_IRS_BF}
D.~Ding, D.~Wu, Y.~Zeng, S.~Jin, and R.~Zhang, ``Environment-aware beam
  selection for {IRS}-aided communication with channel knowledge map,''
  \emph{IEEE Globecom Workshops}, pp. 1--6, 2021.

\bibitem{2078}
T.~S. G.~B. V.~Va, J.~Choi and R.~W. Heath, ``Inverse multipath fingerprinting
  for millimeter wave {V2I} beam alignment,'' \emph{IEEE Trans. Veh. Technol.},
  vol.~67, no.~5, pp. 4042--4058, May 2017.

\bibitem{2076}
A.~M. L.~H. K.~Satyanarayana, M. El-Hajjar, ``Deep learning aided
  fingerprint-based beam alignment for mmwave vehicular communication,''
  \emph{IEEE Trans. Veh. Technol.}, vol.~68, no.~11, pp. 10\,858--10\,871,
  2019.

\bibitem{3053}
{A. Karttunen, A. F. Molisch, S. Hur, J. Park, and C. J. Zhang}, ``Spatially
  consistent street-by-street path loss model for 28-{GHz} channels in micro
  cell urban environments,'' \emph{IEEE Trans. Wirel. Commun.}, vol.~16,
  no.~11, pp. 7538--7550, Nov. 2017.

\bibitem{3054}
{S. Jaeckel, L. Raschkowski, K. Borner, L. Thiele, F. Burkhardt, and E.
  Eberlein}, ``{QuaDRiGa-quasi deterministic radio channel generator, user
  manual and documentation},'' \emph{Fraunhofer Heinrich Hertz Institute,
  Techn. Rep. v2.6.1, 2021}.

\bibitem{2093}
Y.~Z. K.~Li, P.~Li and J.~Xu, ``Channel knowledge map for environment-aware
  communications: {EM} algorithm for map construction,'' \emph{IEEE Wireless
  Commun. Netw. Conf.}, pp. 1659--1664, 2022.

\bibitem{3015}
{A. Duel-Hallen}, ``{Fading channel prediction for mobile radio adaptive
  transmission systems},'' \emph{Proc. IEEE}, vol.~95, no.~12, pp.
  2299�C--2313, Dec. 2007.

\bibitem{3016}
{W. Jiang and H. D. Schotten}, ``{Neural network-based fading channel
  prediction: A comprehensive overview},'' \emph{IEEE Access}, vol.~7, pp.
  118\,112�C--118\,124, Sep. 2019.

\bibitem{3005}
{T. Nitsche, A. B. Flores, E. W. Knightly, and J. Widmer}, ``{Steering with
  eyes closed: Mm-wave beam steering without in-band measurement},''
  \emph{Proc. IEEE Conf. Comput. Commun. (INFOCOM)}, pp. 2416--2424, Apr. 2015.

\bibitem{3006}
{M. Hashemi, C. E. Koksal, and N. B. Shroff}, ``{Out-of-band millimeter wave
  beamforming and communications to achieve low latency and high energy
  efficiency in 5G systems},'' \emph{IEEE Trans. Commun.}, vol.~66, no.~2, pp.
  875--888, Feb. 2018.

\bibitem{3008}
{N. Jalden, H. Asplund, and J. Medbo}, ``{Channel extrapolation based on
  wideband MIMO measurements},'' \emph{in Proc. 6th Eur. Conf. Antennas Propag.
  (EUCAP), Prague, Czech Republic}, pp. 442--446, Mar. 2012.

\bibitem{3007}
{D. Vasisht, S. Kumar, H. Rahul, and D. Katabi}, ``{Eliminating channel
  feedback in next-generation cellular networks},'' \emph{in Proc. ACM
  SIGCOMM}, pp. 398--411, Aug. 2016.

\bibitem{3009}
{F. Rottenberg, T. Choi, P. Luo, C. J. Zhang, and A. F. Molisch}, ``Performance
  analysis of channel extrapolation in {FDD} massive {MIMO} systems,''
  \emph{IEEE Trans. Wireless Commun.}, vol.~19, no.~4, pp. 2728--2741, Apr.
  2020.

\bibitem{3010}
{H. Choi and J. Choi}, ``{Downlink Extrapolation for FDD Multiple Antenna
  Systems Through Neural Network Using Extracted Uplink Path Gains},''
  \emph{IEEE Access}, vol.~8, pp. 67\,100--67\,111, 2020.

\bibitem{3013}
{A. Ali, N. Gonzalez-Prelcic, and R. W. Heath, Jr.}, ``{Millimeter wave
  beam-selection using out-of-band spatial information},'' \emph{IEEE Trans.
  Wireless Commun.}, vol.~17, no.~2, pp. 1038�C--1052, Feb. 2018.

\bibitem{3014}
------, ``{Spatial covariance estimation for millimeter wave hybrid systems
  using out-of-band information},'' \emph{IEEE Trans. Wireless Commun.},
  vol.~18, no.~12, pp. 5471�C--5485, Dec. 2019.

\bibitem{2124}
M.~Alrabeia and A.~Alkhateeb, ``Deep learning for {mmWave} beam and blockage
  prediction using sub-6 {GHz} channels,'' \emph{IEEE Trans. Commun.}, vol.~68,
  no.~9, pp. 5504--5518, 2020.

\bibitem{3017}
{Y. Han, Q. Liu, C.-K. Wen, S. Jin, and K.-K. Wong}, ``{FDD} massive {MIMO}
  based on efficient downlink channel reconstruction,'' \emph{IEEE Trans.
  Commun.}, vol.~67, no.~6, pp. 4020�C--4034, Jun. 2019.

\bibitem{3018}
{Y. Yang, F. Gao, G. Y. Li, and M. Jian}, ``Deep learning-based downlink
  channel prediction for {FDD} massive {MIMO} system,'' \emph{IEEE Commun.
  Letters}, vol.~23, no.~11, pp. 1994�C--1998, Nov. 2019.

\bibitem{2127}
M.~Alrabeia and A.~Alkhateeb, ``Deep learning for {TDD and FDD massive MIMO}:
  Mapping channels in space and frequency,'' \emph{53th Asilomar Conf. Sig.
  Sys. Comput.}, pp. 1465--1470, 2019.

\bibitem{3003}
{Z. Chen et al.}, ``{DoA and DoD estimation and hybrid beamforming for
  radar-aided mmWave MIMO vehicular communication systems},'' \emph{Electron.},
  vol.~7, no.~3, Mar. 2018.

\bibitem{3002}
{A. Ali, N. Gonz��lez-Prelcic, and A. Ghosh}, ``Millimeter wave {V2I}
  beamtraining using base-station mounted radar,'' \emph{Proc. IEEE Radar
  Conf}, pp. 1--5, Apr. 2019.

\bibitem{2112}
U.~Demirhan and A.~Alkhateeb, ``Radar aided 6g beam prediction: Deep learning
  algorithms and real-world demonstration,'' \emph{IEEE Wireless Commun. Netw.
  Conf.}, pp. 2655--2660, 2022.

\bibitem{2119}
------, ``Radar aided proactive blockage prediction in real-world millimeter
  wave systems.'' \emph{available online at https://arxiv.org/abs/2111.14805},
  2021.

\bibitem{2116}
G.~C. S.~Jiang and A.~Alkhateeb, ``{LiDAR} aided future beam prediction in
  real-world millimeter wave {V2I} communications,'' \emph{available online at
  https://arxiv.org/abs/2203.05548}, 2022.

\bibitem{2118}
C.~C. S.~Wu and A.~Alkhateeb, ``{LiDAR}-aided mobile blockage prediction in
  real-world millimeter wave systems,'' \emph{WCNC}, pp. 2631--2636, 2022.

\bibitem{2129}
N.~G.-P. M.~Dias, A.~Klautau and R.~W. Heath, ``Position and {LIDAR}-aided
  mmwave beam selection using deep learning,'' \emph{IEEE Int. Workshop Sig.
  Proc. Advances Wireless Commun.}, pp. 1--5, 2019.

\bibitem{2131}
G.~B. A.~Klautau, N. Gonzalez-Prelcic and R.~W. Heath, ``{LIDAR} data for deep
  learning-based {mmWave} beam-selection,'' \emph{IEEE Wireless Commun. Lett.},
  vol.~8, no.~3, pp. 909--912, 2019.

\bibitem{2114}
{G. Charan, A. Hredzak, C. Stoddard, B. Berrey, M. Seth, H. Nunez, and A.
  Alkhateeb}, ``Towards real-world {6G} drone communication: Position and
  camera aided beam prediction,'' \emph{available online at
  https://arxiv.org/abs/2205.12187}, 2022.

\bibitem{2117}
G.~Charan and A.~Alkhateeb, ``Computer vision aided blockage prediction in
  real-world millimeter wave deployments,'' \emph{available online at
  https://arxiv.org/abs/2203.01907}, 2022.

\bibitem{2120}
M.~A. G.~Charan and A.~Alkhateeb, ``Vision-aided {6G} wireless communications:
  Blockage prediction and proactive handoff,'' \emph{IEEE Trans. Veh. Tech.},
  vol.~70, no.~10, pp. 10\,193--10\,208, 2021.

\bibitem{2121}
------, ``Vision-aided dynamic blockage prediction for 6g wireless
  communication networks,'' \emph{IEEE Int. Conf.e Commun. Workshops}, pp.
  1--6, 2021.

\bibitem{126}
O.~O. Koyluoglu, H.~El~Gamal, L.~Lai, and V.~H. Poor, ``Interference alignment
  for secrecy,'' {A}rXiv pre-print cs.IT/0810.1187v1.
  http://arxiv.org/abs/0810.1187v1.

\bibitem{2073}
Q.~D. Vo and P.~De, ``A survey of fingerprint-based outdoor localization,''
  \emph{IEEE Commun. Surv. Tuts.}, vol.~18, no.~1, pp. 491--506, 2015.

\bibitem{3019}
{S. He and S. H. G. Chan}, ``{Wi-Fi} fingerprint-based indoor positioning:
  recent advances and comparisons,'' \emph{IEEE Commun. Surveys \& Tutorials},
  pp. 466�C--490, 1st Quarter 2016.

\bibitem{3020}
{Z. Yang, Z. Zhou, and Y. Liu}, ``From {RSSI} to {CSI}: Indoor localization via
  channel response,'' \emph{ACM Comput. Surveys}, vol.~46, no.~2, Nov. 2013.

\bibitem{2007}
{C. S, S. Medjkouh, E. Gonulta, T. Goldstein, and O. Tirkkonen}, ``Channel
  charting: Locating users within the radio environment using channel state
  information,'' \emph{IEEE Access}, vol.~6, pp. 47\,682--47\,698, 2018.

\bibitem{110}
Y.-G. Lim, Y.~J. Cho, M.~S. Sim, Y.~Kim, C.-B. Chae, and R.~A. Valenzuela,
  ``Map-based millimeter-wave channel models: An overview, data for {B5G}
  evaluation and machine learning,'' \emph{{IEEE} Wireless Commun.}, vol.~27,
  no.~4, pp. 54--62, Aug. 2020.

\bibitem{111}
Q.~Zhu, K.~Mao, M.~Song, X.~Chen, B.~Hua, W.~Zhong, and X.~Ye, ``Map-based
  channel modeling and generation for {U2V} mmwave communication,''
  \emph{{IEEE} Trans. Veh. Technol.}, vol.~71, no.~8, pp. 8004--8015, Aug.
  2022.

\bibitem{85}
J.~Johansson, W.~A. Hapsari, S.~Kelley, and G.~Bodog, ``Minimization of drive
  tests in {3GPP} release 11,'' \emph{{IEEE} Commun. Mag.}, vol.~50, no.~11,
  pp. 36--43, Nov. 2012.

\bibitem{87}
A.~Galindo-Serrano, B.~Sayrac, S.~Ben~Jemaa, J.~Riihij\"arvi, and
  P.~M\"ah\"onen, ``Harvesting {MDT} data: Radio environment maps for coverage
  analysis in cellular networks,'' in \emph{Proc. Int. Conf. Cogn. Radio
  Oriented Wireless Netw.}, 2013, pp. 37--42.

\bibitem{14}
S.-J. Kim, E.~Dall'Anese, and G.~B. Giannakis, ``Cooperative spectrum sensing
  for cognitive radios using kriged kalman filtering,'' \emph{{IEEE} J. Sel.
  Topics Signal Process.}, vol.~5, no.~1, pp. 24--36, Feb. 2011.

\bibitem{36}
E.~Dall'Anese, S.-J. Kim, and G.~B. Giannakis, ``Channel gain map tracking via
  distributed {Kriging},'' \emph{{IEEE} Trans. Veh. Technol.}, vol.~60, no.~3,
  pp. 1205--1211, Mar. 2011.

\bibitem{42}
S.~Chouvardas, S.~Valentin, M.~Draief, and M.~Leconte, ``A method to
  reconstruct coverage loss maps based on matrix completion and adaptive
  sampling,'' in \emph{Proc. {IEEE} Int. Conf. Acoust., Speech Signal Process.
  (ICASSP)}, 2016, pp. 6390--6394.

\bibitem{58}
T.~Cover and P.~Hart, ``Nearest neighbor pattern classification,'' \emph{{IEEE}
  Trans. Inf. Theory}, vol.~13, no.~1, pp. 21--27, Jan. 1967.

\bibitem{ChiDel:B09}
J.-P. Chiles and P.~Delfiner, \emph{Geostatistics: modeling spatial
  uncertainty}.\hskip 1em plus 0.5em minus 0.4em\relax John Wiley \& Sons,
  2009, vol. 497.

\bibitem{FanGij:B18}
J.~Fan and I.~Gijbels, \emph{Local polynomial modelling and its
  applications}.\hskip 1em plus 0.5em minus 0.4em\relax Routledge, 2018.

\bibitem{SunChe:C22}
H.~Sun and J.~Chen, ``Regression assisted matrix completion for reconstructing
  a propagation field with application to source localization,'' in
  \emph{{{Int. Conf. on Acoustics, Speech, and Signal Process. (ICASSP)}}.},
  Jun. 2022, pp. 1--6.

\bibitem{15}
A.~B.~H. Alaya-Feki, S.~B. Jemaa, B.~Sayrac, P.~Houze, and E.~Moulines,
  ``Informed spectrum usage in cognitive radio networks: Interference
  cartography,'' in \emph{Proc. {IEEE} Int. Symp. Pers., Indoor Mobile Radio
  Commun. (PIMRC)}, 2008, pp. 1--5.

\bibitem{96}
K.~Sato and T.~Fujii, ``Kriging-based interference power constraint: Integrated
  design of the radio environment map and transmission power,'' \emph{{IEEE}
  Trans. Cogn. Commun. Netw.}, vol.~3, no.~1, pp. 13--25, Mar. 2017.

\bibitem{31}
D.~M. Gutierrez-Estevez, I.~F. Akyildiz, and E.~A. Fadel, ``Spatial coverage
  cross-tier correlation analysis for heterogeneous cellular networks,''
  \emph{{IEEE} Trans. Veh. Technol.}, vol.~63, no.~8, pp. 3917--3926, Oct.
  2014.

\bibitem{100}
B.~A. Jayawickrama, E.~Dutkiewicz, G.~Fang, I.~Oppermann, and M.~Mueck,
  ``Downlink power allocation algorithm for licence-exempt {LTE} systems using
  {Kriging} and compressive sensing based spectrum cartography,'' in
  \emph{Proc. {IEEE} GLOBECOM}, 2013, pp. 3766--3771.

\bibitem{19}
J.~A. Bazerque and G.~B. Giannakis, ``Nonparametric basis pursuit via sparse
  kernel-based learning: A unifying view with advances in blind methods,''
  \emph{{IEEE} Signal Process. Mag.}, vol.~30, no.~4, pp. 112--125, Jul. 2013.

\bibitem{21}
C.~Carmeli, E.~De~Vito, A.~Toigo, and V.~Umanità, ``Vector valued reproducing
  kernel hilbert spaces and universality,'' \emph{Analysis and Applications},
  vol.~1, pp. 19--61, 2010.

\bibitem{23}
B.~Sch\"{o}lkopf, \emph{\BIBforeignlanguage{eng}{Learning with kernels: support
  vector machines, regularization, optimization, and beyond / Bernhard
  Scholkopf, Alexander J. Smola.}}, ser. Adaptive computation and machine
  learning.\hskip 1em plus 0.5em minus 0.4em\relax Cambridge: MIT Press, 2002.

\bibitem{Cressie:B15}
N.~Cressie, \emph{Statistics for spatial data}.\hskip 1em plus 0.5em minus
  0.4em\relax John Wiley \& Sons, 2015.

\bibitem{20}
D.~Romero, S.-J. Kim, G.~B. Giannakis, and R.~L\'{o}pez-Valcarce, ``Learning
  power spectrum maps from quantized power measurements,'' \emph{{IEEE} Trans.
  Signal Process.}, vol.~65, no.~10, pp. 2547--2560, May 2017.

\bibitem{95}
R.~Nikbakht, A.~Jonsson, and A.~Lozano, ``Dual-kernel online reconstruction of
  power maps,'' in \emph{Proc. {IEEE} GLOBECOM}, 2018, pp. 1--5.

\bibitem{107}
J.~Chen and U.~Mitra, ``Unimodality-constrained matrix factorization for
  non-parametric source localization,'' \emph{{IEEE} Trans. Signal Process.},
  vol.~67, no.~9, pp. 2371--2386, Sep. 2019.

\bibitem{SunChe:J22}
H.~Sun and J.~Chen, ``Propagation map reconstruction via interpolation assisted
  matrix completion,'' \emph{{IEEE} Trans. Signal Process.}, vol.~70, 2022,
  early access.

\bibitem{ZhaFuWanZha:J20}
G.~Zhang, X.~Fu, J.~Wang, X.-L. Zhao, and M.~Hong, ``Spectrum cartography via
  coupled block-term tensor decomposition,'' \emph{{IEEE} Trans. Signal
  Process.}, vol.~68, pp. 3660--3675, 2020.

\bibitem{SunChe:C21}
H.~Sun and J.~Chen, ``Grid optimization for matrix-based source localization
  under inhomogeneous sensor topology,'' in \emph{{{Int. Conf. on Acoustics,
  Speech, and Signal Process. (ICASSP)}}.}, Toronto, Canada, Jun. 2021, pp.
  5110--5114.

\bibitem{CanRec:J12}
E.~Candes and B.~Recht, ``Exact matrix completion via convex optimization,''
  \emph{Communications of the ACM}, vol.~55, no.~6, pp. 111--119, 2012.

\bibitem{PimBosNow:J16}
D.~L. Pimentel-Alarc{\'o}n, N.~Boston, and R.~D. Nowak, ``A characterization of
  deterministic sampling patterns for low-rank matrix completion,'' \emph{IEEE
  Journal of Selected Topics in Signal Processing}, vol.~10, no.~4, pp.
  623--636, 2016.

\bibitem{30}
I.~Popescu, I.~Nafomita, P.~Constantinou, A.~Kanatas, and N.~Moraitis, ``Neural
  networks applications for the prediction of propagation path loss in urban
  environments,'' in \emph{Proc. {IEEE} VTS Veh. Technol. Conf., Spring},
  vol.~1, 2001, pp. 387--391 vol.1.

\bibitem{li2023automatic}
Q.~Li, X.~Liao, A.~Li, and S.~Valaee, ``Automatic indoor radio map construction
  and localization via multipath fingerprint extrapolation,'' \emph{IEEE Trans.
  on Wireless Comm.}, 2023.

\bibitem{108}
Y.~Teganya and D.~Romero, ``Deep completion autoencoders for radio map
  estimation,'' \emph{{IEEE} Trans. Wireless Commun.}, vol.~21, no.~3, pp.
  1710--1724, Mar. 2022.

\bibitem{ShrFuHon:J22}
S.~Shrestha, X.~Fu, and M.~Hong, ``Deep spectrum cartography: Completing radio
  map tensors using learned neural models,'' \emph{{IEEE} Trans. Signal
  Process.}, vol.~70, pp. 1170--1184, 2022.

\bibitem{ThrZibChr:J20}
J.~Thrane, D.~Zibar, and H.~L. Christiansen, ``Model-aided deep learning method
  for path loss prediction in mobile communication systems at 2.6 ghz,''
  \emph{IEEE Access}, vol.~8, pp. 7925--7936, 2020.

\bibitem{HamMaBaxMat:J13}
B.~R. Hamilton, X.~Ma, R.~J. Baxley, and S.~M. Matechik, ``Propagation modeling
  for radio frequency tomography in wireless networks,'' \emph{{IEEE} J. Sel.
  Areas Commun.}, vol.~8, no.~1, pp. 55--65, 2013.

\bibitem{43}
D.~Lee, S.-J. Kim, and G.~B. Giannakis, ``Channel gain cartography for
  cognitive radios leveraging low rank and sparsity,'' \emph{{IEEE} Trans.
  Wireless Commun.}, vol.~16, no.~9, pp. 5953--5966, Sep. 2017.

\bibitem{39}
D.~Romero, D.~Lee, and G.~B. Giannakis, ``Blind radio tomography,''
  \emph{{IEEE} Trans. Signal Process.}, vol.~66, no.~8, pp. 2055--2069, Apr.
  2018.

\bibitem{CheEsrGesMit:C17}
J.~Chen, O.~Esrafilian, D.~Gesbert, and U.~Mitra, ``Efficient algorithms for
  air-to-ground channel reconstruction in {UAV}-aided communications,'' in
  \emph{IEEE Globecom Workshops}, 2017, pp. 1--6.

\bibitem{ZhaChe:C20}
B.~Zhang and J.~Chen, ``Constructing radio maps for {UAV} communications via
  dynamic resolution virtual obstacle maps,'' in \emph{{IEEE SPAWC}.}, 2020,
  pp. 1--5.

\bibitem{106}
W.~Liu and J.~Chen, ``Geography-aware radio map reconstruction for {UAV}-aided
  communications and localization,'' in \emph{Proc. {IEEE} Int. Conf. Commun.
  (ICC)}, 2021, pp. 1--6.

\bibitem{2063}
P.~Agrawal and N.~Patwari, ``Correlated link shadow fading in multihop wireless
  networks,'' \emph{IEEE Trans. Wireless Commun.}, vol.~8, no.~8, pp.
  4024--4036, Aug. 2009.

\bibitem{LiuChe:J23}
W.~Liu and J.~Chen, ``{UAV}-aided radio map construction exploiting environment
  semantics,'' \emph{{IEEE} Trans. Wireless Commun.}, 2023, early access.

\bibitem{ZenChe:C22}
P.~Zeng and J.~Chen, ``{UAV}-aided joint radio map and {3D} environment
  reconstruction using deep learning approaches,'' in \emph{{IEEE ICC}.}, Jun.
  2022, pp. 5341--5346.

\bibitem{BurLiBiz:D22}
\BIBentryALTinterwordspacing
F.~Burmeister, Z.~Li, and I.~Bizon, ``High-resolution radio environment map
  data set for indoor office environment,'' 2022. [Online]. Available:
  \url{https://dx.doi.org/10.21227/waxd-9525}
\BIBentrySTDinterwordspacing

\bibitem{YuMaoZhoSun:D23}
\BIBentryALTinterwordspacing
N.~Yu, S.~Mao, C.~Zhou, G.~Sun, Z.~Shi, and J.~Chen, ``{DroneRFa}: A
  large-scale dataset of drone radio frequency signals for detecting
  low-altitude drones,'' 2023. [Online]. Available:
  \url{https://jeit.ac.cn/web/data/getData?dataType=Dataset3}
\BIBentrySTDinterwordspacing

\bibitem{WanWen:D23}
\BIBentryALTinterwordspacing
W.~Chen, W.~Liu, and J.~Chen, ``A dataset for {3D} air-to-ground radio map
  construction,'' 2023. [Online]. Available:
  \url{https://github.com/chenwangqian-dr/3DRadioMap_Dataset.git}
\BIBentrySTDinterwordspacing

\bibitem{YapLevKut:D22}
\BIBentryALTinterwordspacing
C.~Yapar, R.~Levie, G.~Kutyniok, and G.~Caire, ``Dataset of pathloss and {ToA}
  radio maps with localization application,'' 2022. [Online]. Available:
  \url{https://dx.doi.org/10.21227/0gtx-6v30}
\BIBentrySTDinterwordspacing

\bibitem{LuoZhouYan:D23}
\BIBentryALTinterwordspacing
J.~Luo, B.~Zhou, Y.~Yu, P.~Zhang, X.~Peng, J.~Ma, P.~Zhu, J.~Lu, and W.~Tong,
  ``Sensiverse: A dataset for {ISAC} study,'' 2023. [Online]. Available:
  \url{https://sensiverse.github.io}
\BIBentrySTDinterwordspacing

\bibitem{ChaAndVit:D22}
\BIBentryALTinterwordspacing
A.~Chaves-Villota and C.~A. Viteri-Mera, ``{DeepREM}: Deep-learning-based radio
  environment map estimation from sparse measurements,'' 2022. [Online].
  Available: \url{https://doi.org/10.5281/zenodo.7839447}
\BIBentrySTDinterwordspacing

\bibitem{KinKopHae:Data08}
T.~King, S.~Kopf, T.~Haenselmann, C.~Lubberger, and W.~Effelsberg, ``{CRAWDAD}
  dataset mannheim/compass (v. 2008-04-11),'' Downloaded from
  \url{https://crawdad.org/mannheim/compass/20080411/signalstrength}, Apr.
  2008, traceset: signalstrength.

\bibitem{VerFunRaj:J16}
A.~Verdin, C.~Funk, B.~Rajagopalan, and W.~Kleiber, ``Kriging and local
  polynomial methods for blending satellite-derived and gauge precipitation
  estimates to support hydrologic early warning systems,'' \emph{IEEE Trans.
  Geosci. Remote Sens.}, vol.~54, no.~5, pp. 2552--2562, 2016.

\bibitem{JuaGonGeo:J11}
J.~A. Bazerque, G.~Mateos, and G.~B. Giannakis, ``Group-lasso on splines for
  spectrum cartography,'' \emph{{IEEE} Trans. Signal Process.}, vol.~59,
  no.~10, pp. 4648--4663, 2011.

\bibitem{SunXueYu:J18}
W.~Sun, M.~Xue, H.~Yu, H.~Tang, and A.~Lin, ``Augmentation of fingerprints for
  indoor wifi localization based on gaussian process regression,'' \emph{IEEE
  Transactions on Vehicular Technology}, vol.~67, no.~11, pp. 10\,896--10\,905,
  2018.

\bibitem{TanNeh:J11}
G.~Tang and A.~Nehorai, ``Lower bounds on the mean-squared error of low-rank
  matrix reconstruction,'' \emph{{IEEE} Trans. Signal Process.}, vol.~59,
  no.~10, pp. 4559--4571, 2011.

\bibitem{RonFisBro:J15}
O.~Ronneberger, P.~Fischer, and T.~Brox, ``U-net: Convolutional networks for
  biomedical image segmentation,'' in \emph{Proc. of International Conference
  on Medical Image Computing and Computer-Assisted Intervention--MICCAI},
  Munich, Germany, Oct. 2015, pp. 234--241.

\bibitem{61}
J.~Chen, U.~Yatnalli, and D.~Gesbert, ``Learning radio maps for uav-aided
  wireless networks: A segmented regression approach,'' in \emph{Proc. {IEEE}
  Int. Conf. Commun. (ICC)}, 2017, pp. 1--6.

\bibitem{6884811}
J.~Xu and R.~Zhang, ``Energy beamforming with one-bit feedback,'' \emph{IEEE
  Trans. on Signal Processing}, vol.~62, no.~20, pp. 5370--5381, Oct. 2014.

\bibitem{6578560}
Y.~Noam and A.~J. Goldsmith, ``The one-bit null space learning algorithm and
  its convergence,'' \emph{IEEE Trans. on Signal Processing}, vol.~61, no.~24,
  pp. 6135--6149, Dec. 2013.

\bibitem{7961152}
K.~Venugopal, A.~Alkhateeb, N.~Gonz¨¢lez~Prelcic, and R.~W. Heath, ``Channel
  estimation for hybrid architecture-based wideband millimeter wave systems,''
  \emph{IEEE J. Sel. Areas Commun.}, vol.~35, no.~9, pp. 1996--2009, Sep. 2017.

\bibitem{9499127}
Y.~Zhou, G.~Liu, J.~Li, Y.~Li, S.~Ye, and L.~Li, ``A high-efficiency beam
  sweeping algorithm for doa estimation in the hybrid analog-digital
  structure,'' \emph{IEEE Wireless Commun. Lett.}, vol.~10, no.~10, pp.
  2323--2327, Oct. 2021.

\bibitem{4026}
J.~Lee, G.-T. Gil, and Y.~H. Lee, ``Channel estimation via orthogonal matching
  pursuit for hybrid mimo systems in millimeter wave communications,''
  \emph{IEEE Trans. Commun.}, vol.~64, no.~6, pp. 2370--2386, Jun. 2016.

\bibitem{4027}
N.~Garcia, H.~Wymeersch, E.~G. Str{\"o}m, and D.~Slock, ``Location-aided
  mm-wave channel estimation for vehicular communication,'' in
  \emph{SPAWC}.\hskip 1em plus 0.5em minus 0.4em\relax IEEE, 2016, pp. 1--5.

\bibitem{zhang2018wireless}
Z.~Zhang, H.~Pang, A.~Georgiadis, and C.~Cecati, ``Wireless power transfer—an
  overview,'' \emph{IEEE Trans. on Ind. Electron.}, vol.~66, no.~2, pp.
  1044--1058, Feb. 2018.

\bibitem{yan2019comprehensive}
C.~Yan, L.~Fu, J.~Zhang, and J.~Wang, ``A comprehensive survey on {UAV}
  communication channel modeling,'' \emph{IEEE Access}, vol.~7, pp.
  107\,769--107\,792, 2019.

\bibitem{zeng2020uav}
Y.~Zeng, I.~Guvenc, R.~Zhang, G.~Geraci, and D.~W. Matolak, \emph{{UAV}
  Communications for {5G} and Beyond}.\hskip 1em plus 0.5em minus 0.4em\relax
  John Wiley \& Sons, 2020.

\bibitem{CKM_Utilization_Shuowen_UAV}
S.~Zhang and R.~Zhang, ``{Radio map based 3D path planning for
  cellular-connected UAV},'' \emph{IEEE Trans. Wireless Commun.}, vol.~20,
  no.~3, pp. 1975--1989, Mar. 2020.

\bibitem{CKM_Utilization_Yuwei_UAV_RL}
Y.~Huang, X.~Mo, J.~Xu, L.~Qiu, and Y.~Zeng, ``Online maneuver design for
  {UAV}-enabled {NOMA} systems via reinforcement learning,'' \emph{IEEE WCNC},
  pp. 1--6, 2020.

\bibitem{CKM_Utilization_Haoyun_UAV}
H.~Li, P.~Li, J.~Xu, J.~Chen, and Y.~Zeng, ``Channel knowledge map
  {(CKM)}-assisted multi-{UAV} wireless network: {CKM} construction and {UAV}
  placement,'' {A}rXiv pre-print cs.IT/2207.01931.
  https://arxiv.org/abs/2207.01931.

\bibitem{8664604}
W.~Cui, K.~Shen, and W.~Yu, ``Spatial deep learning for wireless scheduling,''
  \emph{IEEE J. on Sel. Areas Commun.}, vol.~37, no.~6, pp. 1248--1261, Jun.
  2019.

\bibitem{Qi2006Loc}
H.~S. Y.~Qi, H.~Kobayashi, ``Analysis of wireless geolocation in a
  non-line-of-sight environment,'' \emph{IEEE Trans. Wireless Commun.}, vol.~5,
  no.~3, pp. 672--681, 2006.

\bibitem{Long2022}
Y.~Long, Y.~Zeng, X.~Xu, and Y.~Huang, ``Environment-aware wireless
  localization enabled by channel knowledge map,'' in \emph{Proc. IEEE
  Globecom}, 2022.

\bibitem{Yapar2020RealtimeLU}
C.~Yapar, R.~Levie, G.~Kutyniok, and G.~Caire, ``Real-time localization using
  radio maps,'' \emph{ArXiv}, vol. abs/2006.05397, 2020.

\bibitem{ShiQiICC}
S.~Zeng, X.~Xu, Y.~Zeng, and F.~Liu, ``{CKM}-assisted {LoS} identification and
  predictive beamforming for cellular-connected {UAV},'' in \emph{Proc. IEEE
  ICC}, 2023.

\bibitem{Xu2023arXiv}
X.~Xu and Y.~Zeng, ``How much data is needed for channel knowledge map
  construction?'' 2023, [Online] Available:
  \url{http://arxiv.org/abs/2312.06966}.

\bibitem{4012}
R.~Ade and P.~Deshmukh, ``Methods for incremental learning: a survey,''
  \emph{International Journal of Data Mining \& Knowledge Management Process},
  vol.~3, no.~4, p. 119, 2013.

\bibitem{4013}
Y.~Engel, S.~Mannor, and R.~Meir, ``The kernel recursive least-squares
  algorithm,'' \emph{IEEE Trans. on signal processing}, vol.~52, no.~8, pp.
  2275--2285, Aug. 2004.

\bibitem{2138}
J.~S. J.~Wang, H. T.~Shen and J.~Ji, ``Hashing for similarity search: A
  survey,'' \emph{available online at https://arxiv.org/abs/1408.2927}, 2014.

\bibitem{4033}
{Z. Dai, D. Wu, Z. Dong, et al}, ``Prototyping and experimental results for
  environment-aware millimeter wave beam alignment via channel knowledge map,''
  \emph{submitted to Trans. Veh. Technol.}, 2023.

\bibitem{DigitalTwin_Niyato}
{L. U. Khan, W. Saad, D. Niyato, Z. Han, and C. S. Hong},
  ``Digital-twin-enabled 6g: Vision, architectural trends, and future
  directions,'' \emph{IEEE Commun. Mag.}, vol.~60, no.~1, pp. 74--80, 2022.

\bibitem{SemanticCom1}
\BIBentryALTinterwordspacing
``{Semantic communications: Principles and challenges}.'' [Online]. Available:
  \url{arXiv preprint arXiv:2201.01389}
\BIBentrySTDinterwordspacing

\bibitem{SemanticCom2}
{Q. Lan, D. Wen, Z. Zhang, Q. Zeng, X. Chen, P. Popovski, and K. Huang}, ``What
  is semantic communication? a view on conveying meaning in the era of machine
  intelligence,'' \emph{J. Commun. Inform. Netw.}, vol.~6, no.~4, pp.
  336–--371, 2021.

\bibitem{SemanticJie}
{J. Ren, Z. Zhang, J. Xu, G. Chen, Y, Sun, P, Zhang, and S. Cui}, ``Knowledge
  base enabled semantic communication: A generative perspective,'' \emph{arXiv
  preprint arXiv:2311.12443}.

\end{thebibliography}

\begin{IEEEbiography}[{\includegraphics[width=1in,height=1.25in,clip,keepaspectratio]{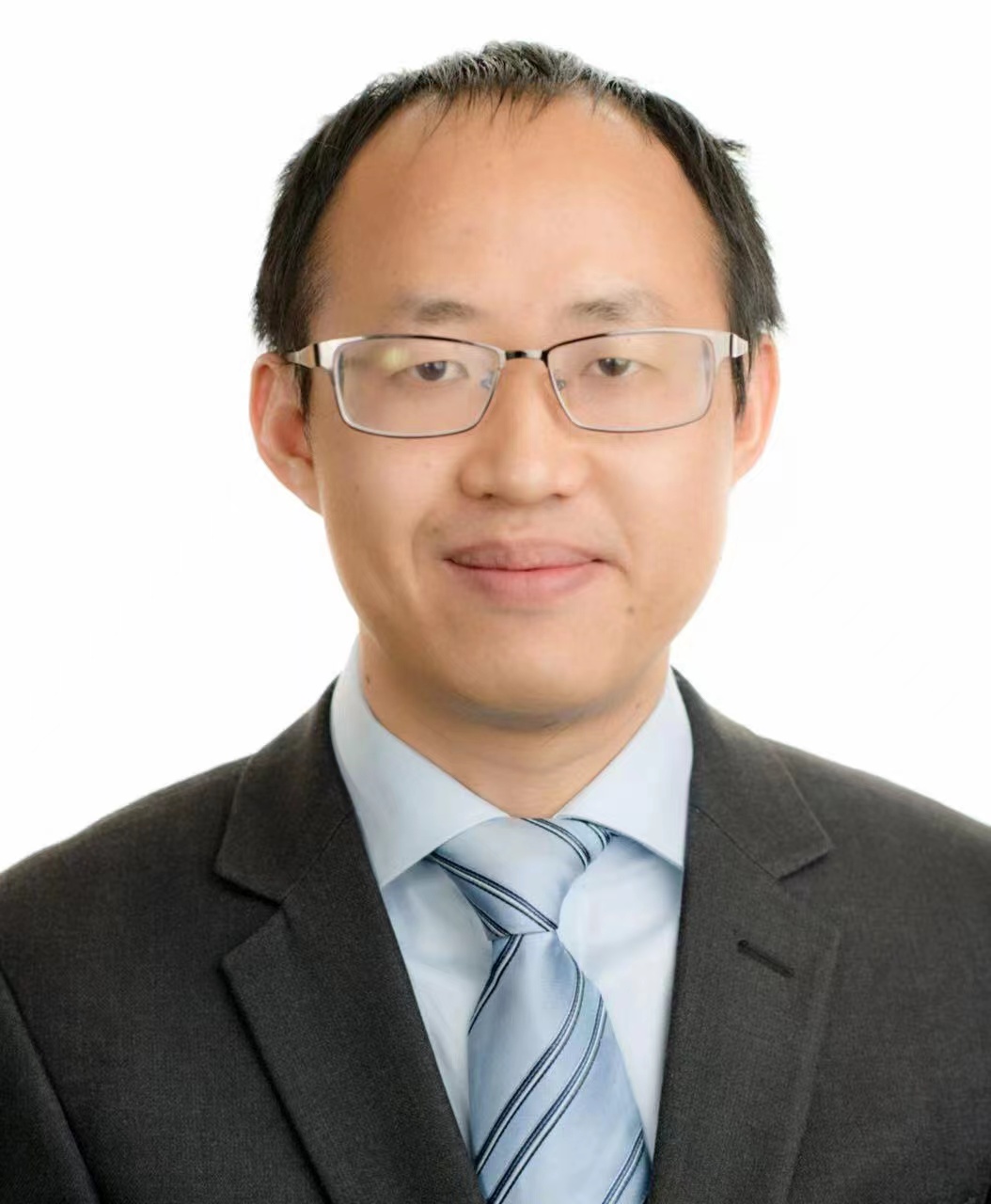}}]{Yong Zeng}
	(S'12-M'14-SM'22) is a Full Professor with the National Mobile Communications Research Laboratory, Southeast University, China, and also with the Purple Mountain Laboratories, Nanjing, China. He received both the Bachelor of Engineering (First-Class Honours) and Ph.D. degrees from Nanyang Technological University, Singapore. From 2013 to 2018, he was a Research Fellow and Senior Research Fellow at the Department of Electrical and Computer Engineering, National University of Singapore. From 2018 to 2019, he was a Lecturer at the School of Electrical and Information Engineering, the University of Sydney, Australia. 
	Dr. Zeng was listed as Highly Cited Researcher by Clarivate Analytics for five consecutive years (2019-2023). He is the recipient of the Australia Research Council (ARC) Discovery Early Career Researcher Award (DECRA), 2020 IEEE Marconi Prize Paper Award in Wireless Communications, 2018 IEEE Communications Society Asia-Pacific Outstanding Young Researcher Award, 2020 \& 2017 IEEE Communications Society Heinrich Hertz Prize Paper Award, 2021 IEEE ICC Best Paper Award, and 2021 China Communications Best Paper Award. He serves as an Editor for IEEE Transactions on Communications, IEEE Communications Letters and IEEE Open Journal of Vehicular Technology, Leading Guest Editor for IEEE Wireless Communications on ``Integrating UAVs into 5G and Beyond'' and China Communications on ``Network-Connected UAV Communications''. He is the Symposium Chair for IEEE Globecom 2021 Track on Aerial Communications, the workshop co-chair for ICC 2018-2023 workshop on UAV communications, the tutorial speaker for Globecom 2018/2019 and ICC 2019 tutorials on UAV communications. He has published more than 160 papers, which have been cited by more than 24,000 times based on Google Scholar.
	
\end{IEEEbiography}

\begin{IEEEbiography}[{\includegraphics[width=1in,height=1.25in,clip,keepaspectratio]{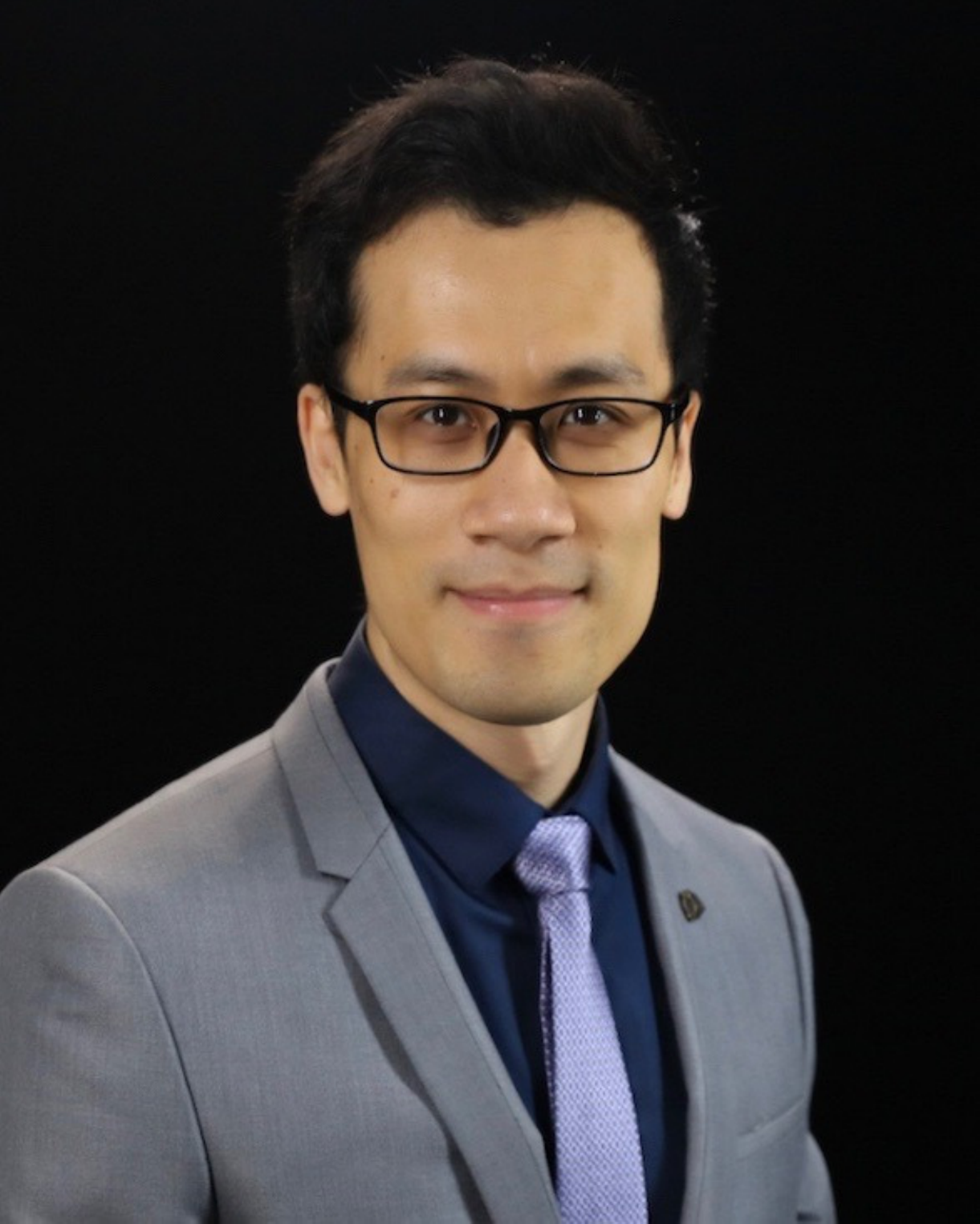}}]{Junting Chen}
	 (S'11--M'16) received the Ph.D.\ degree in Electronic and Computer Engineering from the Hong Kong University of Science and Technology (HKUST), Hong Kong SAR China, in 2015, and the B.Sc.\ degree in Electronic Engineering from Nanjing University, Nanjing, China, in 2009. From 2014--2015, he was a visiting student with the Wireless Information and Network Sciences Laboratory at MIT, Cambridge, MA, USA.  
	
	He is an Assistant Professor with the School of Science and Engineering and the Future Network of Intelligence Institute (FNii) at The Chinese University of Hong Kong, Shenzhen (CUHK--Shenzhen), Guangdong, China. Prior to joining CUHK--Shenzhen, he was a Postdoctoral Research Associate with the Ming Hsieh Department of Electrical Engineering, University of Southern California (USC), Los Angeles, CA, USA, from 2016--2018, and with the Communication Systems Department of EURECOM, Sophia--Antipolis, France, from 2015--2016. His research interests include channel estimation, MIMO beamforming, machine learning, and optimization for wireless communications and localization. His current research focuses on radio map sensing, construction, and application for wireless communications. Dr. Chen was a recipient of the HKTIIT Post-Graduate Excellence Scholarships in 2012. He was nominated as the Exemplary Reviewer of {\scshape IEEE Wireless Communications Letters} in 2018. His paper received the Charles Kao Best Paper Award from WOCC 2022.

\end{IEEEbiography}

\begin{IEEEbiography}[{\includegraphics[width=1in,height=1.25in,clip,keepaspectratio]{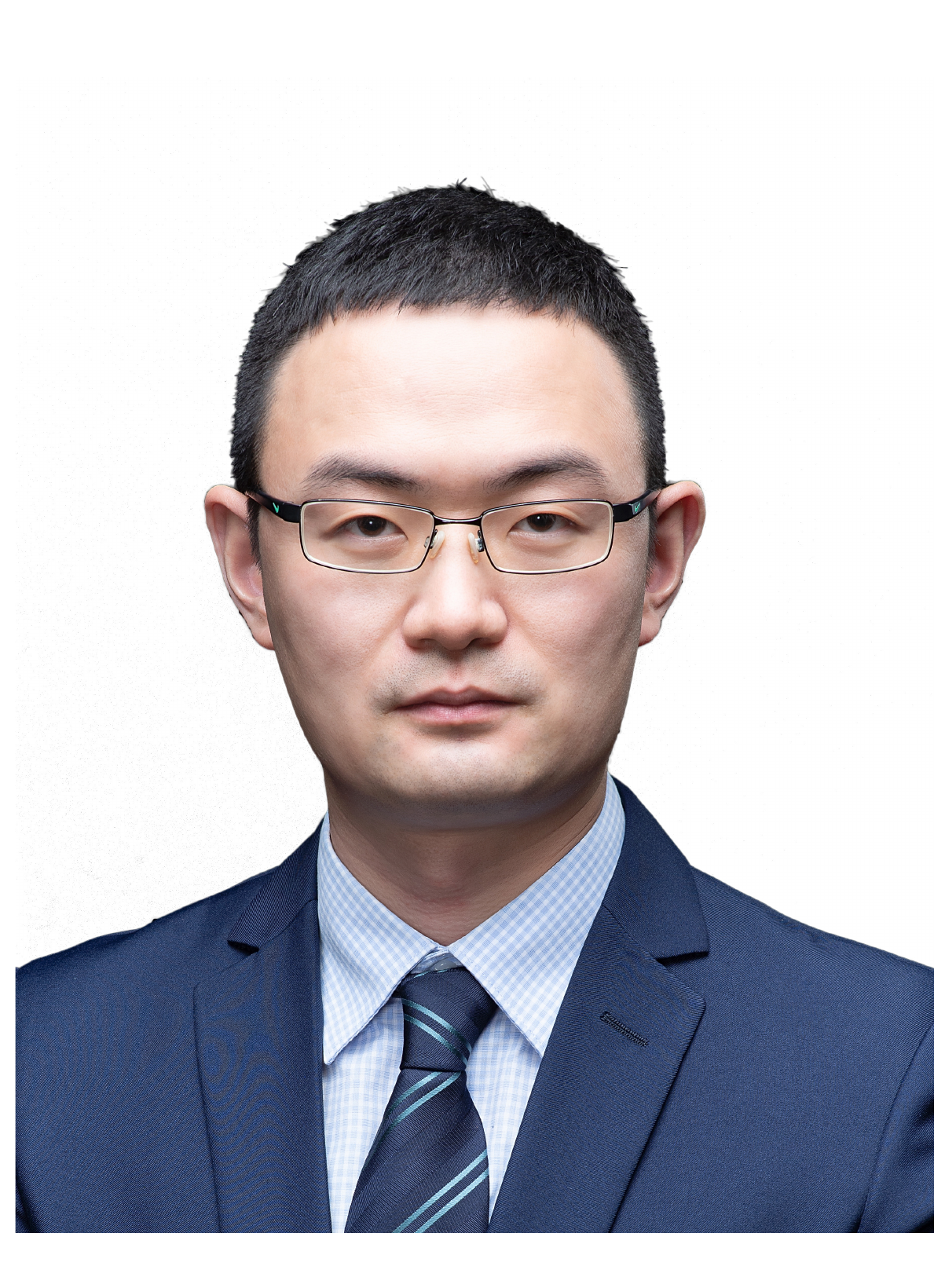}}]{Jie Xu}
(Senior Member, IEEE) received the B.E. and Ph.D. degrees from the University of Science and Technology of China in 2007 and 2012, respectively. From 2012 to 2014, he was a Research Fellow with the Department of Electrical and Computer Engineering, National University of Singapore. From 2015 to 2016, he was a Post-Doctoral Research Fellow with the Engineering Systems and Design Pillar, Singapore University of Technology and Design. From 2016 to 2019, he was a Professor with the School of Information Engineering, Guangdong University of Technology, China. He is currently an Associate Professor (Tenured) with the School of Science and Engineering, The Chinese University of Hong Kong, Shenzhen, China. His research interests include wireless communications, wireless information and power transfer, UAV communications, edge computing and intelligence, and integrated sensing and communication (ISAC). He was a recipient of the 2017 IEEE Signal Processing Society Young Author Best Paper Award, the IEEE/CIC ICCC 2019 Best Paper Award, the 2019 IEEE Communications Society Asia-Pacific Outstanding Young Researcher Award, and the 2019 Wireless Communications Technical Committee Outstanding Young Researcher Award. He is the Symposium Co-Chair of the IEEE GLOBECOM 2019 Wireless Communications Symposium, the workshop co-chair of several IEEE ICC and GLOBECOM workshops, the Tutorial Co-Chair of the IEEE/CIC ICCC 2019, the Vice Chair of the IEEE Wireless Communications Technical Committee (WTC), and the Vice Co-chair of the IEEE Emerging Technology Initiative (ETI) on ISAC. He served or is serving as an Associate Editor-in-Chief of the IEEE Transactions on Mobile Computing, an Editor of the IEEE Transactions on Wireless Communications, IEEE Transactions on Communications, IEEE Wireless Communications Letters, and Journal of Communications and Information Networks, an Associate Editor of IEEE Access, and a Guest Editor of the IEEE Wireless Communications, IEEE Journal on Selected Areas in Communications, IEEE Internet of Things Magazine, Science China Information Sciences, and China Communications. He is a Distinguished Lecturer of IEEE Communications Society.

\end{IEEEbiography}

\begin{IEEEbiography}[{\includegraphics[width=1in,height=1.25in,clip,keepaspectratio]{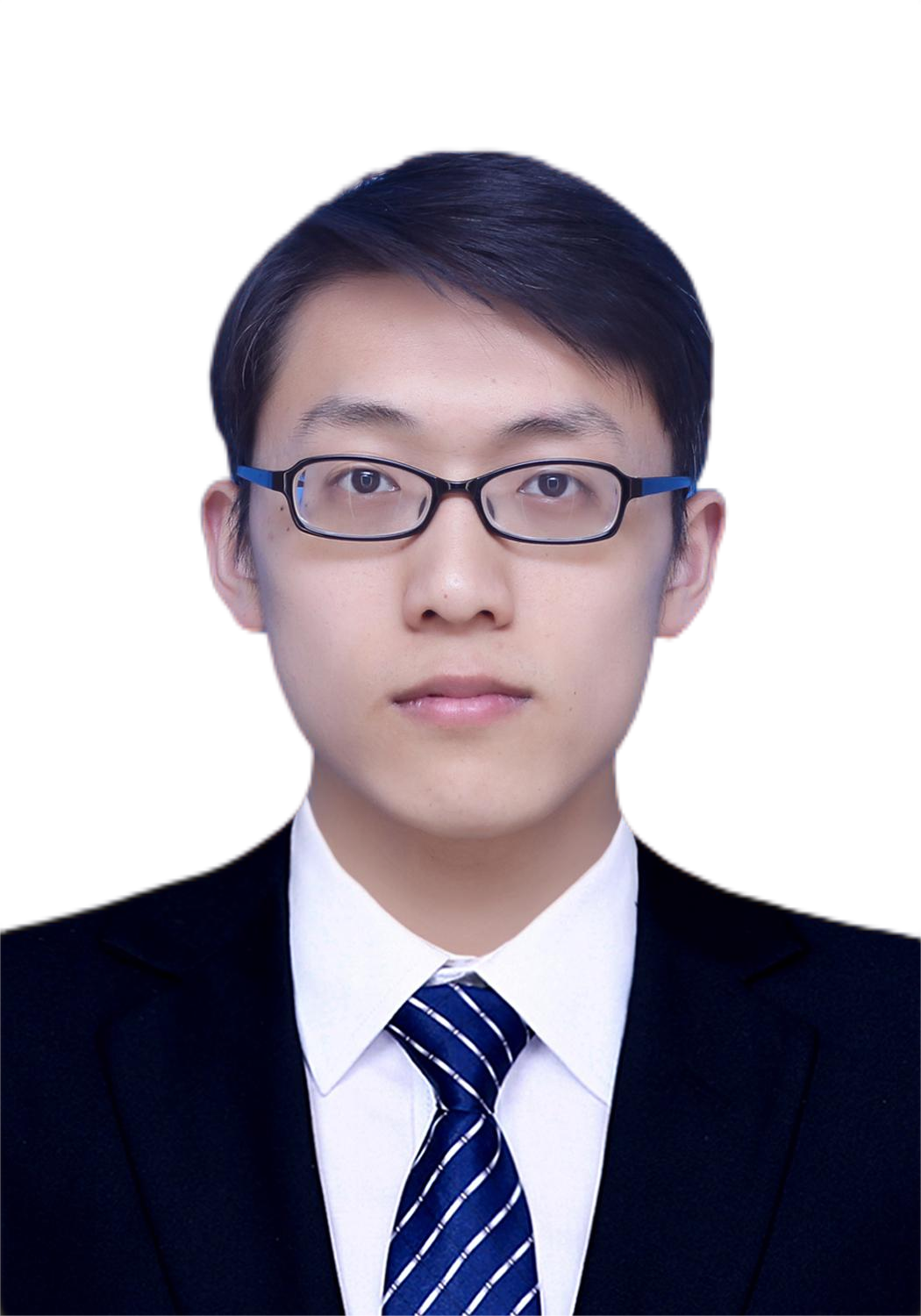}}]{Di Wu}
	received the B.S. degree in communication engineering from Nanjing University of Aeronautics and Astronautics, Nanjing, China, in 2020. He is currently pursuing the Ph.D. degree at the National Mobile Communications Research Laboratory, Southeast University, Nanjing, China. His research interests lies in environment-aware massive MIMO millimeter wave communication via channel knowledge map (CKM).
\end{IEEEbiography}

\begin{IEEEbiography}[{\includegraphics[width=1in,height=1.25in,clip,keepaspectratio]{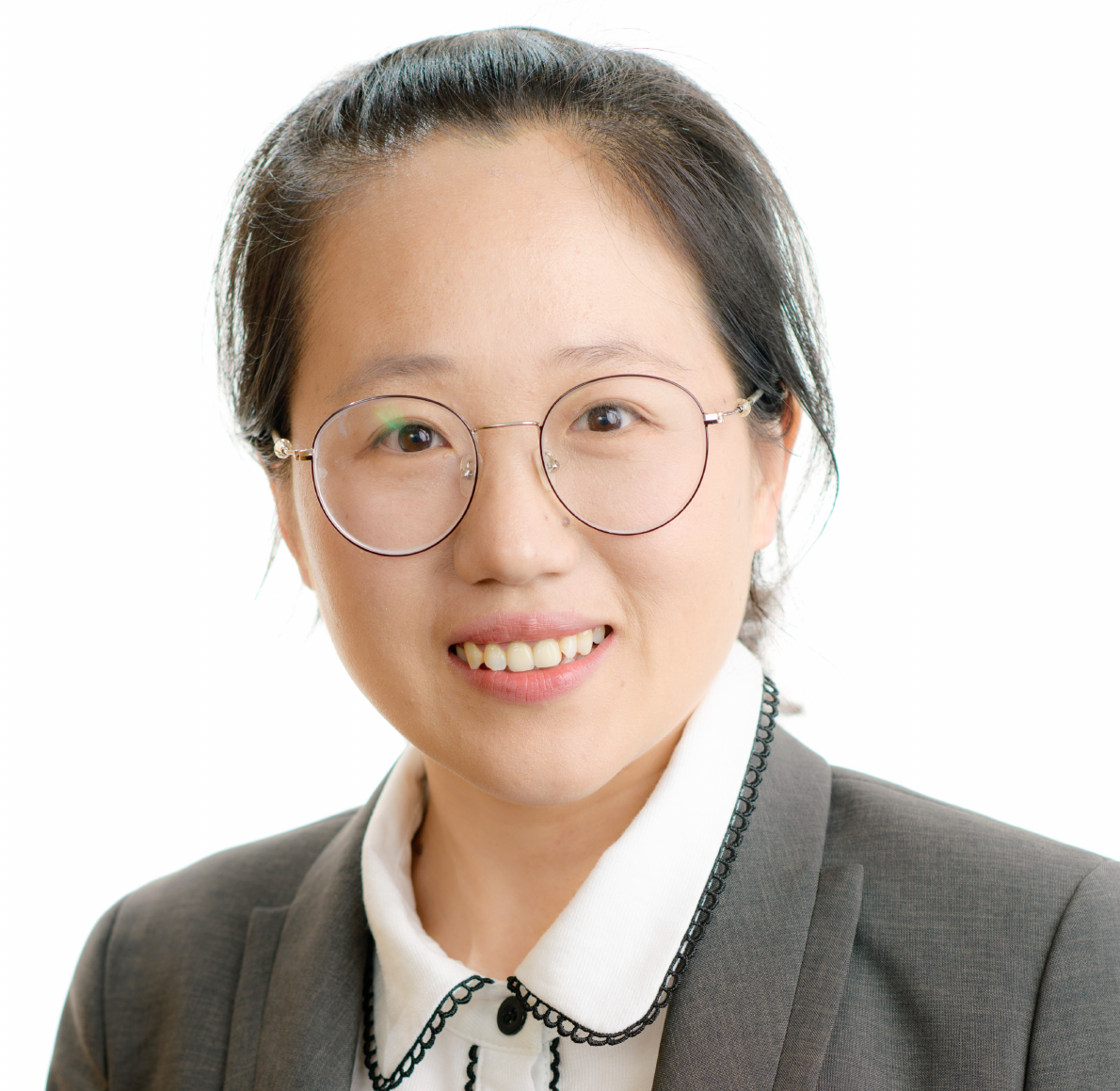}}]{Xiaoli Xu}
	is with the School of Information Science and Engineering, Southeast University, China. She received the Bachelor of Engineering (First-Class Honours) and Ph.D. degrees from Nanyang Technological University, Singapore, in 2009 and 2015, respectively. From 2015 to 2018, she was a Research Fellow at the Nanyang Technological University. From 2018 to 2019, she was a Postdoctoral Research Associate at the University of Sydney, Australia. Her research interests include network coding, semantic communications, and low-latency wireless communications.
\end{IEEEbiography}

\begin{IEEEbiography}[{\includegraphics[width=1in,height=1.25in,clip,keepaspectratio]{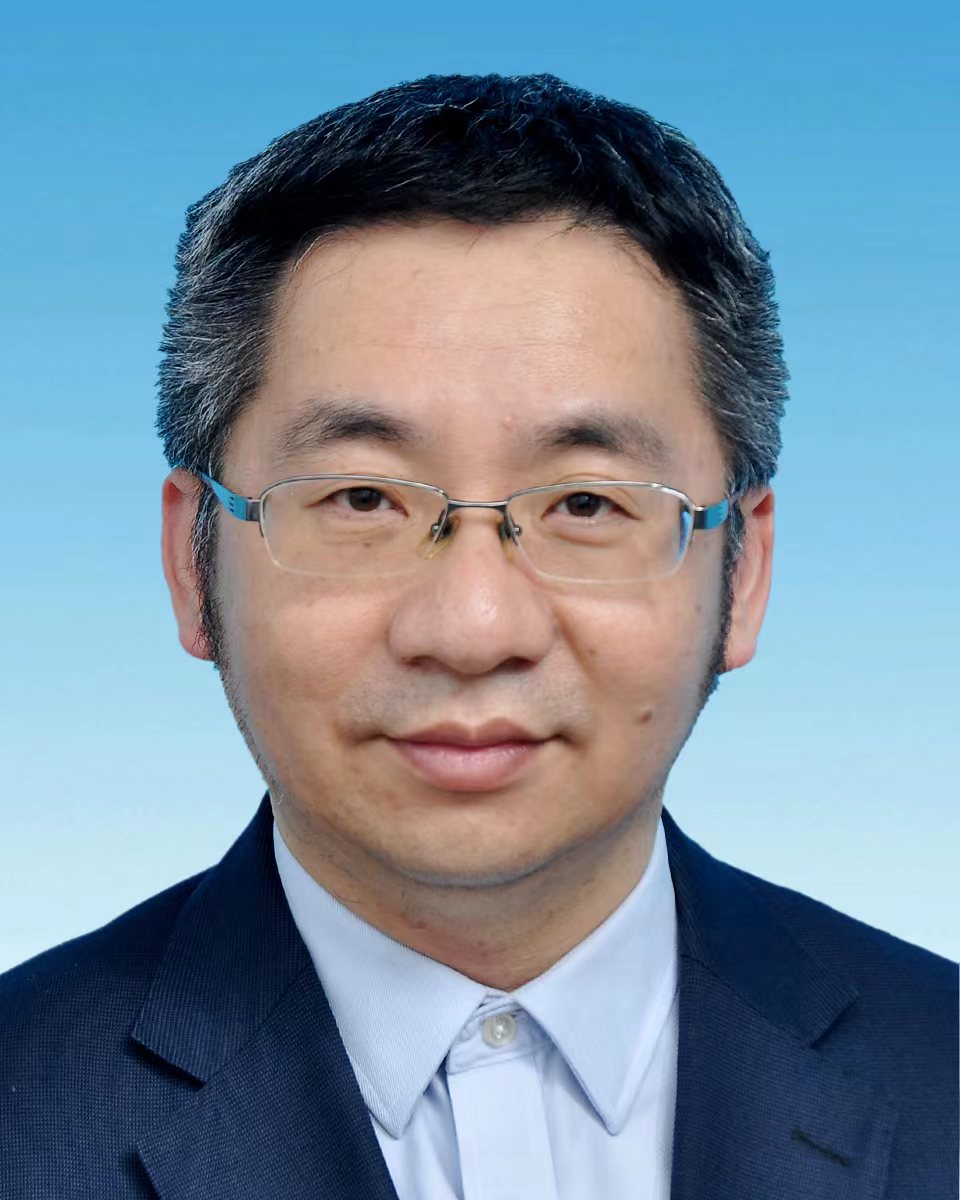}}]{Shi Jin}
	(S'06-M'07-SM'17) received the B.S. degree in communications engineering from Guilin University of Electronic Technology, Guilin, China, in 1996, the M.S. degree from Nanjing University of Posts and Telecommunications, Nanjing, China, in 2003, and the Ph.D. degree in information and communications engineering from Southeast University, Nanjing, in 2007. From June 2007 to October 2009, he was a Research Fellow at the Adastral Park Research Campus, University College London, London, U.K. He is currently affiliated with the faculty of the National Mobile Communications Research Laboratory, Southeast University. His research interests include wireless communications, random matrix theory, and information theory. He serves as an Area Editor for the IEEE Transactions on Communications and IET Electronics Letters. He was previously an Associate Editor for the IEEE Transactions on Wireless Communications, IEEE Communications Letters, and IET Communications. Dr. Jin and his co-authors were awarded the 2011 IEEE Communications Society Stephen O. Rice Prize Paper Award in the field of communication theory, a 2022 Best Paper Award, and a 2010 Young Author Best Paper Award by the IEEE Signal Processing Society.
\end{IEEEbiography}

\begin{IEEEbiography}[{\includegraphics[width=1in,height=1.25in,clip,keepaspectratio]{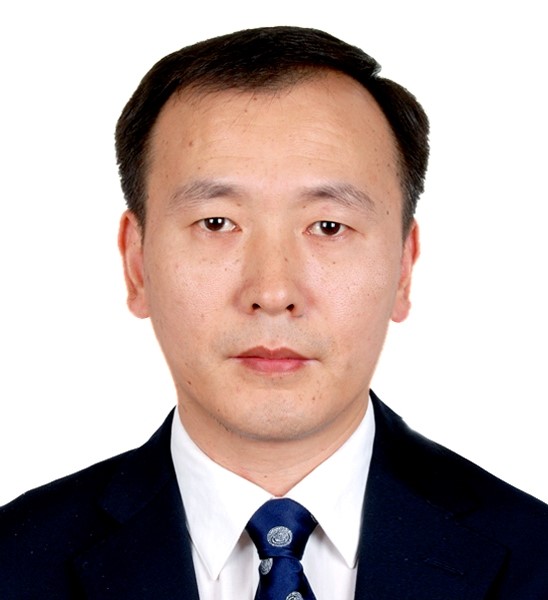}}]{Xiqi Gao}
	 (S'92-AM'96-M'02-SM'07-F'15) received the Ph.D. degree in electrical engineering from Southeast University, Nanjing, China, in 1997.
	He joined the Department of Radio Engineering, Southeast University, in April 1992. Since May 2001, he has been a professor of information systems and communications. From September 1999 to August 2000, he was a visiting scholar at Massachusetts Institute of Technology, Cambridge, and Boston University, Boston, MA. From August 2007 to July 2008, he visited the Darmstadt University of Technology, Darmstadt, Germany, as a Humboldt scholar. His current research interests include broadband multicarrier communications, massive MIMO wireless communications, satellite communications, optical wireless communications, information theory and signal processing for wireless communications. From 2007 to 2012, he served as an Editor for the IEEE Transactions on Wireless Communications. From 2009 to 2013, he served as an Associate Editor for the IEEE Transactions on Signal Processing. From 2015 to 2017, he served as an Editor for the IEEE Transactions on Communications.
	Dr. Gao received the Science and Technology Awards of the State Education Ministry of China in 1998, 2006 and 2009, the National Technological Invention Award of China in 2011, the Science and Technology Award of Jiangsu Province of China in 2014, and the 2011 IEEE Communications Society Stephen O. Rice Prize Paper Award in the field of communications theory.
	
\end{IEEEbiography}

\begin{IEEEbiography}[{\includegraphics[width=1in,height=1.25in,clip,keepaspectratio]{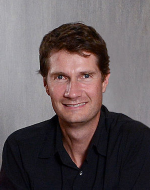}}]{David Gesbert}
	 (Fellow, IEEE) received the Ph.D. degree from the Ecole Nationale Superieure des Telecommunications, France, in 1997. From 1997 to 1999, he was with the Information Systems Labora- tory, Stanford University. He was then a Founding Engineer with Iospan Wireless Inc. and a Stanford spin off pioneering MIMO-OFDM (Intel). Before joining EURECOM in 2004, he was with the Depart- ment of Informatics, University of Oslo, as an Adjunct Professor. Since 2015, he holds the ERC Advanced grant “PERFUME” on the topic of smart device Communications in future wireless networks. He has published about 350 articles and 25 patents, some of them winning the 2019 ICC Best Paper Award, the 2015 IEEE Best Tutorial Paper Award (Communications Society), the 2012 SPS Signal Processing Magazine Best Paper Award, the 2004 IEEE Best Tutorial Paper Award (Communications Society), the 2005 Young Author Best Paper Award for Signal Processing Society journals, and paper awards at conferences 2011 IEEE SPAWC, 2004 ACM MSWiM. He is a Board Member of the OpenAirInterface (OAI) Software Alliance. He has been a Technical Program Co-chair of ICC2017. He was named a Thomson-Reuters Highly Cited Researchers in Computer Science. Since early 2019, he heads the Huawei-funded Chair on Adwanced Wireless Systems Towards 6G Networks. He sits on the Advisory Board of HUAWEI European Research Institute. In 2020, he was awarded funding by the French Interdisciplinary Institute on Artificial Intelligence for a Chair in the area of AI for the future IoT. 
\end{IEEEbiography}

\begin{IEEEbiography}[{\includegraphics[width=1in,height=1.25in,clip,keepaspectratio]{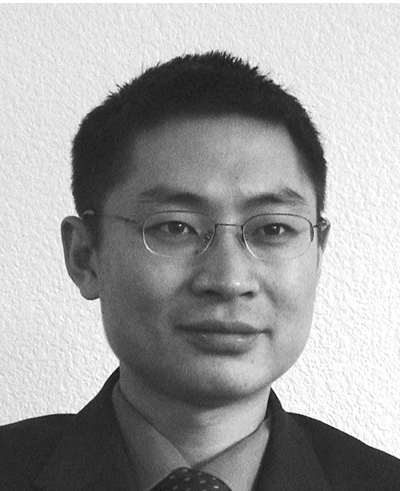}}]{Shuguang Cui}
	 (S'99-M'05-SM'12-F'14) received his Ph.D in Electrical Engineering from Stanford University, California, USA, in 2005. Afterwards, he has been working as assistant, associate, full, Chair Professor in Electrical and Computer Engineering at the Univ. of Arizona, Texas A\&M University, UC Davis, and CUHK at Shenzhen respectively. He has also served as the Executive Dean for the School of Science and Engineering at CUHK, Shenzhen, the Executive Vice Director at Shenzhen Research Institute of Big Data, and the Director for Future Network of Intelligence Institute (FNii). His current research interests focus on the merging between AI and communication neworks. He was selected as the Thomson Reuters Highly Cited Researcher and listed in the Worlds’ Most Influential Scientific Minds by ScienceWatch in 2014. He was the recipient of the IEEE Signal Processing Society 2012 Best Paper Award. He has served as the general co-chair and TPC co-chairs for many IEEE conferences. He has also been serving as the area editor for IEEE Signal Processing Magazine, and associate editors for IEEE Transactions on Big Data, IEEE Transactions on Signal Processing, IEEE JSAC Series on Green Communications and Networking, and IEEE Transactions on Wireless Communications. He has been the elected member for IEEE Signal Processing Society SPCOM Technical Committee (2009~2014) and the elected Chair for IEEE ComSoc Wireless Technical Committee (2017~2018). He is a member of the Steering Committee for IEEE Transactions on Big Data and the Chair of the Steering Committee for IEEE Transactions on Cognitive Communications and Networking. He is also the Vice Chair of the IEEE VT Fellow Evaluation Committee and a member of the IEEE ComSoc Award Committee. He was elected as an IEEE Fellow in 2013, an IEEE ComSoc Distinguished Lecturer in 2014, and IEEE VT Society Distinguished Lecturer in 2019. In 2020, he won the IEEE ICC best paper award, ICIP best paper finalist, the IEEE Globecom best paper award. In 2021, he won the IEEE WCNC best paper award. In 2023, he won the IEEE Marconi Best Paper Award, got elected as a Fellow of  both Canadian Academy of Engineering and the Royal Society of Canada, and starts to serve as the Editor-in-Chief for IEEE Transactions on Mobile Computing. 
\end{IEEEbiography}

\begin{IEEEbiography}[{\includegraphics[width=1in,height=1.25in,clip,keepaspectratio]{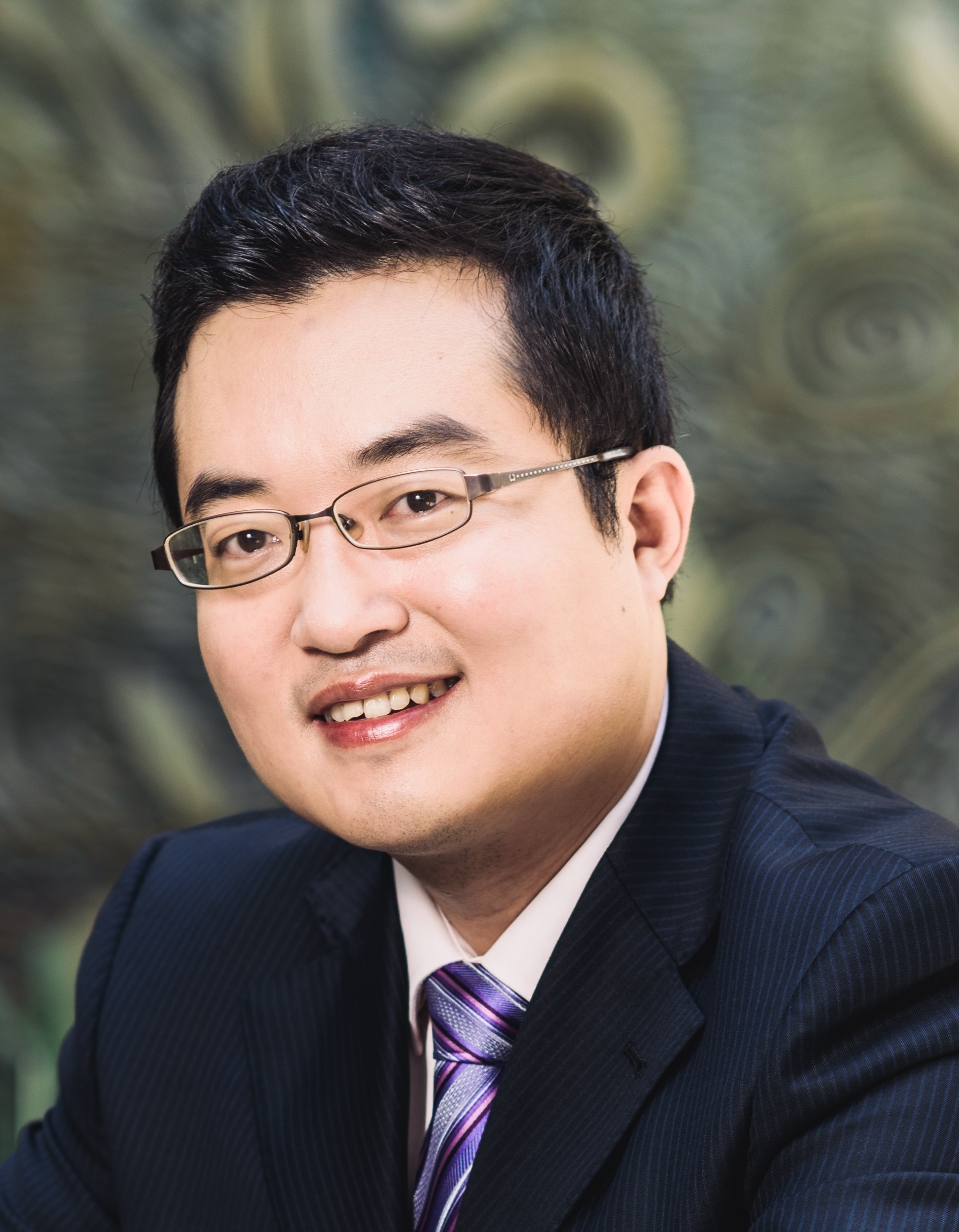}}]{Rui Zhang}
	(S'00-M'07-SM'15-F'17) received the B.Eng. (first-class Hons.) and M.Eng. degrees from the National University of Singapore, Singapore, and the Ph.D. degree from the Stanford University, Stanford, CA, USA, all in electrical engineering.
	From 2007 to 2009, he worked as a researcher at the Institute for Infocomm Research, ASTAR, Singapore. In 2010, he joined the Department of Electrical and Computer Engineering of National University of Singapore, where he was appointed as a Provost’s Chair Professor in 2020. Since 2022, he has joined the School of Science and Engineering, The Chinese University of Hong Kong, Shenzhen, as a Principal's Diligence Chair Professor. He has published over 300 journal papers and over 200 conference papers. He has been listed as a Highly Cited Researcher by Thomson Reuters/Clarivate Analytics since 2015. His current research interests include UAV/satellite communications, wireless power transfer, intelligent reflecting surface, reconfigurable MIMO, integrated sensing and communications.
	He was the recipient of the 6th IEEE Communications Society Asia-Pacific Region Best Young Researcher Award in 2011, the Young Researcher Award of National University of Singapore in 2015, the Wireless Communications Technical Committee Recognition Award in 2020, and the IEEE Signal Processing and Computing for Communications (SPCC) Technical Recognition Award in 2020. His coauthored papers have received 17 IEEE Best Journal Paper Awards, including the IEEE Marconi Prize Paper Award in Wireless Communications in 2015 and 2020, the IEEE Signal Processing Society Best Paper Award in 2016, the IEEE Communications Society Heinrich Hertz Prize Paper Award in 2017, 2020 and 2022, the IEEE Communications Society Stephen O. Rice Prize in 2021, etc. He served for over 30 international conferences as the TPC co-chair or an organizing committee member. He was an elected member of the IEEE Signal Processing Society SPCOM Technical Committee from 2012 to 2017 and SAM Technical Committee from 2013 to 2015, and served as the Vice Chair of the IEEE Communications Society Asia-Pacific Board Technical Affairs Committee from 2014 to 2015. He was a Distinguished Lecturer of IEEE Signal Processing Society and IEEE Communications Society from 2019 to 2020. He served as an Editor for the IEEE TRANSACTIONS ON WIRELESS COMMUNICATIONS from 2012 to 2016, the IEEE JOURNAL ON SELECTED AREAS IN COMMUNICATIONS: Green Communications and Networking Series from 2015 to 2016, the IEEE TRANSACTIONS ON SIGNAL PROCESSING from 2013 to 2017, the IEEE TRANSACTIONS ON GREEN COMMUNICATIONS AND NETWORKING from 2016 to 2020, and the IEEE TRANSACTIONS ON COMMUNICATIONS from 2017 to 2022. He served as a member of the Steering Committee of the IEEE Wireless Communications Letters from 2018 to 2021. He is a Fellow of the Academy of Engineering Singapore.
	
\end{IEEEbiography}

\end{document}